# RAY VELOCITY DERIVATIVES IN ANISOTROPIC ELASTIC MEDIA.

# PART II – POLAR ANISOTROPY

*Igor Ravve (corresponding author) and Zvi Koren, Emerson*

igor.ravve@emerson.com , zvi.koren@emerson.com

## ABSTRACT

In Part I of this study, we obtained the ray (group) velocity gradients and Hessians with respect to the ray locations, directions and the anisotropic model parameters, at nodal points along ray trajectories, considering general anisotropic (triclinic) media and both, quasi-compressional and quasi-shear waves. Ray velocity derivatives for anisotropic media with higher symmetries were considered particular cases of general anisotropy. In this part, Part II, we follow the computational workflow presented in Part I, formulating the ray velocity derivatives directly for polar anisotropic (transverse isotropy with tilted axis of symmetry, TTI) media for the coupled qP and qSV waves and for SH waves. The acoustic approximation for qP waves is considered a special case. The medium properties, normally specified at regular three-dimensional fine grid points, are the five material parameters: the axial compressional and shear velocities and the three Thomsen parameters, and two geometric parameters: the polar angles defining the local direction of the medium symmetry axis. All the parameters are assumed spatially (smoothly) varying, where their gradients and Hessians can be reliably computed. Two case examples are considered; the first represents compacted shale/sand rocks (with positive anellipticity) and the second, unconsolidated sand rocks with strong negative anellipticity (manifesting a qSV triplication). The ray velocity derivatives obtained in this part are first tested by comparing them with the corresponding numerical (finite difference) derivatives. Additionally, we show that exactly the same results (ray velocity derivatives) can be obtained if we transform the given polar anisotropic model parameters (five material and two geometric) into the twenty-one stiffness tensor components of a general anisotropic (triclinic) medium, and apply the theory derived in Part I. Since in many practical wave/ray-based applications in polar anisotropic media only the spatial derivatives of the axial compressional velocity are taken into account, we analyze the effect (sensitivity) of the spatial derivatives of the other parameters on the ray velocity and its derivatives (which, in turn, define the corresponding traveltime derivatives along the ray).

## 1. INTRODUCTION

Considering general anisotropic elastic media parametrized with the spatially varying stiffness properties and all wave types, in Part I of this study we presented a generic approach for explicitly (analytically) formulating the first and second partial derivatives of the ray velocity magnitude, $v_{\text{ray}}\left[\mathbf{C}(\mathbf{x}),\mathbf{r}\right]$, with respect to (wrt) the ray location vector $\mathbf{x}$, direction vector $\mathbf{r}$, and the components of the stiffness matrices $\mathbf{C}(\mathbf{x})$. These gradients and Hessians are the corner stones of the sensitivity kernels of traveltimes and amplitudes of waves/rays in seismic modeling and inversion methods. In this part of the study, we apply the theory presented in Part I explicitly for polar anisotropic (transverse isotropy with tilted axis of symmetry, TTI) media, which considerably simplifies the implementations and improves the computational performance. Note that a similar approach can be further applied to lower anisotropic symmetry materials such as tilted orthorhombic and tilted monoclinic, but these topics are beyond the scope of this paper, to be further developed in our future research.

Starting with the kinematic and dynamic variational formulations for stationary rays, in Part I we introduced the arclength-related Lagrangian $L[\mathbf{C}(\mathbf{x}),\mathbf{r}]$, its corresponding (e.g., through the Legendre transform) arclength-related Hamiltonian $H[\mathbf{C}(\mathbf{x}),\mathbf{p}]$, and an (auxiliary) time-scaled (reference) Hamiltonian $H^{\bar{\tau}}[\mathbf{C}(\mathbf{x}),\mathbf{p}]$, which are used in chain rule operations for computing the above mentioned gradients and Hessians of the ray velocity magnitudes. The slowness vector $\mathbf{p}$ implicitly depends on the ray coordinates $\mathbf{x}$ and the ray direction $\mathbf{r}$. Hence, the workflow for computing the ray velocity derivatives includes four basic steps: a) performing the slowness inversion, implicitly stated as $\mathbf{p}=\mathbf{p}[\mathbf{C}(\mathbf{x}),\mathbf{r}]$, b) computing the derivatives of the reference Hamiltonian $H^{\bar{\tau}}(\mathbf{x},\mathbf{p})$, c) computing the derivatives of the arclength-related Hamiltonian $H(\mathbf{x},\mathbf{p})$, and d) computing the ray velocity derivatives.

All four steps have been elaborated for general anisotropic media (up to triclinic) in Part I of this study, where the higher symmetries have been considered particular cases. In this part, Part II, we propose an approach, where we directly adjust the equations for the specific anisotropic symmetry. The beauty of the proposed approach is that the last two steps in the above-mentioned workflow are generic and remain the same in both approaches, as the computational formulae related to these steps are independent of any particular anisotropic symmetry. Only the first two steps, the slowness inversion and the computation of the derivatives of the reference Hamiltonian, should be adjusted to the specific anisotropic medium type under consideration, and this adjustment essentially improves the implementation performance.

There are eight types of anisotropic symmetries in continuous elastic materials (e.g., Cowin and Mehrabadi, 1985; Bona et al., 2004, 2007, 2008), including the isotropic and triclinic media. In this study we consider polar anisotropic (TTI) model which is commonly used in seismic methods for approximating wave propagation in realistic structural sedimentary (e.g., shale and sand; compacted or unconsolidated) layers.

The slowness inversion (i.e., obtaining the set of feasible slowness vectors from a given unique ray direction vector) in anisotropic media has been studied by several researchers (e.g., Musgrave, 1954a, 1954b, 1970; Fedorov, 1968; Helbig, 1994; Vavryčuk, 2006, Grechka, 2017; Zhang & Zhou, 2018). A brief overview of these studies is provided in the first part of Appendix A. Ravve and Koren (2021b) proposed a new efficient approach to the slowness inversion in polar anisotropy based on the inherent property of this symmetry class, where the three vectors: the phase velocity (or the slowness), the ray velocity and the medium axis of symmetry are coplanar. This leads to a system of two bivariate polynomials of degrees 3 and 4, and under the commonly accepted assumption that the angle between the slowness and the ray direction vectors cannot exceed $90^o$, the system is then reduced to a univariate polynomial of degree six. The method is briefly described in the second part of Appendix A.

To the best of our knowledge, unlike the case of computing the ray velocity (first) derivatives wrt the model parameters (e.g., the Fréchet first derivatives used in seismic tomography), there are very few studies related to the (first and second) spatial and directional derivatives of the ray velocity magnitude of seismic body waves (e.g., for two-point ray bending methods), and in none of them the derivatives are presented directly in the ray (group) angle domain. These types of derivatives are required, for example, for the implementation of the Eigenray method (a special ray bending method) recently presented by Koren and Ravve (2021) and Ravve and Koren (2021a).

For example, in the context of ray-based tomography in polar anisotropic media, Zhou and Greenhalgh (2008) suggested a method to compute the ray velocity derivatives wrt the medium properties and the polar angles of the symmetry axis. To bypass the challenges of directly computing the ray velocity derivatives wrt the model parameters, they use the relation of the ray velocity components to the phase velocity magnitude $v_{\text{phs}}$ (that depends on the phase angle $\vartheta_{\text{phs}}$ between the slowness vector and the medium symmetry axis), and its derivative wrt the phase angle, $v'_{\text{phs}} = dv_{\text{phs}}/d\theta_{\text{phs}}$. The axial and normal components of the ray velocity, $v_{\text{ray}}^{\text{ax}}$ and $v_{\text{ray}}^{\text{nr}}$, respectively, are given by (e.g., Crampin, 1981; Tsvankin, 2012),

$$v_{\text{ray}}^{\text{nr}} = v_{\text{phs}} \sin \vartheta_{\text{phs}} + v'_{\text{phs}} \cos \vartheta_{\text{phs}} \quad,$$
$$v_{\text{ray}}^{\text{ax}} = v_{\text{phs}} \cos \vartheta_{\text{phs}} - v'_{\text{phs}} \sin \vartheta_{\text{phs}} \quad, \qquad v_{\text{ray}} = \sqrt{v_{\text{phs}}^2 + v'^2_{\text{phs}}} \quad . \qquad (1.1)$$

The phase velocity magnitude, in turn, depends on the medium properties and the phase angle, and may be formulated for the coupled qP and qSV waves as (e.g., Tsvankin, 2012),

$$\frac{v_{\text{phs}}^2(\vartheta_{\text{phs}})}{v_P^2} = 1 - \frac{f}{2} + \varepsilon \sin^2 \vartheta_{\text{phs}} \pm \frac{1}{2}\sqrt{\left(f + 2\varepsilon \sin^2 \vartheta_{\text{phs}}\right)^2 - 2f(\varepsilon - \delta)\sin^2 2\vartheta_{\text{phs}}} \quad , \qquad (1.2)$$

where $v_P$ is the axial compressional wave velocity, $f = 1 - v_S^2/v_P^2$ is the shear wave velocity factor, with $v_S$ the axial shear wave velocity, and $\delta$ and $\varepsilon$ the Thomsen polar anisotropic parameters. Note that the $(+)$ upper sign in equation 1.2 is related to qP waves and the $(-)$ lower sign to qSV waves.

In principle, one can use equations 1.1 and 1.2 to further compute the ray velocity derivatives wrt the model parameters, via the phase velocity and its phase-angle derivative, applying a chain rule.

However, in this case, the ray velocity derivatives become functions of the phase (slowness) direction rather than the required ray direction. Our approach is different; it includes several important new items (aspects) with definite advantages. First, the derivatives of the ray velocity are computed directly in the ray angle or ray direction domain; second, we provide the first and second derivatives of the ray velocity magnitude not only wrt the model properties $\mathbf{m}$, but also wrt the ray locations and directions $\mathbf{x}$ and $\mathbf{r}$, respectively; third, we provide a complete set of the velocity gradients and Hessians, including the mixed location/direction Hessian, which are essential, for example, for kinematic and dynamic two-point ray bending methods (e.g., the Eigenray method) and for ray-based tomography. Finally, all the derivatives are formulated in the global frame (Cartesian coordinate system) without the need to apply any global-to-local or local-to-global rotations.

In summary, given the anisotropic model parameters $\mathbf{m}(\mathbf{x})$ and a ray direction unit vector $\mathbf{r}$, the following objects are computed,

$$\begin{array}{ll} \mathbf{p}, v_{\text{ray}} & \text{slowness vector and ray velocity magnitude} \\ \nabla_{\mathbf{x}} v_{\text{ray}}, \nabla_{\mathbf{r}} v_{\text{ray}}, \nabla_{\mathbf{m}} v_{\text{ray}} & \text{gradients} \\ \nabla_{\mathbf{x}} \nabla_{\mathbf{x}} v_{\text{ray}}, \nabla_{\mathbf{r}} \nabla_{\mathbf{r}} v_{\text{ray}}, \underbrace{\nabla_{\mathbf{x}} \nabla_{\mathbf{r}} v_{\text{ray}} = \left( \nabla_{\mathbf{r}} \nabla_{\mathbf{x}} v_{\text{ray}} \right)^T}_{\text{mixed}}, \nabla_{\mathbf{m}} \nabla_{\mathbf{m}} v_{\text{ray}} & \text{Hessians} \end{array} \quad (1.3)$$

We start with the derivation (definition) of the reference Hamiltonians for the coupled qP-qSV waves, and for SH waves. We then discuss the sign of the Hamiltonian for each wave type; this may be considered as a continuation of the sign-rule discussion for general anisotropy provided in Part I. The next sections are devoted to the gradient and Hessian of the reference Hamiltonian wrt the model properties, $H_{\mathbf{m}}^{\bar{\tau}}$ and $H_{\mathbf{mm}}^{\bar{\tau}}$, and then to the spatial and slowness gradients and Hessians of the reference Hamiltonian, $H_{\mathbf{x}}^{\bar{\tau}}, H_{\mathbf{p}}^{\bar{\tau}}, H_{\mathbf{xx}}^{\bar{\tau}}, H_{\mathbf{pp}}^{\bar{\tau}}$, including the mixed Hessian, $H_{\mathbf{px}}^{\bar{\tau}}$. We then formulate the relations between these derivatives and the corresponding derivatives of the arclength-related Hamiltonian, leading eventually to the required derivatives of the ray velocity magnitude.

Any high symmetry medium (i.e., any anisotropy excluding triclinic media), can be defined in its "crystal" reference frame (a term suggested by Pšenčík and Farra, 2017; referred to also as a local frame). Generally, this frame is defined by the three Euler angles: zenith, azimuth and spin, or (in some cases preferably) by the quaternion rotations: Given the three Euler angles, one can compute the rotational matrix, or an equivalent single rotation angle about the tilted axis, which, in turn, yields the quaternion components. The inverse solution provides the Euler angles given the quaternion components. Overall, the Euler rotations result in trigonometric equations, which can lead to a "gimbal" lock, where the quaternion rotations result in algebraic equations, with no such issues. However, the polar anisotropy reference frame is insensitive to the spin rotation (about the crystal symmetry axis), and therefore, only the two angles, zenith and azimuth, suffice. Note that there are different kinds of material parameterizations for a polar anisotropic material in its crystal frame. In addition to the five independent stiffness tensor components, the most known and widely used are the Thomsen (1986) parameterization, Alkhalifah (1998), and the weak anisotropy (WA) parameterization (e.g., Farra and Pšenčík, 2016; Farra et al., 2016).

In this study, the polar anisotropic material parameters include the axial compressional velocity $v_P$, parameter $f = 1 - v_S^2 / v_P^2$ (e.g., Tsvankin, 2012), where $v_S$ is the axial shear velocity, and the three Thomsen parameters, $\delta, \varepsilon, \gamma$. Note that out of the five material properties, only four parameters, $v_P, f, \delta, \varepsilon$ are needed to describe the coupled qP and qSV waves, and three parameters, $v_P, \delta, \varepsilon$ (with $f = 1$) are needed for the acoustic approximation of qP waves. For SH waves, it is suitable to use the axial shear velocity $v_S$ directly, as a single parameter, rather than as a combination of two parameters, $v_S = v_P\sqrt{1-f}$. Therefore, only two material properties, $v_S$ and $\gamma$, are needed to describe SH waves.

Hence, in our study, the properties of the polar anisotropic (TTI) model (combined into array **m**) are divided into two groups: Five material (elastic) properties of the model (combined into array **c**, which includes in addition to $v_P$ and $v_S$, the three Thomsen parameters $\delta, \varepsilon, \gamma$), and two geometric properties: the two angles, $\theta_{ax}$ and $\psi_{ax}$ (combined into array **a**), that define the direction of the symmetry axis. It is convenient to present the symmetry axis direction as a unit-length vector **k**; obviously, vector **k** of length 3 and array **a** of length 2 are dependent. Vector **k** is an auxiliary one, and its introduction simplifies the derivations. Unlike arrays **c**, **a** and **m** (where $\mathbf{m} = \mathbf{c} \| \mathbf{a}$ and symbol $\|$ means concatenation of the two arrays; in this case with lengths $n_\mathbf{c} = 5$ and $n_\mathbf{a} = 2$), which are just collections of scalar values, the symmetry axis direction **k** is a real physical vector that allows vector operations. Each of the spatially varying model parameters of array **m** is normally given on a 3D regular fine grid and it is assumed that their spatial distributions are smooth enough, such that their spatial gradients and Hessians can be reliably (numerically) computed. We also need to compute the gradient and the Hessian of the reference Hamiltonian wrt the axis of symmetry direction vector, $H_\mathbf{k}^{\bar{\tau}}$ and $H_\mathbf{kk}^{\bar{\tau}}$; the two mixed gradients related to the symmetry axis direction are also required, $H_\mathbf{mk}^{\bar{\tau}}$ and $H_\mathbf{pk}^{\bar{\tau}}$. Note that a derivative wrt vector **k** means a set of partial derivatives wrt its three Cartesian components.

For completeness, we also consider the particular (and widely used) case of acoustic approximation (AC) for qP waves and provide the corresponding ray velocity derivatives for this case. We show that although the accuracy of the AC ray velocity magnitude is very high (even in relatively strong polar anisotropic media), the accuracy of the above-mentioned derivatives (in particular, the second derivatives) based on the AC may become questionable in some extreme cases.

Remark: Since in this part of the study we work mostly with the reference Hamiltonian $H^{\bar{\tau}}\left[\mathbf{m}(\mathbf{x}),\mathbf{p}\right]$, the word "reference" will be omitted, with a few exceptions when both the reference and the arclength-related Hamiltonians are considered in the same paragraph.

The final part of this study is dedicated to the actual (analytical and numerical) computations of the above-mentioned ray velocity derivatives. We use two real case examples of a point along a ray traveling in an inhomogeneous polar anisotropic model (where in both examples, the material parameters and the symmetry axis direction angles vary smoothly in space), representing compacted shale/sand sediments (referred to as model 1) and unconsolidated shallow sand rocks (model 2), where in the latter case, for the given ray direction, the slowness inversion yields a shear wave triplication. The ray velocity derivatives are obtained for the coupled qP and qSV waves (quasi compressional and quasi-shear waves, respectively, polarized in the axial symmetry plane), and independently for the pure shear waves polarized in the direction normal to the axial symmetry plane (SH; in the isotropic plane). Additionally, we formulate and compute the acoustic approximation for the qP ray velocity derivatives and discuss their accuracy (validity). The resulted analytic derivatives for all cases are first compared with the corresponding numerical (finite-difference) derivatives. To demonstrate the compatibility of the two approaches (of Part I and II), we show that the exact analytic derivatives can be also obtained when considering the polar anisotropic model as a particular case of general anisotropy. This requires first converting the parameters of polar anisotropy with the tilted symmetry axis to the twenty-one stiffness components of general (triclinic) anisotropy, and then applying the generic formulae of Part I.

Note that in many practical wave/ray-based applications in polar anisotropic (TTI) media, the spatial derivatives of the Thomsen parameters, the ratio between the axial shear and compressional wave velocities (parameter $f$), and/or those of the symmetry axis direction angles are ignored, and only the derivatives of the axial compressional velocity are accounted for. In our test, we analyze the relative contribution of the model parameters' spatial derivatives to the ray velocity gradients and Hessians.

## 2. HAMILTONIANS OF POLAR ANISOTROPY FOR ALL WAVE TYPES

The crystal frame $(x_1\ x_2\ x_3)$ is a local coordinate frame rotated wrt the global frame $(x\ y\ z)$ by the polar angles of the symmetry axis: zenith $\theta_{ax}$ and azimuth $\psi_{ax}$. In the crystal frame, where the local vertical axis is the medium axis of symmetry, the stiffness matrix of the polar anisotropic elastic material has five independent components,

$$\mathbf{C} = \begin{bmatrix} C_{11} & C_{12} & C_{13} & & & \\ C_{12} & C_{11} & C_{13} & & & \\ C_{13} & C_{13} & C_{33} & & & \\ & & & C_{44} & & \\ & & & & C_{44} & \\ & & & & & C_{66} \end{bmatrix} \quad , \quad C_{12} = C_{11} - 2C_{66} \quad . \tag{2.1}$$

Without losing generality, assume that the propagation occurs (locally) in the $x_1 x_3$ plane, where $p_2 = 0$. With the density-normalized fourth-order stiffness tensor $\tilde{\mathbf{C}}$, the Christoffel matrix reads,

$$\mathbf{\Gamma} = \mathbf{p}\tilde{\mathbf{C}}\mathbf{p} = \begin{bmatrix} C_{11} p_1^2 + C_{44} p_3^2 & 0 & (C_{13} + C_{44}) p_1 p_3 \\ 0 & C_{66} p_1^2 + C_{44} p_3^2 & 0 \\ (C_{13} + C_{44}) p_1 p_3 & 0 & C_{44} p_1^2 + C_{33} p_3^2 \end{bmatrix} . \quad (2.2)$$

The Christoffel matrix splits into two blocks: a $2 \times 2$ block of coupled quasi-compressional (qP) and quasi-shear (qSV) waves, polarized in the "vertical" (axial) plane, and a scalar component (in the second row and column) for SH shear waves, polarized in the "horizontal" (normal; isotropic) plane. The vanishing reference Hamiltonian $H^{\bar{\tau}}$ can be defined from the Christoffel matrix $\mathbf{\Gamma}$, whose components are listed in equation 2.2. According to this split, the Hamiltonian is factorized into two terms related to the coupled qP-qSV and to SH waves,

$$H^{\bar{\tau}} = \det(\mathbf{\Gamma} - \mathbf{I}) = H^{\bar{\tau}}_{PSV} \cdot H^{\bar{\tau}}_{SH} = 0 \quad , \quad (2.3)$$

where,

$$\begin{aligned} H^{\bar{\tau}}_{PSV} &= -C_{11} C_{44} p_1^4 - C_{33} C_{44} p_3^4 - \left( C_{11} C_{33} - C_{13}^2 - 2 C_{13} C_{44} \right) p_1^2 p_3^2 \\ &\quad + (C_{11} + C_{44}) p_1^2 + (C_{33} + C_{44}) p_3^2 - 1 = 0 \quad , \end{aligned} \quad (2.4)$$

and,

$$H^{\bar{\tau}}_{SH} = C_{66} p_1^2 + C_{44} p_3^2 - 1 = 0 \quad . \quad (2.5)$$

In order to extend the validity of the reference Hamiltonians in equations 2.4 and 2.5 for any reference frame, we introduce the concept of the axial and the normal slowness components $p_{ax} = p_3$ and $p_{nr} = p_1$, where $p_{ax}$ and $p_{nr}$ are physical scalar invariants, independent of any specific reference frame. For this, however, the crystal-frame stiffness matrix components $C_{ij}$ should be replaced by parameters which are also independent of the reference frame (physical invariants), like the Thomsen (1986) or Alkhalifah (1998) parameters. As indicated, the model properties' array **m** in this study is represented as a concatenation of two arrays of lengths 5 and 2,

$$\mathbf{c} = \{v_P \quad f \quad \delta \quad \varepsilon \quad \gamma\} \quad , \quad \mathbf{a} = \{\theta_{ax} \quad \psi_{ax}\} \quad \text{and} \quad \mathbf{m} = \mathbf{c} \| \mathbf{a} \quad , \quad (2.6)$$

where the dependency of the Thomsen parameters on the stiffness tensor components and the inverse relationships for the stiffness components vs. the Thomsen parameters are,

$$\begin{array}{ll}
(a) & (b) \\
v_P = \sqrt{C_{33}} \, , & \dfrac{C_{11}}{v_P^2} = 1 + 2\varepsilon \, , \\
f = 1 - \dfrac{v_S^2}{v_P^2} = \dfrac{C_{33} - C_{44}}{C_{33}} \, , & \dfrac{C_{13}}{v_P^2} = \sqrt{f(f + 2\delta)} - (1 - f) \, , \\
\varepsilon = \dfrac{C_{11} - C_{33}}{2C_{33}} \, , & \dfrac{C_{33}}{v_P^2} = 1 \, , \\
\delta = \dfrac{(C_{13} + C_{44})^2 - (C_{33} - C_{44})^2}{2C_{33}(C_{33} - C_{44})} \, , & \dfrac{C_{44}}{v_P^2} = 1 - f \, , \\
\gamma = \dfrac{C_{66} - C_{44}}{2C_{44}} \, , & \dfrac{C_{66}}{v_P^2} = (1 - f)(1 + 2\gamma) \, .
\end{array} \quad (2.7)$$

Equation set 2.7b has been obtained from 2.7a with the commonly used assumption $C_{13} + C_{44} > 0$, which excludes anomalous polarizations in the axial plane (Helbig and Schoenberg 1987; Schoenberg and Helbig 1997). This leads to the plus sign before the square root in the equation for $C_{13}$.

Introduction of the Thomsen parameterization (equation 2.7b) into the Hamiltonians of PSV and SH waves (equations 2.4 and 2.5, respectively), leads to,

$$H^{\bar{\tau}}_{PSV}(\mathbf{x},\mathbf{p}) = -\left[(1-f)\left(p^2_{nr}+p^2_{ax}\right)^2 + 2(\varepsilon - f\delta)p^2_{nr}\left(p^2_{nr}+p^2_{ax}\right) - 2f(\varepsilon - \delta)p^4_{nr}\right]v^4_P \qquad (2.8)$$
$$+ \left[(2-f)\left(p^2_{nr}+p^2_{ax}\right) + 2\varepsilon p^2_{nr}\right]v^2_P - 1 = 0 ,$$

and,

$$H^{\bar{\tau}}_{SH}(\mathbf{x},\mathbf{p}) = (1-f)(1+2\gamma)p^2_{nr}v^2_P + (1-f)p^2_{ax}v^2_P - 1 = 0 \qquad (2.9)$$

For the SH-wave Hamiltonian, $H^{\bar{\tau}}_{SH}$, it is convenient to introduce the axial shear velocity, $v_S$. According to the definition of the shear velocity factor $f$ (Tsvankin, 2012), $v_S = v_P\sqrt{1-f}$, and the SH Hamiltonian of equation 2.9 simplifies to,

$$H^{\bar{\tau}}_{SH}(\mathbf{x},\mathbf{p}) = (1+2\gamma)p^2_{nr}v^2_S + p^2_{ax}v^2_S - 1 = 0 \qquad (2.10)$$

Thus, $H^{\bar{\tau}}_{SH}$ depends on two material properties only, $v_S$ and $\gamma$.

Remark: Although the Thomsen parameters at each grid point are physical scalars independent of the reference frame, still, they describe the medium properties in its crystal frame.

The Hamiltonians in equations 2.8 and 2.10 are arranged in coordinate-free notations. To compute the axial and normal slowness invariants, we introduce the unit-length axis direction vector $\mathbf{k}$, whose components in the global frame are defined by the Euler direction angles: zenith $\theta_{ax}$ and azimuth $\psi_{ax}$,

$$\mathbf{k} = \{\sin\theta_{ax}\cos\psi_{ax} \quad \sin\theta_{ax}\sin\psi_{ax} \quad \cos\theta_{ax}\} \qquad (2.11)$$

Hence, the slowness invariants, $p_{ax}$ and $p_{nr}$, can be presented in terms of the scalar product of the physical vectors: the symmetry axis direction vector $\mathbf{k}$ and the slowness vector $\mathbf{p}$,

$$p_{ax} = \mathbf{k}\cdot\mathbf{p}, \quad p_{nr} = \sqrt{\mathbf{p}\cdot\mathbf{p} - (\mathbf{k}\cdot\mathbf{p})^2} \qquad (2.12)$$

This leads to,

$$H^{\bar{\tau}}_{PSV}(\mathbf{x},\mathbf{p}) = -(1-f)(\mathbf{p}\cdot\mathbf{p})^2 v_P^4 - 2(\varepsilon - f\delta)\left[\mathbf{p}\cdot\mathbf{p} - (\mathbf{k}\cdot\mathbf{p})^2\right]\mathbf{p}\cdot\mathbf{p} v_P^4$$
$$+ 2f(\varepsilon - \delta)\left[\mathbf{p}\cdot\mathbf{p} - (\mathbf{k}\cdot\mathbf{p})^2\right]^2 v_P^4 + (2-f)\mathbf{p}\cdot\mathbf{p} v_P^2 + 2\varepsilon\left[\mathbf{p}\cdot\mathbf{p} - (\mathbf{k}\cdot\mathbf{p})^2\right] v_P^2 - 1 = 0 ,$$
(2.13)

and,

$$H^{\bar{\tau}}_{SH}(\mathbf{x},\mathbf{p}) = (1+2\gamma)\left[\mathbf{p}\cdot\mathbf{p} - (\mathbf{k}\cdot\mathbf{p})^2\right] v_S^2 + (\mathbf{k}\cdot\mathbf{p})^2 v_S^2 - 1 = 0 .$$
(2.14)

Equations 2.13 and 2.14 expand into,

$$H^{\bar{\tau}}_{PSV}(\mathbf{x},\mathbf{p}) = -(1+2\varepsilon)(1-f)(\mathbf{p}\cdot\mathbf{p})^2 v_P^4 + 2\left[\varepsilon(1-f) - f(\varepsilon-\delta)\right](\mathbf{k}\cdot\mathbf{p})^2 (\mathbf{p}\cdot\mathbf{p}) v_P^4$$
$$+ 2f(\varepsilon - \delta)(\mathbf{k}\cdot\mathbf{p})^4 v_P^4 + \left[(1-f) + (1+2\varepsilon)\right](\mathbf{p}\cdot\mathbf{p}) v_P^2 - 2\varepsilon(\mathbf{k}\cdot\mathbf{p})^2 v_P^2 - 1 = 0 ,$$
(2.15)

and,

$$H^{\bar{\tau}}_{SH}(\mathbf{x},\mathbf{p}) = (1+2\gamma)(\mathbf{p}\cdot\mathbf{p}) v_S^2 - 2\gamma(\mathbf{k}\cdot\mathbf{p})^2 v_S^2 - 1 = 0 ,$$
(2.16)

where we recall that the material parameters of the medium $\mathbf{c}$, and the symmetry axis direction $\mathbf{k}$, are spatially varying. Equations 2.15 and 2.16 provide the polar anisotropic (TTI) Hamiltonians in the global frame, making it possible to obtain the ray velocity derivatives directly in the global frame as well.

### 3. SIGN OF THE HAMILTONIAN

The (reference) Hamiltonian is a vanishing function representing the Christoffel equation, and thus multiplying it by a positive or negative factor has no effect on the solution of this equation. For the inverse problem of finding the slowness vector, the sign of the Hamiltonian is inessential as well, because the governing equation of the inversion (along with the vanishing Hamiltonian) manifests the collinearity of the slowness gradient of the reference Hamiltonian and the ray direction vector, $H^{\bar{\tau}}_\mathbf{p} \times \mathbf{r} = 0$, which is equivalent to $H^{\bar{\tau}}_\mathbf{p} = \alpha \mathbf{r}$, $\alpha \neq 0$. Still, the two collinear vectors may prove to have the same or opposite directions, i.e., parameter $\alpha$, with the units of velocity, may be positive or negative.

However, the sign is important for the Hamiltonian derivatives. An improper sign (after scaling the reference Hamiltonian $H^{\bar{\tau}}(\mathbf{x},\mathbf{p})$ to the arclength-related one, $H(\mathbf{x},\mathbf{p}) = H^{\bar{\tau}} / \sqrt{H^{\bar{\tau}}_{\mathbf{p}} H^{\bar{\tau}}_{\mathbf{p}}}$ equation 4.2 in Part I), yields $H_{\mathbf{p}} = -\mathbf{r}$ instead of $H_{\mathbf{p}} = +\mathbf{r}$, which is an evidence of a wrong Hamiltonian sign. In Appendix B, we analyze the sign of the Hamiltonians for all wave types $H^{\bar{\tau}}_P, H^{\bar{\tau}}_{SV}, H^{\bar{\tau}}_{SH}$ and conclude that,

$$H^{\bar{\tau}}_P = +H^{\bar{\tau}}_{PSV}, \qquad H^{\bar{\tau}}_{SV} = -H^{\bar{\tau}}_{PSV} = -H^{\bar{\tau}}_P \qquad , \tag{3.1}$$

where $H^{\bar{\tau}}_{PSV}$ is the Hamiltonian of the coupled qP-qSV waves listed in equation 2.15. The sign of the Hamiltonian $H^{\bar{\tau}}_{SH}$ in equation 2.16 is correct. Therefore, in this paper, the Hamiltonian derivatives are computed for qP and SH waves only, where for qSV waves, we consider $\nabla H^{\bar{\tau}}_{SV} = -\nabla H^{\bar{\tau}}_P$, and the "nabla" symbol with no subscript means that this relation holds for all types of first and second derivatives. Note, however, that the values obtained for $\nabla H^{\bar{\tau}}_P$ and $\nabla H^{\bar{\tau}}_{SV}$ are not just opposite numbers, they have different magnitudes because they refer to either qP or the qSV slowness vectors introduced into $\nabla H^{\bar{\tau}}_P$ and $\nabla H^{\bar{\tau}}_{SV}$, respectively.

## 4. HAMILTONIAN GRADIENT WRT THE MATERIAL PROPERTIES

The gradient of the (reference) Hamiltonians wrt the model parameters (that include both material and geometric counterparts) are written as,

$$H_{P,\mathbf{m}}^{\bar{\tau}} = \left\{ \frac{\partial H_P^{\bar{\tau}}}{\partial v_P} \quad \frac{\partial H_P^{\bar{\tau}}}{\partial f} \quad \frac{\partial H_P^{\bar{\tau}}}{\partial \delta} \quad \frac{\partial H_P^{\bar{\tau}}}{\partial \varepsilon} \quad \frac{\partial H_P^{\bar{\tau}}}{\partial \theta_{ax}} \quad \frac{\partial H_P^{\bar{\tau}}}{\partial \psi_{ax}} \right\}, \quad n_c + 2 = 6,$$

$$H_{SH,\mathbf{m}}^{\bar{\tau}} = \left\{ \frac{\partial H_{SH}^{\bar{\tau}}}{\partial v_S} \quad \frac{\partial H_{SH}^{\bar{\tau}}}{\partial \gamma} \quad\quad\quad\quad \frac{\partial H_{SH}^{\bar{\tau}}}{\partial \theta_{ax}} \quad \frac{\partial H_{SH}^{\bar{\tau}}}{\partial \psi_{ax}} \right\}, \quad n_c + 2 = 4,$$

(4.1)

where, $H_{\mathbf{m}}^{\bar{\tau}} = H_{\mathbf{c}}^{\bar{\tau}} \parallel H_{\mathbf{a}}^{\bar{\tau}}$. In this section, we only compute the derivatives with respect to the material components, $H_{\mathbf{c}}^{\bar{\tau}}$, while in the next section, the derivatives are computed wrt the geometric components $H_{\mathbf{a}}^{\bar{\tau}}$. The material gradient of the qP-qSV Hamiltonian can be obtained from equation 2.15, and the components of this gradient are,

$$\frac{\partial H_P^{\bar{\tau}}}{\partial v_P} = -4(1+2\varepsilon)(1-f)(\mathbf{p}\cdot\mathbf{p})^2 v_P^3 + 8\left[\varepsilon(1-f) - f(\varepsilon-\delta)\right](\mathbf{k}\cdot\mathbf{p})^2 (\mathbf{p}\cdot\mathbf{p}) v_P^3$$
$$+ 8f(\varepsilon-\delta)(\mathbf{k}\cdot\mathbf{p})^4 v_P^3 + 2\left[(1-f) + (1+2\varepsilon)\right](\mathbf{p}\cdot\mathbf{p}) v_P - 4\varepsilon(\mathbf{k}\cdot\mathbf{p})^2 v_P,$$

$$\frac{\partial H_P^{\bar{\tau}}}{\partial f} = +(1+2\varepsilon)(\mathbf{p}\cdot\mathbf{p})^2 v_P^4 - 2(2\varepsilon-\delta)(\mathbf{k}\cdot\mathbf{p})^2 (\mathbf{p}\cdot\mathbf{p}) v_P^4$$
$$+ 2(\varepsilon-\delta)(\mathbf{k}\cdot\mathbf{p})^4 v_P^4 - (\mathbf{p}\cdot\mathbf{p}) v_P^2,$$

(4.2)

$$\frac{\partial H_P^{\bar{\tau}}}{\partial \delta} = +2f(\mathbf{k}\cdot\mathbf{p})^2 \left[(\mathbf{p}\cdot\mathbf{p}) - (\mathbf{k}\cdot\mathbf{p})^2\right] v_P^4,$$

$$\frac{\partial H_P^{\bar{\tau}}}{\partial \varepsilon} = 2f\left[(\mathbf{p}\cdot\mathbf{p}) - (\mathbf{k}\cdot\mathbf{p})^2\right]^2 v_P^4 - 2\left[(\mathbf{p}\cdot\mathbf{p}) - (\mathbf{k}\cdot\mathbf{p})^2\right]\left[(\mathbf{p}\cdot\mathbf{p}) v_P^2 - 1\right] v_P^2.$$

The gradient of the SH Hamiltonian can be obtained from equation 2.16,

$$\frac{\partial H_{SH}^{\bar{\tau}}}{\partial v_S} = +2\left[(1+2\gamma)(\mathbf{p}\cdot\mathbf{p}) - 2\gamma(\mathbf{k}\cdot\mathbf{p})^2\right] v_P, \quad \frac{\partial H_{SH}^{\bar{\tau}}}{\partial \gamma} = +2\left[(\mathbf{p}\cdot\mathbf{p}) - (\mathbf{k}\cdot\mathbf{p})^2\right] v_S^2. \quad (4.3)$$

## 5. HAMILTONIAN GRADIENT WRT THE GEOMETRIC PROPERTIES

To obtain the Hamiltonian gradient wrt the polar angles, we apply the chain rule for a multivariate function, multiplying the Hamiltonian gradient wrt the symmetry axis direction vector, and the gradient of the axis components wrt the two polar angles,

$$H_{\mathbf{a}}^{\bar{\tau}} = H_{\mathbf{k}}^{\bar{\tau}} \cdot \mathbf{k_a} \quad , \tag{5.1}$$

where $H_{\mathbf{k}}^{\bar{\tau}}$ is a vector of length 3, and $\mathbf{k_a}$ is a matrix of dimension $3 \times 2$ containing the derivatives of the components of the symmetry axis direction vector $\mathbf{k}$ wrt its polar angles. The gradient of the qP-qSV Hamiltonian and that of the SH Hamiltonian wrt the axis direction are, respectively,

$$H_{P,\mathbf{k}}^{\bar{\tau}}(\mathbf{x},\mathbf{p}) = 4\left[\varepsilon(1-f) - f(\varepsilon-\delta)\right](\mathbf{k}\cdot\mathbf{p})(\mathbf{p}\cdot\mathbf{p})\mathbf{p}v_P^4 \\ + 8f(\varepsilon-\delta)(\mathbf{k}\cdot\mathbf{p})^3 \mathbf{p}v_P^4 - 4\varepsilon(\mathbf{k}\cdot\mathbf{p})\mathbf{p}v_P^2 \quad , \tag{5.2}$$

and,

$$H_{SH,\mathbf{k}}^{\bar{\tau}}(\mathbf{x},\mathbf{p}) = -4\gamma(\mathbf{k}\cdot\mathbf{p})\mathbf{p}v_S^2 \quad . \tag{5.3}$$

Using the components of the axial direction vector $\mathbf{k}$ listed in equation 2.9, we obtain the gradient of this vector with respect to the symmetry axis direction angles,

$$\mathbf{k_a}(3,2) = \begin{bmatrix} +\cos\theta_{ax}\cos\psi_{ax} & -\sin\theta_{ax}\sin\psi_{ax} \\ +\cos\theta_{ax}\sin\psi_{ax} & +\sin\theta_{ax}\cos\psi_{ax} \\ -\sin\theta_{ax} & 0 \end{bmatrix} \quad . \tag{5.4}$$

The first column of this matrix includes the derivatives of vector $\mathbf{k}$ wrt the zenith angle $\theta_{ax}$, and the second column – wrt the azimuth $\psi_{ax}$. The rows include the Cartesian components.

Now we can obtain the gradients of the qP-qSV and SH Hamiltonians wrt all model properties (both material and geometric), whose components are listed in equation 4.1,

$$\underbrace{H_{\mathbf{m}}^{\bar{\tau}}}_{1\times(n_{\mathbf{c}}+2)} = \underbrace{H_{\mathbf{c}}^{\bar{\tau}}}_{1\times n_{\mathbf{c}}} \| \underbrace{H_{\mathbf{a}}^{\bar{\tau}}}_{1\times 2} \quad , \quad \underbrace{H_{\mathbf{a}}^{\bar{\tau}}}_{1\times 2} = \underbrace{H_{\mathbf{k}}^{\bar{\tau}}}_{1\times 3} \cdot \underbrace{\mathbf{k}_{\mathbf{a}}}_{3\times 2} \quad , \tag{5.5}$$

where recall that symbol $\|$ means concatenation of two arrays (in this case, with lengths $n_{\mathbf{c}}$ and $2$); the resulting array includes all the components of the model at a given point and has length $n_{\mathbf{c}} + 2$, where $n_{\mathbf{c}} = 4$ for qP and qSV waves, $n_{\mathbf{c}} = 3$ for the acoustic approximation of qP waves, and $n_{\mathbf{c}} = 2$ for SH waves.

## 6. SPATIAL GRADIENT OF THE HAMILTONIAN

The Hamiltonian depends on the spatial coordinates indirectly, implicitly through the medium properties $\mathbf{m}(\mathbf{x})$; therefore, its gradient obeys the chain rule for a multivariate function,

$$\underbrace{H_{\mathbf{x}}^{\bar{\tau}}(\mathbf{x},\mathbf{p})}_{1\times 3} = \underbrace{H_{\mathbf{m}}^{\bar{\tau}}}_{1\times(n_{\mathbf{c}}+2)} \cdot \underbrace{\mathbf{m}_{\mathbf{x}}}_{(n_{\mathbf{c}}+2)\times 3} \quad \text{where} \quad \underbrace{\mathbf{m}_{\mathbf{x}}}_{(n_{\mathbf{c}}+2)\times 3} = \underbrace{\mathbf{c}_{\mathbf{x}}}_{n_{\mathbf{c}}\times 3} \| \underbrace{\mathbf{a}_{\mathbf{x}}}_{2\times 3} \quad . \tag{6.1}$$

Vector $H_{\mathbf{m}}^{\bar{\tau}}(1, n_{\mathbf{c}}+2)$ is a gradient of the Hamiltonian wrt all (material and geometric) model properties, obtained in equation 5.5; $\mathbf{c}_{\mathbf{x}}(n_{\mathbf{c}}, 3)$ and $\mathbf{a}_{\mathbf{x}}(2, 3)$ are the two matrices containing the spatial gradients of the physical and geometric properties, respectively. We assume that the spatial gradients of the model parameters,

$$\mathbf{c_x}(5,3) = \begin{bmatrix} \partial v_P / \partial x_1 & \partial v_P / \partial x_2 & \partial v_P / \partial x_3 \\ \partial f / \partial x_1 & \partial f / \partial x_2 & \partial f / \partial x_3 \\ \partial \delta / \partial x_1 & \partial \delta / \partial x_2 & \partial \delta / \partial x_3 \\ \partial \varepsilon / \partial x_1 & \partial \varepsilon / \partial x_2 & \partial \varepsilon / \partial x_3 \\ \partial \gamma / \partial x_1 & \partial \gamma / \partial x_2 & \partial \gamma / \partial x_3 \end{bmatrix}, \qquad (6.2)$$

and,

$$\mathbf{a_x}(2,3) = \begin{bmatrix} \partial \theta_{ax} / \partial x_1 & \partial \theta_{ax} / \partial x_2 & \partial \theta_{ax} / \partial x_3 \\ \partial \psi_{ax} / \partial x_1 & \partial \psi_{ax} / \partial x_2 & \partial \psi_{ax} / \partial x_3 \end{bmatrix}, \qquad (6.3)$$

can be reliably computed. Recall that all the five parameters involved in matrix $\mathbf{c_x}$ are never used simultaneously. As mentioned, two, three or four parameters out of the five are applied in each case. The resulting spatial gradient of the Hamiltonian, $H_{\mathbf{x}}^{\bar{\tau}}$, is a vector of length 3.

## 7. HAMILTONIAN HESSIAN WRT THE MATERIAL PROPERTIES

It is convenient to split the Hamiltonian Hessian into four blocks,

$$H_{\mathbf{mm}}^{\tau} = \begin{bmatrix} H_{\mathbf{cc}}^{\tau} & H_{\mathbf{ca}}^{\tau} \\ H_{\mathbf{ac}}^{\tau} & H_{\mathbf{aa}}^{\tau} \end{bmatrix}, \qquad (7.1)$$

where the upper and lower diagonal blocks, $H_{\mathbf{cc}}^{\tau}$ and $H_{\mathbf{aa}}^{\tau}$, respectively, are symmetric and related to the material and geometric properties, respectively, while the off-diagonal blocks, $H_{\mathbf{ac}}^{\bar{\tau}}(2, n_{\mathbf{c}})$ and $H_{\mathbf{ca}}^{\bar{\tau}}(n_{\mathbf{c}}, 2)$, are mixed (i.e., related to both, material and geometric properties of the model); the mixed blocks are transposed of each other. Numbers in brackets indicate the dimensions of the matrices. In this section, we derive only the Hamiltonian Hessian wrt the material properties (the upper left block), while the other blocks are derived in the next two sections. For the qP-qSV Hamiltonian, the Hessian can be obtained from equation 4.2 (the unspecified components are zeros),

$$\frac{\partial^2 H_P^{\bar{\tau}}}{\partial v_P^2} = -12(1+2\varepsilon)(1-f)(\mathbf{p}\cdot\mathbf{p})^2 v_P^2 + 24\left[\varepsilon(1-f) - f(\varepsilon-\delta)\right](\mathbf{k}\cdot\mathbf{p})^2 (\mathbf{p}\cdot\mathbf{p}) v_P^2$$

$$+ 24 f(\varepsilon-\delta)(\mathbf{k}\cdot\mathbf{p})^4 v_P^2 + 2\left[(1-f)+(1+2\varepsilon)\right](\mathbf{p}\cdot\mathbf{p}) - 4\varepsilon(\mathbf{k}\cdot\mathbf{p})^2 \quad ,$$

$$\frac{\partial^2 H_P^{\bar{\tau}}}{\partial v_P \, \partial f} = +4(1+2\varepsilon)(\mathbf{p}\cdot\mathbf{p})^2 v_P^3 - 8(2\varepsilon-\delta)(\mathbf{k}\cdot\mathbf{p})^2 (\mathbf{p}\cdot\mathbf{p}) v_P^3$$

$$+ 8(\varepsilon-\delta)(\mathbf{k}\cdot\mathbf{p})^4 v_P^3 - 2(\mathbf{p}\cdot\mathbf{p}) v_P \quad ,$$

$$\frac{\partial^2 H_P^{\bar{\tau}}}{\partial v_P \, \partial \delta} = +8f\left[(\mathbf{p}\cdot\mathbf{p}) - (\mathbf{k}\cdot\mathbf{p})^2\right](\mathbf{k}\cdot\mathbf{p})^2 v_P^3 \quad ,$$

$$\frac{\partial^2 H_P^{\bar{\tau}}}{\partial v_P \, \partial \varepsilon} = +8f\left[(\mathbf{p}\cdot\mathbf{p}) - (\mathbf{k}\cdot\mathbf{p})^2\right]^2 v_P^3 - 4\left[(\mathbf{p}\cdot\mathbf{p}) - (\mathbf{k}\cdot\mathbf{p})^2\right]\left[2(\mathbf{p}\cdot\mathbf{p}) v_P^2 - 1\right] v_P \quad ,$$

$$\frac{\partial^2 H_P^{\bar{\tau}}}{\partial f \, \partial \delta} = +2\ \left[(\mathbf{p}\cdot\mathbf{p}) - (\mathbf{k}\cdot\mathbf{p})^2\right](\mathbf{k}\cdot\mathbf{p})^2 v_P^4 \quad ,$$

$$\frac{\partial^2 H_P^{\bar{\tau}}}{\partial f \, \partial \varepsilon} = +2\ \left[(\mathbf{p}\cdot\mathbf{p}) - (\mathbf{k}\cdot\mathbf{p})^2\right]^2 v_P^4 \quad . \tag{7.2}$$

For the SH Hamiltonian, the Hessian can be obtained from equation 4.3 (again, the unspecified components are zeros),

$$\frac{\partial^2 H_{SH}^{\bar{\tau}}}{\partial v_S^2} = +2\left[(1+2\gamma)(\mathbf{p}\cdot\mathbf{p}) - 2\gamma(\mathbf{k}\cdot\mathbf{p})^2\right] \quad , \quad \frac{\partial^2 H_{SH}^{\bar{\tau}}}{\partial v_S \, \partial \gamma} = +4\left[(\mathbf{p}\cdot\mathbf{p}) - (\mathbf{k}\cdot\mathbf{p})^2\right] v_P \quad . \tag{7.3}$$

## 8. HAMILTONIAN HESSIAN WRT THE GEOMETRIC PROPERTIES

Recall that the Hamiltonian depends on the symmetry axis direction vector, and the latter, in turn, depends on its two polar angles, $\theta_{ax}$ and $\psi_{ax}$. Applying the chain rule for the second derivative of a multivariate function, we obtain,

$$\underbrace{H^{\bar{\tau}}_{\mathbf{aa}}}_{2\times 2} = \underbrace{\mathbf{k}_{\mathbf{a}}^T}_{2\times 3} \cdot \underbrace{H^{\bar{\tau}}_{\mathbf{kk}}}_{3\times 3} \cdot \underbrace{\mathbf{k}_{\mathbf{a}}}_{3\times 2} + \underbrace{H^{\bar{\tau}}_{\mathbf{k}}}_{1\times 3} \cdot \underbrace{\mathbf{k}_{\mathbf{aa}}}_{3\times 2\times 2} \quad . \tag{8.1}$$

The result is a $2\times 2$ symmetric matrix – the Hamiltonian Hessian wrt the polar angles $\theta_{ax}$ and $\psi_{ax}$. The formula (equation 8.1) is generic for any wave type, but the gradient, $H^{\bar{\tau}}_{\mathbf{k}}$, and the Hessian, $H^{\bar{\tau}}_{\mathbf{kk}}$, of the Hamiltonian wrt the symmetry axis direction vector differ for the qP-qSV and the SH wave Hamiltonians. The gradients are listed in equations 5.2 and 5.3. These equations also make it possible to establish the two Hessians,

$$\begin{aligned} H^{\bar{\tau}}_{P,\mathbf{kk}}(\mathbf{x},\mathbf{p}) = & 4\left[\varepsilon(1-f)-f(\varepsilon-\delta)\right](\mathbf{p}\cdot\mathbf{p})(\mathbf{p}\otimes\mathbf{p})v_P^4 \\ & + 24f(\varepsilon-\delta)(\mathbf{k}\cdot\mathbf{p})^2(\mathbf{p}\otimes\mathbf{p})v_P^4 - 4\varepsilon(\mathbf{p}\otimes\mathbf{p})v_P^2 \quad , \end{aligned} \tag{8.2}$$

and,

$$H^{\bar{\tau}}_{SH,\mathbf{kk}}(\mathbf{x},\mathbf{p}) = -4\gamma(\mathbf{p}\otimes\mathbf{p})v_S^2 \quad . \tag{8.3}$$

Matrix $\mathbf{k}_{\mathbf{a}}$ of dimensions $3\times 2$ is given in equation 5.4. Three-dimensional array $\mathbf{k}_{\mathbf{aa}}, 3\times 2\times 2$ represents a "Hessian of vector $\mathbf{k}$ wrt array $\mathbf{a}$", where $\mathbf{a} = \{\theta_{ax} \ \psi_{ax}\}$; in other words, $[\mathbf{k}_{\mathbf{aa}}]_{ijk} = \partial^2 k_i / (\partial a_j \partial a_k)$. This array is symmetric wrt the last two indices. It can be computed from equation 2.9 that lists the components of the axis direction vector $\mathbf{k}$,

$$\mathbf{k_{aa}} = \begin{bmatrix} \begin{bmatrix} -\sin\theta_{ax}\cos\psi_{ax} & -\cos\theta_{ax}\sin\psi_{ax} \\ -\cos\theta_{ax}\sin\psi_{ax} & -\sin\theta_{ax}\cos\psi_{ax} \end{bmatrix} \\ \begin{bmatrix} -\sin\theta_{ax}\sin\psi_{ax} & +\cos\theta_{ax}\cos\psi_{ax} \\ +\cos\theta_{ax}\cos\psi_{ax} & -\sin\theta_{ax}\sin\psi_{ax} \end{bmatrix} \\ \begin{bmatrix} -\cos\theta_{ax} & 0 \\ 0 & 0 \end{bmatrix} \end{bmatrix} . \tag{8.4}$$

## 9. MIXED HESSIAN OF THE HAMILTONIAN WRT THE MATERIAL AND GEOMETRIC PROPERTIES

In this section, we derive the mixed Hessian of the Hamiltonian, $H_{\mathbf{ac}}^{\bar{\tau}}(2, n_{\mathbf{c}})$ related to both, the material and geometric properties of the Hamiltonian; this block is shown in equation 7.1. We start from equation 5.1, $H_{\mathbf{a}}^{\bar{\tau}} = H_{\mathbf{k}}^{\bar{\tau}} \cdot \mathbf{k_a}$. The left-hand side of this equation is a vector of length 2, and the right-hand side is a product of vector $H_{\mathbf{k}}^{\bar{\tau}}$ of length 3 and matrix $\mathbf{k_a}$ of dimensions $3 \times 2$. This matrix is given in equation 5.4; as we see, its components are independent of the model parameters. We apply a general algebraic law for arbitrary vector $\mathbf{a}$ and matrix $\mathbf{B}$,

$$\nabla(\mathbf{aB}) = \underbrace{\mathbf{B}^T \nabla \mathbf{a} + \mathbf{a} \nabla \mathbf{B}}_{\text{second-order tensor}} , \tag{9.1}$$

and in our case, this leads to,

$$\underset{2 \times n_{\mathbf{c}}}{H_{\mathbf{ac}}^{\bar{\tau}}} = \underset{2 \times 3}{\mathbf{k}_{\mathbf{a}}^{T}} \cdot \underset{3 \times n_{\mathbf{c}}}{H_{\mathbf{kc}}^{\bar{\tau}}} . \tag{9.2}$$

Thus, we need to compute the components of matrix $H_{\mathbf{kc}}^{\bar{\tau}}$,

$$\frac{\partial^2 H^{\bar{\tau}}}{\partial \mathbf{k}\, \partial c_i} = \frac{\partial H^{\bar{\tau}}_{\mathbf{k}}(\mathbf{x},\mathbf{p})}{\partial c_i}, \qquad \begin{array}{l} i = 1, \ldots 4 \text{ for } H^{\bar{\tau}}_P, \\ i = 1, \ldots 3 \text{ for } H^{\bar{\tau}}_{SH}. \end{array} \tag{9.3}$$

For the qP-qSV Hamiltonian, we apply equation 5.2 and obtain,

$$\frac{\partial H^{\bar{\tau}}_{P,\mathbf{k}}(\mathbf{x},\mathbf{p})}{\partial v_P} = 16\big[\varepsilon(1-f) - f(\varepsilon - \delta)\big](\mathbf{k}\cdot\mathbf{p})(\mathbf{p}\cdot\mathbf{p})\mathbf{p}\,v_P^3$$
$$+ 32 f(\varepsilon - \delta)(\mathbf{k}\cdot\mathbf{p})^3 \mathbf{p}\,v_P^3 - 8\varepsilon(\mathbf{k}\cdot\mathbf{p})\mathbf{p}\,v_P,$$
$$\frac{\partial H^{\bar{\tau}}_{P,\mathbf{k}}(\mathbf{x},\mathbf{p})}{\partial f} = -8(\varepsilon - \delta)\big[(\mathbf{p}\cdot\mathbf{p}) - (\mathbf{p}\cdot\mathbf{k})^2\big](\mathbf{p}\cdot\mathbf{k})\mathbf{p}\,v_P^4 - 4\delta(\mathbf{p}\cdot\mathbf{p})(\mathbf{p}\cdot\mathbf{k})\mathbf{p}\,v_P^4, \tag{9.4}$$
$$\frac{\partial H^{\bar{\tau}}_{P,\mathbf{k}}(\mathbf{x},\mathbf{p})}{\partial \delta} = 4f\big[(\mathbf{p}\cdot\mathbf{p}) - 2(\mathbf{p}\cdot\mathbf{k})^2\big](\mathbf{p}\cdot\mathbf{k})\mathbf{p}\,v_P^4,$$
$$\frac{\partial H^{\bar{\tau}}_{P,\mathbf{k}}(\mathbf{x},\mathbf{p})}{\partial \varepsilon} = 4\big[(1 - 2f)(\mathbf{p}\cdot\mathbf{p})v_P^2 + 2f(\mathbf{p}\cdot\mathbf{k})^2 v_P^2 - 1\big](\mathbf{p}\cdot\mathbf{k})\mathbf{p}\,v_P^2.$$

For the SH Hamiltonian, we apply equation 5.3 and obtain,

$$\frac{\partial H^{\bar{\tau}}_{SH,\mathbf{k}}}{\partial v_S} = -8\gamma(\mathbf{k}\cdot\mathbf{p})\mathbf{p}\,v_S, \qquad \frac{\partial H^{\bar{\tau}}_{SH,\mathbf{k}}}{\partial \gamma} = -4(\mathbf{k}\cdot\mathbf{p})\mathbf{p}\,v_S^2. \tag{9.5}$$

Each item on the left-hand side of equation sets 9.4 and 9.5 is a column of a matrix.

### 10. SPATIAL HESSIAN OF THE HAMILTONIAN

To compute the spatial Hessian of the Hamiltonian, we apply the chain rule in a similar way it has been applied in equation 6.1, but now for the second derivatives (Faà di Bruno, 1855, 1857), and adjusted for a multivariate function,

$$\underbrace{H_{\mathbf{xx}}^{\bar{\tau}}(\mathbf{x},\mathbf{p})}_{3\times 3} = \underbrace{\mathbf{m_x^T}}_{3\times(n_\mathbf{c}+2)} \cdot \underbrace{H_{\mathbf{mm}}^{\bar{\tau}}}_{(n_\mathbf{c}+2)\times(n_\mathbf{c}+2)} \cdot \underbrace{\mathbf{m_x}}_{(n_\mathbf{c}+2)\times 3} + \underbrace{H_{\mathbf{m}}^{\bar{\tau}}}_{1\times(n_\mathbf{c}+2)} \cdot \underbrace{\mathbf{m_{xx}}}_{(n_\mathbf{c}+2)\times 3\times 3} \quad , \qquad (10.1)$$

where,

$$\underbrace{\mathbf{m_x}}_{(n_\mathbf{c}+2)\times 3} = \underbrace{\mathbf{c_x}}_{n_\mathbf{c}\times 3} \| \underbrace{\mathbf{a_x}}_{2\times 3} \quad , \qquad \underbrace{\mathbf{m_{xx}}}_{(n_\mathbf{c}+2)\times 3\times 3} = \underbrace{\mathbf{c_{xx}}}_{n_\mathbf{c}\times 3\times 3} \| \underbrace{\mathbf{a_{xx}}}_{2\times 3\times 3} \quad . \qquad (10.2)$$

Note that $\mathbf{c_x}$, $\mathbf{a_x}$ and $\mathbf{m_x}$ are two-dimensional arrays (spatial gradients of one-dimensional arrays). They are not second-order tensors, but collections of vectors. Similarly, $\mathbf{c_{xx}}$, $\mathbf{a_{xx}}$ and $\mathbf{m_{xx}}$ are three-dimensional arrays (spatial Hessians of one-dimensional arrays). They are not third-order tensors, but collections of second-order tensors. However, their multiplication rules are identical to those of the second- and third-order tensors.

Throughout this study, we assume that the spatially varying model parameters are smooth and well-conditioned, and their spatial derivatives: $\mathbf{c_x}$, $\mathbf{a_x}$, $\mathbf{c_{xx}}$, $\mathbf{a_{xx}}$ can be reliably computed. The computation of the gradient $H_{\mathbf{m}}^{\bar{\tau}}$ and the Hessian $H_{\mathbf{mm}}^{\bar{\tau}}$ is carried out according to equations 5.5 and 7.1, respectively.

## 11. SLOWNESS GRADIENT OF THE HAMILTONIAN

Applying the slowness gradient operator $\nabla_\mathbf{p}$ to the Hamiltonians of the qP-qSV and SH-waves listed in equations 2.15 and 2.16, respectively, we obtain the corresponding slowness gradients of the two Hamiltonians,

$$H^{\bar{\tau}}_{P,\mathbf{p}}(\mathbf{x},\mathbf{p}) = -4(1+2\varepsilon)(1-f)(\mathbf{p}\cdot\mathbf{p})\mathbf{p}v_P^4$$
$$+4\left[\varepsilon(1-f)-f(\varepsilon-\delta)\right]\left[(\mathbf{k}\cdot\mathbf{p})^2\mathbf{p}+(\mathbf{k}\cdot\mathbf{p})(\mathbf{p}\cdot\mathbf{p})\mathbf{k}\right]v_P^4 \qquad (11.1)$$
$$+8f(\varepsilon-\delta)(\mathbf{k}\cdot\mathbf{p})^3\mathbf{k}v_P^4 + 2\left[(1-f)+(1+2\varepsilon)\right]\mathbf{p}v_P^2 - 4\varepsilon(\mathbf{k}\cdot\mathbf{p})\mathbf{k}v_P^2,$$

and,

$$H^{\bar{\tau}}_{SH,\mathbf{p}}(\mathbf{x},\mathbf{p}) = 2(1+2\gamma)\mathbf{p}v_S^2 - 4\gamma(\mathbf{k}\cdot\mathbf{p})\mathbf{k}v_S^2 \qquad (11.2)$$

## 12. SLOWNESS HESSIAN OF THE HAMILTONIAN

Applying the slowness gradient operator $\nabla_{\mathbf{p}}$ to the gradients of the Hamiltonians of the qP-qSV and SH-waves listed in equations 11.1 and 11.2, respectively, we obtain the corresponding Hessians of the Hamiltonians,

$$H^{\bar{\tau}}_{P,\mathbf{pp}}(\mathbf{x},\mathbf{p}) = -4(1+2\varepsilon)(1-f)\left[2(\mathbf{p}\otimes\mathbf{p})+(\mathbf{p}\cdot\mathbf{p})\mathbf{I}\right]v_P^4$$
$$+4\left[\varepsilon(1-f)-f(\varepsilon-\delta)\right]\left[2(\mathbf{k}\cdot\mathbf{p})(\mathbf{p}\otimes\mathbf{k}+\mathbf{k}\otimes\mathbf{p})+(\mathbf{p}\cdot\mathbf{p})(\mathbf{k}\otimes\mathbf{k})+(\mathbf{k}\cdot\mathbf{p})^2\mathbf{I}\right]v_P^4 \qquad (12.1)$$
$$+24f(\varepsilon-\delta)(\mathbf{k}\cdot\mathbf{p})^2(\mathbf{k}\otimes\mathbf{k})v_P^4 + 2\left[(1-f)+(1+2\varepsilon)\right]\mathbf{I}v_P^2 - 4\varepsilon(\mathbf{k}\otimes\mathbf{k})v_P^2,$$

and,

$$H^{\bar{\tau}}_{SH,\mathbf{pp}}(\mathbf{x},\mathbf{p}) = 2(1+2\gamma)\mathbf{I}v_S^2 - 4\gamma(\mathbf{k}\otimes\mathbf{k})v_S^2 \qquad (12.2)$$

## 13. MIXED SLOWNESS-SPATIAL HESSIAN OF THE HAMILTONIAN

Applying the spatial gradient operator $\nabla_{\mathbf{x}}$ to the slowness gradient of the Hamiltonian $H^{\bar{\tau}}_{\mathbf{p}}$, we obtain the mixed slowness-spatial Hessian of the Hamiltonian, $H^{\bar{\tau}}_{\mathbf{px}}$. Recall that the Hamiltonian depends on the spatial coordinates indirectly, through the model properties, and thus, we compute the spatial derivatives with the use of the chain rule for a multivariate function,

$$\underbrace{H_{\mathbf{px}}^{\bar{\tau}}}_{3\times 3} = \underbrace{H_{\mathbf{pm}}^{\bar{\tau}}}_{3\times(n_c+2)} \cdot \underbrace{\mathbf{m_x}}_{(n_c+2)\times 3} \quad , \tag{13.1}$$

where the model gradient array, $\mathbf{m_x}(n_c+2,3)$, is a result of concatenation of the material properties' gradient array, $\mathbf{c_x}(n_c,3)$ (where $n$ is 2, 3 or 4, depending on the wave type), and the axis angles' gradient array, $\mathbf{a_x}(2,3)$. In this study, arrays $\mathbf{c_x}$ and $\mathbf{a_x}$ are considered the input data.

It is suitable to split array $H_{\mathbf{pm}}^{\bar{\tau}}(3, n_c+2)$ into two blocks that will be computed separately and then concatenated,

$$\underbrace{H_{\mathbf{pm}}^{\bar{\tau}}}_{3\times(n_c+2)} = \underbrace{H_{\mathbf{pc}}^{\bar{\tau}}}_{3\times n_c} \parallel \underbrace{H_{\mathbf{pa}}^{\bar{\tau}}}_{3\times 2} \quad . \tag{13.2}$$

The mixed "slowness-material" Hessian $H_{\mathbf{pc}}^{\bar{\tau}}$ can be computed for qP-qSV waves as,

$$\frac{\partial H_{P,\mathbf{p}}^{\bar{\tau}}(\mathbf{x},\mathbf{p})}{\partial v_P} = -8\varepsilon(\mathbf{p}\cdot\mathbf{k})\mathbf{k}\,v_P + 4\big[(1-f)+(1+2\varepsilon)\big]\mathbf{p}\,v_P + 32f(\varepsilon-\delta)(\mathbf{p}\cdot\mathbf{k})^3\,\mathbf{k}\,v_P^3$$

$$-16(1+2\varepsilon)(1-f)(\mathbf{p}\cdot\mathbf{p})\mathbf{p}\,v_P^3 + 16\big[\varepsilon(1-f)-f(\varepsilon-\delta)\big]\big[(\mathbf{p}\cdot\mathbf{k})\mathbf{p}+(\mathbf{p}\cdot\mathbf{p})\mathbf{k}\big](\mathbf{p}\cdot\mathbf{k})v_P^3\ ,$$

$$\frac{\partial H_{P,\mathbf{p}}^{\bar{\tau}}(\mathbf{x},\mathbf{p})}{\partial f} = +4(1+2\varepsilon)(\mathbf{p}\cdot\mathbf{p})\mathbf{p}\,v_P^4 - 4(2\varepsilon-\delta)\big[(\mathbf{k}\cdot\mathbf{p})\mathbf{p}+(\mathbf{p}\cdot\mathbf{p})\mathbf{k}\big](\mathbf{k}\cdot\mathbf{p})v_P^4$$

$$+8(\varepsilon-\delta)(\mathbf{k}\cdot\mathbf{p})^3\,\mathbf{k}\,v_P^4 - 2\mathbf{p}\,v_P^2\ , \tag{13.3}$$

$$\frac{\partial H_{P,\mathbf{p}}^{\bar{\tau}}(\mathbf{x},\mathbf{p})}{\partial \delta} = 4f\big[(\mathbf{p}\cdot\mathbf{k})\mathbf{p}+(\mathbf{p}\cdot\mathbf{p})\mathbf{k}\big](\mathbf{p}\cdot\mathbf{k})v_P^4 - 8f(\mathbf{p}\cdot\mathbf{k})^3\,\mathbf{k}\,v_P^4\ ,$$

$$\frac{\partial H_{P,\mathbf{p}}^{\bar{\tau}}(\mathbf{x},\mathbf{p})}{\partial \varepsilon} = -4\big[(\mathbf{p}\cdot\mathbf{k})\mathbf{k}-\mathbf{p}\big]v_P^2 + 8f(\mathbf{p}\cdot\mathbf{k})^3\,\mathbf{k}\,v_P^4$$

$$-8(1-f)(\mathbf{p}\cdot\mathbf{p})\mathbf{p}\,v_P^4 + 4(1-2f)\big[(\mathbf{p}\cdot\mathbf{k})+(\mathbf{p}\cdot\mathbf{p})\mathbf{k}\big](\mathbf{p}\cdot\mathbf{k})v_P^4\ ,$$

and for the SH waves, this mixed Hessian can be computed as,

$$\frac{\partial H_{SH,\mathbf{p}}^{\bar{\tau}}}{\partial v_S} = 4(1+2\gamma)\mathbf{p}\,v_S - 8\gamma(\mathbf{k}\cdot\mathbf{p})\mathbf{k}\,v_S\ ,\qquad \frac{\partial H_{SH,\mathbf{p}}^{\bar{\tau}}}{\partial \gamma} = 4\big[\mathbf{p}-(\mathbf{k}\cdot\mathbf{p})\mathbf{k}\big]v_S^2 \ . \tag{13.4}$$

Each item on the left-hand side of equation sets 13.3 and 13.4 is a column of a matrix.

Eventually, we need the second block on the right-hand side of equation 13.2, $H_{\mathbf{pa}}^{\bar{\tau}}(3,2)$. Recall that the Hamiltonian depends on the material properties (array $\mathbf{c}$) directly, and on the polar angles (array $\mathbf{a}$) indirectly, through the unit-length symmetry axis direction vector $\mathbf{k}$, and this leads to an application of one more chain rule for a multivariate function,

$$\underbrace{H_{\mathbf{pa}}^{\bar{\tau}}}_{3\times 2} = \underbrace{H_{\mathbf{pk}}^{\bar{\tau}}}_{3\times 3} \cdot \underbrace{\mathbf{k_a}}_{3\times 2}\ . \tag{13.5}$$

The components of matrix $\mathbf{k_a}(3,2)$ are given in equation 5.4. The mixed "slowness/axis of symmetry" Hessian can be obtained from equation 11.1 for the slowness gradient of qP-qSV Hamiltonian,

$$\begin{aligned}H^{\bar{\tau}}_{P,\mathbf{pk}}(\mathbf{x},\mathbf{p}) = &\; 4\left[\varepsilon(1-f) - f(\varepsilon-\delta)\right] \\ &\left[2(\mathbf{p}\cdot\mathbf{k})(\mathbf{p}\otimes\mathbf{p}) + (\mathbf{p}\cdot\mathbf{p})(\mathbf{k}\otimes\mathbf{p}) + (\mathbf{p}\cdot\mathbf{k})(\mathbf{p}\cdot\mathbf{p})\mathbf{I}\right]v_P^4 \\ &+ 24f(\varepsilon-\delta)(\mathbf{p}\cdot\mathbf{k})^2(\mathbf{k}\otimes\mathbf{p})v_P^4 + 8f(\varepsilon-\delta)(\mathbf{p}\cdot\mathbf{k})^3\mathbf{I}v_P^4 \\ &- 4\varepsilon(\mathbf{k}\otimes\mathbf{p})v_P^2 - 4\varepsilon(\mathbf{p}\cdot\mathbf{k})\mathbf{I}v_P^2 \quad,\end{aligned} \quad (13.6)$$

and from equation 11.2 for the slowness gradient of the SH Hamiltonian,

$$H^{\bar{\tau}}_{SH,\mathbf{pk}}(\mathbf{x},\mathbf{p}) = -4\gamma\left[\mathbf{k}\otimes\mathbf{p} + (\mathbf{k}\cdot\mathbf{p})\mathbf{I}\right]v_S^2 \quad. \quad (13.7)$$

Note that the Hamiltonian gradient wrt the model parameters, $H^{\bar{\tau}}_{\mathbf{m}}$, the corresponding Hessian, $H^{\bar{\tau}}_{\mathbf{mm}}$, the slowness Hessian, $H^{\bar{\tau}}_{\mathbf{pp}}$, and the mixed Hessian, $H^{\bar{\tau}}_{\mathbf{pm}}$, are needed to compute the gradient and the Hessian of the ray velocity magnitude wrt the model, $\nabla_{\mathbf{m}} v_{\text{ray}}$ and $\nabla_{\mathbf{m}}\nabla_{\mathbf{m}} v_{\text{ray}}$.

## 14. TTI MODEL 1: PARAMETER SETTING AND SLOWNESS INVERSION

Consider a polar anisotropic (TTI) model, whose properties $m_i$ and their relative (normalized) spatial gradients and Hessians are listed in Tables 1 and 2, respectively. The normalized gradient and Hessian are defined by,

$$\overline{\nabla_{\mathbf{x}} m_i} = \frac{\nabla_{\mathbf{x}} m_i}{m_i^o} \quad , \quad \overline{\nabla_{\mathbf{x}}\nabla_{\mathbf{x}} m_i} = \frac{\nabla_{\mathbf{x}}\nabla_{\mathbf{x}} m_i}{m_i^o} \quad , \quad (14.1)$$

where $m_i^o = m_i(\mathbf{x}_o)$ is the nodal value of each parameter $m_i$. Thus, in the proximity of the reference node $\mathbf{x}_o$, the value of the model parameter is assumed a quadratic function of the Cartesian coordinates,

$$\frac{m_i}{m_i^o} = 1 + \overline{\nabla_{\mathbf{x}} m_i} \cdot (\mathbf{x} - \mathbf{x}_o) + \frac{1}{2}(\mathbf{x} - \mathbf{x}_o) \cdot \overline{\nabla_{\mathbf{x}} \nabla_{\mathbf{x}} m_i} \cdot (\mathbf{x} - \mathbf{x}_o) \qquad (14.2)$$

Remark 1: These relative derivatives are presented solely to provide a better "feeling" of how fast the values change wrt their reference point values. In this specific example the values of $m_i$ at the reference (nodal) point and its proximity are far from zero, which makes this normalization possible and reasonable; otherwise, the absolute values of the first and second derivatives, $\nabla_{\mathbf{x}} m_i$ and $\nabla_{\mathbf{x}} \nabla_{\mathbf{x}} m_i$ should be specified directly.

Table 1. Properties of the polar anisotropy and their spatial gradients, model 1

| Component | | Value | Relative gradient, km$^{-1}$ (in global frame) | | |
|---|---|---|---|---|---|
| # | $m_i$ | | $x_1$ | $x_2$ | $x_3$ |
| 1 | $v_P$ | 3.5 km/s | +0.0842 | +0.0612 | –0.0308 |
| 2 | $f$ | 0.78 | +0.1022 | +0.1156 | +0.0923 |
| 3 | $\delta$ | +0.10 | +0.0914 | +0.0718 | +0.0310 |
| 4 | $\varepsilon$ | +0.25 | –0.0763 | –0.1080 | –0.0604 |
| 5 | $\gamma$ | +0.08 | +0.1123 | +0.0330 | –0.0431 |
| 6 | $\theta_{ax}$ | $30^o = \pi/6$ rad | –0.0854 | +0.1211 | –0.1163 |
| 7 | $\psi_{ax}$ | $45^o = \pi/4$ rad | +0.0221 | –0.1099 | +0.0529 |

Table 2. Spatial Hessians of polar anisotropy properties, model 1

| Component | | Relative Hessian, km$^{-2}$ (in global frame) | | | | | |
|---|---|---|---|---|---|---|---|
| # | $m_i$ | $x_1 x_1$ | $x_1 x_2$ | $x_1 x_3$ | $x_2 x_2$ | $x_2 x_3$ | $x_3 x_3$ |
| 1 | $v_P$ | +0.0290 | +0.0368 | −0.0972 | +0.0623 | −0.1077 | −0.0319 |
| 3 | $f$ | +0.0372 | +0.0263 | +0.1012 | −0.0871 | −0.0210 | +0.0414 |
| 3 | $\delta$ | +0.0963 | −0.1211 | −0.0655 | +0.0822 | −0.0553 | −0.0850 |
| 4 | $\varepsilon$ | −0.0281 | −0.0719 | +0.0269 | −0.0357 | +0.0543 | −0.0308 |
| 5 | $\gamma$ | +0.0355 | −0.0421 | +0.0101 | +0.0823 | −0.0311 | −0.0826 |
| 6 | $\theta_{ax}$ | −0.1172 | +0.1231 | +0.0633 | +0.0904 | +0.0308 | −0.0404 |
| 7 | $\psi_{ax}$ | +0.0615 | +0.0320 | −0.0801 | +0.0412 | +0.0765 | −0.0835 |

The symmetry axis direction angles lead to the following Cartesian components of the axis direction,

$$\mathbf{k} = \left[ \frac{\sqrt{2}}{4} \quad \frac{\sqrt{2}}{4} \quad \frac{\sqrt{3}}{2} \right] = [0.35355339 \quad 0.35355339 \quad 0.86602540] \quad . \tag{14.3}$$

The ray velocity direction is chosen to be,

$$\mathbf{r} = [+0.36 \quad +0.48 \quad +0.80] \quad . \tag{14.4}$$

For the given model parameters and the ray direction, we first perform the slowness inversion and obtain a single solution for each wave type: quasi-compressional qP, quasi-shear SV, and shear SH; the solutions are listed in Table 3. There is no qSV shear wave triplication for the given medium and ray velocity direction.

In addition, we obtain a solution for the acoustic approximation of qP waves, $qP_{ac}$, by setting $f = 1$.

Table 3. Results of the slowness vector inversion, model 1

| Mode | $p_1$, s/km | $p_2$, s/km | $p_3$, s/km | $v_{phs}$, km/s | $v_{ray}$, km/s |
|------|-------------|-------------|-------------|-----------------|-----------------|
| $qP_{ac}$ | 0.10254291 | 0.13092751 | 0.23182512 | 3.5049993 | 3.5060563 |
| qP | 0.10254249 | 0.13091618 | 0.23183152 | 3.5050011 | 3.5060621 |
| qSV | 0.21704016 | 0.24831725 | 0.51031286 | 1.6457988 | 1.6513176 |
| SH | 0.21875393 | 0.28185727 | 0.49281150 | 1.6436346 | 1.6439470 |

Remark 2: Normally, the accuracy of the input model parameters and the computed results does not exceed three or four digits. We keep eight digits here, as it is a reproducible benchmark problem. For the same reason, along with the final results, we provide also the most important intermediate values. The units of distances are km, the units of velocities are km/s. The reference Hamiltonian is unitless, and the arclength-related Hamiltonian has the units of slowness, s/km.

### 15. MODEL 1. RAY VELOCITY DERIVATIVES FOR qP WAVES

Gradients and Hessians of the Reference Hamiltonian

The gradient of the reference Hamiltonian wrt the material properties, $\mathbf{c} = \{v_P, f, \delta, \varepsilon\}$, is computed with equation 4.2,

$$H_{\mathbf{c}}^{\bar{\tau}} = \begin{bmatrix} +4.4373311 \cdot 10^{-1} & -1.1765600 \cdot 10^{-5} & +2.1314490 \cdot 10^{-2} & +3.8040622 \cdot 10^{-4} \end{bmatrix} , \quad (15.1)$$

where the first component of $H_\mathbf{c}^{\bar{\tau}}$ has the units of slowness, and the others are unitless. The Hessian of the reference Hamiltonian wrt the medium properties is computed with equation 7.2,

$$H_{\mathbf{cc}}^{\bar{\tau}} = \begin{bmatrix} -1.9158416 \cdot 10^{-2} & +5.6978709 \cdot 10^{-1} & +2.4359417 \cdot 10^{-2} & -1.5446185 \cdot 10^{-2} \\ +5.6978709 \cdot 10^{-1} & 0 & +2.7326269 \cdot 10^{-2} & +3.8618753 \cdot 10^{-4} \\ +2.4359417 \cdot 10^{-2} & +2.7326269 \cdot 10^{-2} & 0 & 0 \\ -1.5446185 \cdot 10^{-2} & +3.8618753 \cdot 10^{-4} & 0 & 0 \end{bmatrix}, \quad (15.2)$$

where the component in the first row and column, $H_{\mathbf{cc}}^{\bar{\tau}}(1,1)$ has the units of slowness squared, the other components in the first row and the first column have the units of slowness, and the remaining components are unitless. The zeros are related to $\frac{\partial^2 H^{\bar{\tau}}}{\partial f^2}, \frac{\partial^2 H^{\bar{\tau}}}{\partial \delta^2}, \frac{\partial^2 H^{\bar{\tau}}}{\partial \varepsilon^2}$, and $\frac{\partial^2 H^{\bar{\tau}}}{\partial \delta \partial \varepsilon}$.

The gradient and Hessian of the reference Hamiltonian wrt the symmetry axis direction vector are computed with equations 5.2 and 8.2, respectively,

$$H_{\mathbf{k}}^{\bar{\tau}} = \begin{bmatrix} -1.1636151 \cdot 10^{-1} & -1.4856218 \cdot 10^{-1} & -2.6307190 \cdot 10^{-1} \end{bmatrix} \text{ (unitless)}, \quad (15.3)$$

$$H_{\mathbf{kk}}^{\bar{\tau}} = \begin{bmatrix} +1.9497624 \cdot 10^{-1} & +2.4893193 \cdot 10^{-1} & +4.4080529 \cdot 10^{-1} \\ +2.4893193 \cdot 10^{-1} & +3.1781875 \cdot 10^{-1} & +5.6278914 \cdot 10^{-1} \\ +4.4080529 \cdot 10^{-1} & +5.6278914 \cdot 10^{-1} & +9.9657938 \cdot 10^{-1} \end{bmatrix} \text{ (unitless)}, \quad (15.4)$$

where the numerical components of the symmetry axis direction are given in equation 14.3. The first and second derivatives of this direction wrt the symmetry axis direction angles (zenith and azimuth), $\mathbf{a} = \{\theta_{ax} \quad \psi_{ax}\}$, are computed with equations 5.4 and 8.4, respectively,

$$\mathbf{k_a} = \begin{bmatrix} +6.1237244 \cdot 10^{-1} & -3.5355339 \cdot 10^{-1} \\ +6.1237244 \cdot 10^{-1} & +3.5355339 \cdot 10^{-1} \\ -5 \cdot 10^{-1} & 0 \end{bmatrix} \text{ radian}^{-1} \quad , \tag{15.5}$$

$$\mathbf{k_{aa}} = \begin{bmatrix} \begin{bmatrix} -3.5355339 \cdot 10^{-1} & -6.1237244 \cdot 10^{-1} \\ -6.1237244 \cdot 10^{-1} & -3.5355339 \cdot 10^{-1} \end{bmatrix} \\ \begin{bmatrix} -3.5355339 \cdot 10^{-1} & +6.1237244 \cdot 10^{-1} \\ +6.1237244 \cdot 10^{-1} & -3.5355339 \cdot 10^{-1} \end{bmatrix} \\ \begin{bmatrix} -8.6602540 \cdot 10^{-1} & 0 \\ 0 & 0 \end{bmatrix} \end{bmatrix} \text{ radian}^{-2} \quad . \tag{15.6}$$

The reference Hamiltonian gradient and Hessian wrt the symmetry axis direction angles are obtained with equations 5.1 and 8.1, respectively,

$$H_{\mathbf{a}}^{\bar{\tau}} = \begin{bmatrix} -3.0696020 \cdot 10^{-2} & -1.1384656 \cdot 10^{-2} \end{bmatrix} \text{ radian}^{-1} \quad , \tag{15.7}$$

$$H_{\mathbf{aa}}^{\bar{\tau}} = \begin{bmatrix} +3.3505997 \cdot 10^{-1} & -1.4686521 \cdot 10^{-2} \\ -1.4686521 \cdot 10^{-2} & +9.5531063 \cdot 10^{-2} \end{bmatrix} \text{ radian}^{-2} \quad . \tag{15.8}$$

The mixed Hessian of the reference Hamiltonian wrt medium properties and symmetry axis direction angles is computed with equation 9.2,

$$H_{\mathbf{ac}}^{\bar{\tau}} = \begin{bmatrix} +1.8565234 \cdot 10^{-2} & -3.9010957 \cdot 10^{-2} & -2.8393439 \cdot 10^{-1} & -9.2103212 \cdot 10^{-3} \\ +6.8855442 \cdot 10^{-3} & -1.4468532 \cdot 10^{-2} & -1.0530666 \cdot 10^{-1} & -3.4159588 \cdot 10^{-3} \end{bmatrix} . \tag{15.9}$$

The first column has the units of slowness divided by radians, and the other components' units are radian$^{-1}$. Next, we assemble the gradient and Hessian of the reference Hamiltonian wrt the model parameters (both material and geometric) according to equations 5.5 and 7.1, respectively.

The spatial gradient and Hessian of the reference Hamiltonian are computed with equations 6.1 and 10.1, respectively,

$$H_{\mathbf{x}}^{\bar{\tau}} = \begin{bmatrix} +1.3212974 \cdot 10^{-1} & +9.4225640 \cdot 10^{-2} & -4.6378730 \cdot 10^{-2} \end{bmatrix} \text{ km}^{-1} \quad , \tag{15.10}$$

$$H_{\mathbf{xx}}^{\bar{\tau}} = \begin{bmatrix} +7.2751789 \cdot 10^{-2} & +7.7488563 \cdot 10^{-2} & -1.4229782 \cdot 10^{-1} \\ +7.7488563 \cdot 10^{-2} & +1.1866113 \cdot 10^{-1} & -1.6670916 \cdot 10^{-1} \\ -1.4229782 \cdot 10^{-1} & -1.6670916 \cdot 10^{-1} & -5.5466724 \cdot 10^{-2} \end{bmatrix} \text{ km}^{-2} \quad . \tag{15.11}$$

The slowness gradient and Hessian of the reference Hamiltonian are computed with equations 11.1 and 12.1, respectively,

$$H_{\mathbf{p}}^{\bar{\tau}} = \begin{bmatrix} +1.9602510 & +2.6137340 & +4.3560698 \end{bmatrix} \text{ km/s} \quad , \tag{15.12}$$

$$H_{\mathbf{pp}}^{\bar{\tau}} = \begin{bmatrix} +19.652092 & -4.7411750 & -7.3413575 \\ -4.7411750 & +16.604434 & -10.465764 \\ -7.3413575 & -10.465764 & +7.1765820 \end{bmatrix} (\text{km/s})^2 \quad . \tag{15.13}$$

The mixed Hessians "slowness-material properties", "slowness-axis of symmetry", and "slowness-geometric properties" of the reference Hamiltonian are computed with equations 13.3, 13.6 and 13.5, respectively,

$$H_{\mathbf{pc}}^{\bar{\tau}} = \begin{bmatrix} +4.7222015 \cdot 10^{-1} & +2.5170615 & +1.425168 \cdot 10^{-1} & -6.6964429 \cdot 10^{-2} \\ +5.3574936 \cdot 10^{-1} & +3.3547044 & +1.2089096 & -5.2182917 \cdot 10^{-2} \\ +1.1133842 & +5.5944109 & -3.7797376 \cdot 10^{-1} & -1.7410672 \cdot 10^{-1} \end{bmatrix} \quad , \tag{15.14}$$

where the first column is unitless, and the three others have the units of velocity,

$$H_{\mathbf{pk}}^{\bar{\tau}} = \begin{bmatrix} -6.8424181 \cdot 10^{-1} & +5.7519181 \cdot 10^{-1} & +1.0185419 \\ +3.8915989 \cdot 10^{-1} & -6.3790958 \cdot 10^{-1} & +8.7981866 \cdot 10^{-1} \\ +1.1453785 & +1.4623386 & +1.4547279 \end{bmatrix} \text{ km/s} \quad , \tag{15.15}$$

$$H_{\mathbf{pa}}^{\bar{\tau}} = \begin{bmatrix} -5.7605014 \cdot 10^{-1} & +4.4527703 \cdot 10^{-1} \\ -5.9223678 \cdot 10^{-1} & -3.6312389 \cdot 10^{-1} \\ +8.6953015 \cdot 10^{-1} & +1.1206231 \cdot 10^{-1} \end{bmatrix} \text{ km}/(\text{radian} \cdot \text{s}) \quad . \tag{15.16}$$

The mixed Hessian "slowness-model properties", $H_{\mathbf{pm}}^{\bar{\tau}}$, is then assembled according to equation 13.2. The mixed "slowness-spatial" Hessian of the reference Hamiltonian, $H_{\mathbf{px}}^{\bar{\tau}}$, is computed with equation 13.1,

$$H_{\mathbf{px}}^{\bar{\tau}} = \begin{bmatrix} +3.7588038 \cdot 10^{-1} & +2.5597898 \cdot 10^{-1} & +1.8533944 \cdot 10^{-1} \\ +4.5753299 \cdot 10^{-1} & +4.2112410 \cdot 10^{-1} & +2.0927746 \cdot 10^{-1} \\ +7.3700854 \cdot 10^{-1} & +7.9037307 \cdot 10^{-1} & +2.3590478 \cdot 10^{-1} \end{bmatrix} \text{ s}^{-1} \quad . \quad (15.17)$$

This completes the computation of the derivatives of the reference Hamiltonian. The resulting gradients are $H_{\mathbf{x}}^{\bar{\tau}}$, $H_{\mathbf{p}}^{\bar{\tau}}$, and $H_{\mathbf{m}}^{\bar{\tau}}$, and the resulting Hessians are $H_{\mathbf{xx}}^{\bar{\tau}}$, $H_{\mathbf{pp}}^{\bar{\tau}}$, $H_{\mathbf{mm}}^{\bar{\tau}}$, $H_{\mathbf{px}}^{\bar{\tau}}$, and $H_{\mathbf{pm}}^{\bar{\tau}}$; all other vectors and matrices are intermediate results.

Gradients and Hessians of the arclength-related Hamiltonian

Next, we compute these three gradients and five Hessians for the arclength-related Hamiltonian, $H(\mathbf{x}, \mathbf{p})$, using the conversion formulae derived in Part I. Since these conversion formulae do not include the medium properties, they are universal and can be equally applied to all anisotropic symmetries.

The spatial gradient and Hessian of the arclength-related Hamiltonian are computed with equations 4.10 and 4.12 of Part I,

$$H_{\mathbf{x}} = \begin{bmatrix} +2.4265528 \cdot 10^{-2} & +1.7304482 \cdot 10^{-2} & -8.5175169 \cdot 10^{-3} \end{bmatrix} \text{ s/km}^2 \quad , \quad (15.18)$$

$$H_{\mathbf{xx}} = \begin{bmatrix} +4.9423420 \cdot 10^{-3} & +7.0998463 \cdot 10^{-3} & -2.6241540 \cdot 10^{-2} \\ +7.0998463 \cdot 10^{-3} & +1.5902803 \cdot 10^{-2} & -3.0297691 \cdot 10^{-2} \\ -2.6241540 \cdot 10^{-2} & -3.0297691 \cdot 10^{-2} & -9.0730785 \cdot 10^{-3} \end{bmatrix} \text{ s/km}^3 \quad . \quad (15.19)$$

The slowness gradient and Hessian of the reference Hamiltonian are computed with equations 4.10 and 4.13 of Part I,

$$H_\mathbf{p} = [+0.36 \quad +0.48 \quad +0.80] = \mathbf{r} \quad (\text{unitless}) \quad , \tag{15.20}$$

$$H_\mathbf{pp} = \begin{bmatrix} +3.7511331 & -6.3658517 \cdot 10^{-1} & -1.0631631 \\ -6.3658517 \cdot 10^{-1} & +3.4212749 & -1.4424407 \\ -1.0631631 & -1.4424407 & +1.8836560 \end{bmatrix} \quad \text{km/s} \quad . \tag{15.21}$$

We will need also the inverse matrix of the slowness Hessian,

$$H_\mathbf{pp}^{-1} = \begin{bmatrix} +4.4590566 \cdot 10^{-1} & +2.7922631 \cdot 10^{-1} & +4.6549785 \cdot 10^{-1} \\ +2.7922631 \cdot 10^{-1} & +6.0649970 \cdot 10^{-1} & +6.2203658 \cdot 10^{-1} \\ +4.6549785 \cdot 10^{-1} & +6.2203658 \cdot 10^{-1} & +1.2699511 \end{bmatrix} \quad \text{s/km} \quad . \tag{15.22}$$

The eigenvalues of the inverse Hessian are all positive, as the slowness surface of compressional waves is convex,

$$\lambda = \begin{bmatrix} +1.8526471 & +2.3644739 \cdot 10^{-1} & +2.3326198 \cdot 10^{-1} \end{bmatrix} \quad \text{s/km} \quad . \tag{15.23}$$

The mixed "slowness-space" Hessian of the arclength Hamiltonian is computed with equation 4.14 of Part I,

$$H_\mathbf{px} = \begin{bmatrix} +1.1369338 \cdot 10^{-2} & -1.0836831 \cdot 10^{-2} & +8.8275821 \cdot 10^{-3} \\ +1.0158733 \cdot 10^{-2} & +2.3586643 \cdot 10^{-3} & +3.7607307 \cdot 10^{-3} \\ +5.1607592 \cdot 10^{-3} & +1.5136887 \cdot 10^{-2} & -1.1975706 \cdot 10^{-2} \end{bmatrix} \quad \text{km}^{-1} \quad . \tag{15.24}$$

The gradient of the arclength-related Hamiltonian wrt the model properties is computed with equation 17.13 of Part I. It is presented here as consisting of two parts,

$$H_\mathbf{m} = H_\mathbf{c} \| H_\mathbf{a} \quad , \tag{15.25}$$

where the material part is,

$$H_{\mathbf{c}} = \begin{bmatrix} +8.1491509 \cdot 10^{-2} & -2.7493568 \cdot 10^{-6} & +3.9136076 \cdot 10^{-3} & +6.9849459 \cdot 10^{-5} \end{bmatrix}, \quad (15.26)$$

and the geometric part is,

$$H_{\mathbf{a}} = \begin{bmatrix} -5.6367353 \cdot 10^{-3} & -2.0905715 \cdot 10^{-3} \end{bmatrix} \text{ s}/(\text{km} \cdot \text{radian}) \quad . \quad (15.27)$$

The first component of the material part, $H_{\mathbf{c}}$, has the units of slowness squared; the other components of this vector have the units of slowness. The Hessian of the arclength-related Hamiltonian wrt the model properties is computed with equation 18.9 of Part I. In this part (Part II), this matrix consists of the four blocks (material, geometric and mixed),

$$H_{\mathbf{mm}} = \begin{bmatrix} H_{\mathbf{cc}} & H_{\mathbf{ca}} \\ H_{\mathbf{ac}} & H_{\mathbf{aa}} \end{bmatrix}, \quad (15.28)$$

where the two diagonal blocks are symmetric, and the two off-diagonal blocks are transposed of each other. The material block reads,

$$H_{\mathbf{cc}} = \begin{bmatrix} -4.2964811 \cdot 10^{-2} & +1.3160941 \cdot 10^{-6} & -1.4006158 \cdot 10^{-3} & -3.3416162 \cdot 10^{-5} \\ +1.3160941 \cdot 10^{-6} & +7.0606624 \cdot 10^{-6} & -7.6839587 \cdot 10^{-6} & -1.8891415 \cdot 10^{-5} \\ -1.4006158 \cdot 10^{-3} & -7.6839587 \cdot 10^{-6} & -4.7314960 \cdot 10^{-4} & +1.3118816 \cdot 10^{-4} \\ -3.3416162 \cdot 10^{-5} & -1.8891415 \cdot 10^{-5} & +1.3118816 \cdot 10^{-4} & +4.8335714 \cdot 10^{-6} \end{bmatrix}, \quad (15.29)$$

where the component $H_{\mathbf{cc}}(1,1)$ has the units of slowness to power three, the other elements of the first row and the first column have the units of slowness squared, and the rest of the components have the units of slowness. The mixed "geometric-material" block reads,

$$H_{\mathbf{ac}} = \begin{bmatrix} +1.7212877 \cdot 10^{-3} & +7.4437213 \cdot 10^{-5} & -5.1945246 \cdot 10^{-2} & -1.8888263 \cdot 10^{-3} \\ +6.3839704 \cdot 10^{-4} & +2.7607526 \cdot 10^{-5} & -1.9265629 \cdot 10^{-2} & -7.0053432 \cdot 10^{-4} \end{bmatrix}, \quad (14.34)$$

where the first column has the units of slowness squared divided by radians, and the other components have the units of slowness divided by radians. The geometric block reads,

$$H_{aa} = \begin{bmatrix} +6.1955323 \cdot 10^{-2} & -2.5403875 \cdot 10^{-3} \\ -2.5403875 \cdot 10^{-3} & +1.7601993 \cdot 10^{-2} \end{bmatrix} \frac{s}{km \cdot radian^2} \quad . \tag{15.30}$$

Eventually, the $6 \times 3$ mixed "model-slowness" Hessian $H_{mp}$ of the arclength-related Hamiltonian is computed with equation 18.10 of Part I,

$$H_{mp} = \begin{bmatrix} +1.5667574 \cdot 10^{-2} & +1.3777136 \cdot 10^{-2} & +3.9666468 \cdot 10^{-2} \\ -6.8679118 \cdot 10^{-6} & -2.7799839 \cdot 10^{-4} & +1.6803458 \cdot 10^{-4} \\ +5.1798463 \cdot 10^{-3} & +1.9449259 \cdot 10^{-1} & -1.1638594 \cdot 10^{-1} \\ +1.7428814 \cdot 10^{-4} & +7.0541632 \cdot 10^{-3} & -4.2637990 \cdot 10^{-3} \\ -1.2038848 \cdot 10^{-1} & -1.2892618 \cdot 10^{-1} & +1.2772736 \cdot 10^{-1} \\ +7.6361026 \cdot 10^{-2} & -7.4165347 \cdot 10^{-2} & +8.7262162 \cdot 10^{-3} \end{bmatrix} \quad . \tag{15.31}$$

The components of the first row have the units of slowness, the components of rows 2, 3, 4 are unitless, and the components of rows 5, 6 have the units $radian^{-1}$.

Gradients and Hessians of the ray velocity magnitude

Until now we have computed all necessary derivatives of the arclength-related Hamiltonian, needed to establish the required gradients and Hessians of the ray velocity magnitude. The computation of the compressional ray velocity yields $v_{ray} = 3.5060621 \, km/s$. The spatial gradient and Hessian of the ray velocity magnitude are computed with equations 7.6 and 8.6 of Part I, respectively,

$$\nabla_x v_{ray} = \begin{bmatrix} +2.9828330 \cdot 10^{-1} & +2.1271485 \cdot 10^{-1} & -1.0470133 \cdot 10^{-1} \end{bmatrix} s^{-1} \quad , \tag{15.32}$$

$$\nabla_x \nabla_x v_{ray} = \begin{bmatrix} +1.0734758 \cdot 10^{-1} & +1.2109813 \cdot 10^{-1} & -3.3941305 \cdot 10^{-1} \\ +1.2109813 \cdot 10^{-1} & +2.1854053 \cdot 10^{-1} & -3.8350808 \cdot 10^{-1} \\ -3.3941305 \cdot 10^{-1} & -3.8350808 \cdot 10^{-1} & -1.0637793 \cdot 10^{-1} \end{bmatrix} (km \cdot s)^{-1} \quad . \tag{15.33}$$

The directional gradient and Hessian of the ray velocity magnitude are computed with equations 9.2 and 10.3 of Part I, respectively,

$$\nabla_{\mathbf{r}} v_{\text{ray}} = \begin{bmatrix} +1.6817488 \cdot 10^{-3} & +7.3626471 \cdot 10^{-2} & -4.4932669 \cdot 10^{-2} \end{bmatrix} \text{ km/s} \quad , \quad (15.34)$$

$$\nabla_{\mathbf{r}} \nabla_{\mathbf{r}} v_{\text{ray}} = \begin{bmatrix} +5.2064468 \cdot 10^{-1} & -1.3019110 \cdot 10^{-1} & -1.5827763 \cdot 10^{-1} \\ -1.3019110 \cdot 10^{-1} & +4.2233608 \cdot 10^{-1} & -2.8684874 \cdot 10^{-1} \\ -1.5827763 \cdot 10^{-1} & -2.8684874 \cdot 10^{-1} & +2.9950002 \cdot 10^{-1} \end{bmatrix} \text{ km/s} \quad , \quad (15.35)$$

and the mixed "spatial-directional" Hessian of the ray velocity magnitude is computed with equation 11.5 of Part I,

$$\nabla_{\mathbf{x}} \nabla_{\mathbf{r}} v_{\text{ray}} = \begin{bmatrix} +1.9621841 \cdot 10^{-2} & +2.3573946 \cdot 10^{-2} & -2.2974196 \cdot 10^{-2} \\ -4.1062108 \cdot 10^{-2} & +2.9615883 \cdot 10^{-3} & +1.6700996 \cdot 10^{-2} \\ +3.0360352 \cdot 10^{-2} & +1.2625893 \cdot 10^{-2} & -2.1237694 \cdot 10^{-2} \end{bmatrix} \text{ s}^{-1} \quad . \quad (15.36)$$

The gradient of the ray velocity wrt the model parameters was computed with equation 17.13 of Part I. We present the gradient as a concatenation of the material and geometric counterparts,

$$\nabla_{\mathbf{m}} v_{\text{ray}} = \nabla_{\mathbf{c}} v_{\text{ray}} \, \| \, \nabla_{\mathbf{a}} v_{\text{ray}} \quad , \quad (15.37)$$

where the material part reads,

$$\nabla_{\mathbf{c}} v_{\text{ray}} = \begin{bmatrix} +1.001732 & -3.3796388 \cdot 10^{-5} & +4.8107908 \cdot 10^{-2} & +8.5862246 \cdot 10^{-4} \end{bmatrix} \quad , \quad (15.38)$$

and the geometric part (symmetry axis direction angles) reads,

$$\nabla_{\mathbf{a}} v_{\text{ray}} = \begin{bmatrix} -6.9289406 \cdot 10^{-2} & -2.569829 \cdot 10^{-2} \end{bmatrix} \text{ km/(s} \cdot \text{radian)} \quad . \quad (15.39)$$

Observations:

The first component of $\nabla_{\mathbf{c}} v_{\text{ray}}$ is unitless, while the other components of this vector have the units of velocity. As expected, parameter $v_P$ (the axial compressional velocity) directly affects the ray velocity magnitude $v_{\text{ray}}$, and the derivative $\partial v_{\text{ray}} / \partial v_P$ in realistic (weakly and moderately anisotropic) cases is normally close to 1; in this example, despite the relatively strong anisotropy, the value of the derivative differs from 1 by $1.73 \cdot 10^{-3}$. Parameter $\partial v_{\text{ray}} / \partial f$ (the second component of array $\nabla_{\mathbf{c}} v_{\text{ray}}$) is fairly small, which is also not a surprise, as it was shown by Alkhalifah (1998) for polar anisotropy and Alkhalifah (2003) for orthorhombic media that the axial shear velocity $v_S$ (or alternatively parameter $f$) has a negligible effect on the propagation of qP waves. It is suitable to apply the normalized characteristic, which proves to be a tiny fraction of percent,

$$\frac{1}{v_{\text{ray}}} \frac{\partial v_{\text{ray}}}{\partial f} = -9.6394154 \cdot 10^{-6} \quad \text{(unitless)} \quad . \tag{15.40}$$

The Hessian of the ray velocity wrt the model parameters $\nabla_{\mathbf{m}} \nabla_{\mathbf{m}} v_{\text{ray}}$ is computed with equation 18.6 of Part I. This $6 \times 6$ matrix (for qP and qSV waves) consists of four blocks: two symmetric diagonal blocks ($\nabla_{\mathbf{c}} \nabla_{\mathbf{c}} v_{\text{ray}}$ and $\nabla_{\mathbf{a}} \nabla_{\mathbf{a}} v_{\text{ray}}$ of dimensions $4 \times 4$ and $2 \times 2$) related to either material or geometric properties of the model, respectively, and two mutually transposed off-diagonal blocks ($2 \times 4$ and $4 \times 2$, $\nabla_{\mathbf{a}} \nabla_{\mathbf{c}} v_{\text{ray}}$ and $\nabla_{\mathbf{c}} \nabla_{\mathbf{a}} v_{\text{ray}}$, respectively) related to both, the material and geometric properties,

$$\nabla_{\mathbf{m}} \nabla_{\mathbf{m}} v_{\text{ray}} = \begin{bmatrix} \nabla_{\mathbf{c}} \nabla_{\mathbf{c}} v_{\text{ray}} & \nabla_{\mathbf{c}} \nabla_{\mathbf{a}} v_{\text{ray}} \\ \nabla_{\mathbf{a}} \nabla_{\mathbf{c}} v_{\text{ray}} & \nabla_{\mathbf{a}} \nabla_{\mathbf{a}} v_{\text{ray}} \end{bmatrix} \quad . \tag{15.41}$$

The material block reads,

$$\nabla_{\mathbf{c}}\nabla_{\mathbf{c}}v_{\text{ray}} = \begin{bmatrix} -5.5511151 \cdot 10^{-16} & -9.6561110 \cdot 10^{-6} & +1.3745117 \cdot 10^{-2} & +2.4532070 \cdot 10^{-4} \\ -9.6561110 \cdot 10^{-6} & +8.6490900 \cdot 10^{-5} & +1.1589334 \cdot 10^{-4} & -2.2455677 \cdot 10^{-4} \\ +1.3745117 \cdot 10^{-2} & +1.1589334 \cdot 10^{-4} & -1.5196709 \cdot 10^{-1} & -3.7248805 \cdot 10^{-3} \\ +2.4532070 \cdot 10^{-4} & -2.2455677 \cdot 10^{-4} & -3.7248805 \cdot 10^{-3} & -1.3509062 \cdot 10^{-4} \end{bmatrix}.$$

(15.42)

The component in the first row and the first column, $\partial^2 v_{\text{ray}} / \partial v_P^2$, has the units of slowness. To evaluate whether its magnitude is large or small, we normalize this number, using the ray velocity $v_{\text{ray}} = 3.5060621 \text{ km/s}$ as a scaling factor. This leads to a unitless characteristic number of $v_{\text{ray}} \partial^2 v_{\text{ray}} / \partial v_P^2 \approx -1.95 \cdot 10^{-15}$, which can be considered comparable to a "machine epsilon" for the eight-byte (double precision) arithmetic that we use; in other words, $\partial^2 v_{\text{ray}} / \partial v_P^2 = 0$. This is an expected result (confirmed also by the finite-difference numerical evaluation of this derivative). Parameter $v_P$ is a scaler for the ray and phase velocities. Should we keep fixed the other (material and geometric) parameters of the model, the wave type, and the ray velocity direction $\mathbf{r}$, then both the ray and phase velocities become proportional to $v_P$. The ray and phase velocities are first-degree homogeneous functions of the axial compressional velocity $v_P$, and the Euler theorem holds (e.g., Buchanan and Yoon, 1999); for this simple case it reduces to,

$$\frac{\partial v_{\text{ray}}}{\partial v_P} = \frac{v_{\text{ray}}}{v_P} \quad \rightarrow \quad v_{\text{ray}} = A v_P, \quad A = \text{const} \quad , \quad (15.43)$$

where "const" means independent of $v_P$ (but still depending, of course, on all the other model parameters and the ray direction). The derivative $\partial v_{\text{ray}} / \partial v_P$ on the left-hand side is the first component of the array in equation 15.38, where $v_P = 3.5 \text{ km/s}$. Introducing the numbers, we make sure that equality 15.43 is honored; for the phase velocity, a similar property follows directly from equation 1.2.

The other components in the first row and the first column of the matrix in equation 15.42 (except the vanishing $\partial^2 v_{ray} / \partial v_P^2$) are unitless; the rest of the components in this matrix have the units of velocity. The mixed block reads,

$$\nabla_{\mathbf{a}}\nabla_{\mathbf{c}} v_{ray} = \begin{bmatrix} -1.9796973 \cdot 10^{-3} & +7.4958122 \cdot 10^{-4} & -5.2399585 \cdot 10^{-1} & -1.9020481 \cdot 10^{-2} \\ -7.3423686 \cdot 10^{-3} & +2.7800723 \cdot 10^{-4} & -1.9434136 \cdot 10^{-1} & -7.0543807 \cdot 10^{-3} \end{bmatrix}.$$

(15.44)

The first row of the mixed block has the units $\text{radian}^{-1}$, while the other components have the units $\text{km}/(\text{s} \cdot \text{radian})$. Finally, the geometric block reads,

$$\nabla_{\mathbf{a}}\nabla_{\mathbf{a}} v_{ray} = \begin{bmatrix} +6.2753676 \cdot 10^{-1} & -3.4126970 \cdot 10^{-2} \\ -3.4126970 \cdot 10^{-2} & +1.837392 \cdot 10^{-1} \end{bmatrix} \frac{\text{km}}{\text{s} \cdot \text{radian}^2} \quad . \quad (15.45)$$

Validation test

In order to check the correctness of all the ray velocity derivatives, obtained analytically, we compare their values with numerical values obtained by the finite differences method. The numerical directional derivatives of the ray velocity, approximating the gradient and the Hessians $\nabla_{\mathbf{r}}\mathbf{v}_{ray}, \nabla_{\mathbf{r}}\nabla_{\mathbf{r}} v_{ray}$ and the mixed Hessian $\nabla_{\mathbf{x}}\nabla_{\mathbf{r}} v_{ray}$, were computed wrt the zenith and azimuth angles, $\theta_{ray}$ and $\psi_{ray}$, of the ray direction vector, and then were related analytically to the Cartesian components of this direction, as explained in Appendix F of Part I. The numerical approximations for the ray velocity derivatives were computed with the spatial resolution $10^{-4}$ km and angular resolution $10^{-5}$ radians for the spatial and directional gradients and Hessians of the ray velocity magnitude, including the mixed Hessian. When computing the numerical ray velocity derivatives wrt the model parameters, $\nabla_{\mathbf{m}} v_{\text{ray}}$ and $\nabla_{\mathbf{m}} \nabla_{\mathbf{m}} v_{\text{ray}}$, the resolution was $8 \cdot 10^{-4}$ times the absolute value of the parameter. The (unitless) relative errors of the numerical approximations of the ray velocity derivatives for qP waves are listed below.

- For the spatial gradient and Hessian,

$$E(\nabla_{\mathbf{x}} v_{\text{ray}}) = \begin{bmatrix} +4.41 \cdot 10^{-11} & +1.18 \cdot 10^{-11} & -1.06 \cdot 10^{-11} \end{bmatrix} \quad , \tag{15.46}$$

$$E(\nabla_{\mathbf{x}} \nabla_{\mathbf{x}} v_{\text{ray}}) = \begin{bmatrix} -1.91 \cdot 10^{-6} & -3.31 \cdot 10^{-7} & +1.87 \cdot 10^{-9} \\ -3.31 \cdot 10^{-7} & -4.62 \cdot 10^{-7} & -8.77 \cdot 10^{-9} \\ +1.87 \cdot 10^{-9} & -8.77 \cdot 10^{-9} & +1.67 \cdot 10^{-6} \end{bmatrix} \quad . \tag{15.47}$$

- For the directional gradient and Hessian,

$$E(\nabla_{\mathbf{r}} v_{\text{ray}}) = \begin{bmatrix} +4.73 \cdot 10^{-8} & +1.13 \cdot 10^{-10} & +9.07 \cdot 10^{-10} \end{bmatrix} \quad , \tag{15.48}$$

$$E(\nabla_{\mathbf{r}} \nabla_{\mathbf{r}} v_{\text{ray}}) = \begin{bmatrix} +1.82 \cdot 10^{-6} & +5.80 \cdot 10^{-5} & -2.59 \cdot 10^{-5} \\ +5.80 \cdot 10^{-5} & -2.14 \cdot 10^{-6} & -1.37 \cdot 10^{-5} \\ -2.59 \cdot 10^{-5} & -1.37 \cdot 10^{-5} & -1.41 \cdot 10^{-5} \end{bmatrix} \quad . \tag{15.49}$$

- For the mixed Hessian,

$$E(\nabla_{\mathbf{x}} \nabla_{\mathbf{r}} v_{\text{ray}}) = \begin{bmatrix} +1.20 \cdot 10^{-5} & -1.97 \cdot 10^{-5} & -7.54 \cdot 10^{-6} \\ -3.25 \cdot 10^{-6} & +1.59 \cdot 10^{-5} & -5.29 \cdot 10^{-6} \\ +1.24 \cdot 10^{-5} & -3.32 \cdot 10^{-5} & -3.89 \cdot 10^{-6} \end{bmatrix} \quad . \tag{15.50}$$

- For the material, $\nabla_{\mathbf{c}} v_{\text{ray}}$, and geometric, $\nabla_{\mathbf{a}} v_{\text{ray}}$, parts of the model-related gradient $\nabla_{\mathbf{m}} v_{\text{ray}}$,

$$E\left(\nabla_{\mathbf{c}} v_{\text{ray}}\right) = \begin{bmatrix} +1.51 \cdot 10^{-13} & +6.50 \cdot 10^{-7} & +1.58 \cdot 10^{-8} & -7.90 \cdot 10^{-10} \end{bmatrix} \quad , \tag{15.51}$$

$$E\left(\nabla_{\mathbf{a}} v_{\text{ray}}\right) = \begin{bmatrix} +2.71 \cdot 10^{-7} & +1.39 \cdot 10^{-7} \end{bmatrix} \quad . \tag{15.52}$$

- For the material, $\nabla_{\mathbf{c}} \nabla_{\mathbf{c}} v_{\text{ray}}$, geometric, $\nabla_{\mathbf{a}} \nabla_{\mathbf{a}} v_{\text{ray}}$, and mixed, $\nabla_{\mathbf{a}} \nabla_{\mathbf{c}} v_{\text{ray}}$, parts of the model-related Hessian $\nabla_{\mathbf{m}} \nabla_{\mathbf{m}} v_{\text{ray}}$,

$$E\left(\nabla_{\mathbf{c}} \nabla_{\mathbf{c}} v_{\text{ray}}\right) = \begin{bmatrix} - & +7.41 \cdot 10^{-6} & +7.68 \cdot 10^{-8} & -1.11 \cdot 10^{-6} \\ +7.41 \cdot 10^{-6} & +7.31 \cdot 10^{-7} & -4.24 \cdot 10^{-5} & +2.41 \cdot 10^{-5} \\ +7.68 \cdot 10^{-8} & -4.24 \cdot 10^{-5} & +1.10 \cdot 10^{-6} & +5.86 \cdot 10^{-6} \\ -1.11 \cdot 10^{-6} & +2.41 \cdot 10^{-5} & +5.86 \cdot 10^{-6} & +9.72 \cdot 10^{-6} \end{bmatrix} \quad , \tag{15.53}$$

$$E\left(\nabla_{\mathbf{a}} \nabla_{\mathbf{c}} v_{\text{ray}}\right) = \begin{bmatrix} +2.72 \cdot 10^{-7} & +9.17 \cdot 10^{-6} & -4.80 \cdot 10^{-7} & +7.67 \cdot 10^{-6} \\ +1.47 \cdot 10^{-7} & +1.09 \cdot 10^{-5} & -3.42 \cdot 10^{-7} & +5.43 \cdot 10^{-6} \end{bmatrix} \quad , \tag{15.54}$$

$$E\left(\nabla_{\mathbf{a}} \nabla_{\mathbf{a}} v_{\text{ray}}\right) = \begin{bmatrix} +1.07 \cdot 10^{-7} & +1.08 \cdot 10^{-6} \\ +1.08 \cdot 10^{-6} & +4.77 \cdot 10^{-8} \end{bmatrix} \quad . \tag{15.55}$$

A dash has been placed to the upper left corner of $E\left(\nabla_{\mathbf{c}} \nabla_{\mathbf{c}} v_{\text{ray}}\right)$ because the relative error cannot be estimated for a parameter whose exact theoretical value is zero; however, the numerical (finite difference) estimate was also exact zero (i.e., the result of subtracting close values in the finite-difference approximation of the second derivative $\partial^2 v_{\text{ray}} / \partial v_P^2$ proved to be below the machine epsilon for double-precision arithmetic).

We performed one more validation test. In the proposed workflow, we used the reference and arclength-related Hamiltonians to compute the spatial gradient and Hessian of the ray velocity. However, this gradient and Hessian can also be obtained through a different chain of operations that involves the derivatives of the ray velocity wrt the model properties, $\nabla_{\mathbf{m}} v_{\text{ray}}$ and $\nabla_{\mathbf{m}} \nabla_{\mathbf{m}} v_{\text{ray}}$, and the spatial derivatives of the model properties, $\mathbf{m_x}$ and $\mathbf{m_{xx}}$,

$$\begin{aligned} \nabla_{\mathbf{x}} v_{\text{ray}} &= \nabla_{\mathbf{m}} v_{\text{ray}} \cdot \mathbf{m_x} \\ \nabla_{\mathbf{x}} \nabla_{\mathbf{x}} v_{\text{ray}} &= \mathbf{m_x}^T \cdot \nabla_{\mathbf{m}} \nabla_{\mathbf{m}} v_{\text{ray}} \cdot \mathbf{m_x} + \nabla_{\mathbf{m}} \cdot \mathbf{m_{xx}} \end{aligned}, \quad (15.56)$$

where (as mentioned in Part 1) $\nabla_{\mathbf{m}} v_{\text{ray}}$ is a 1D array (not a physical vector), $\nabla_{\mathbf{m}} \nabla_{\mathbf{m}} v_{\text{ray}}$ is a 2D array (not a tensor and not even a collection of vectors), $\mathbf{m_x}$ is another 2D array representing a collection of vectors, and $\mathbf{m_{xx}}$ is a 3D array representing a collection of second-order tensors. However, the multiplications in the above equation set are performed as if all these items were vectors and tensors of the corresponding orders; this allows a shorthand notation where summations and multiplications are hidden. Applying the chain rule of equation 15.56 does not show any apparent advantages, because the gradient and Hessian of the ray velocity wrt the model parameters, $\nabla_{\mathbf{m}} v_{\text{ray}}$ and $\nabla_{\mathbf{m}} \nabla_{\mathbf{m}} v_{\text{ray}}$, respectively, have also been computed with the use of the derivatives of the reference and the arclength-related Hamiltonians; this equation is therefore only used for double-checking the correctness of the resulting spatial derivatives. These tests revealed full coincidence of the spatial gradient and Hessian of the ray velocity, $\nabla_{\mathbf{x}} v_{\text{ray}}$ and $\nabla_{\mathbf{x}} \nabla_{\mathbf{x}} v_{\text{ray}}$, computed with our proposed workflow, and with equation 15.56; the relative errors for all components were of order $10^{-16}$, i.e., within the machine accuracy.

Sensitivity of the ray velocity derivatives to the spatial derivatives of the model parameters

Our next test checks the significance (the effect) of the spatial gradients and Hessians of the model parameters $f, \delta, \varepsilon, \theta_{ax}$ and $\psi_{ax}$ (all model parameters except the axial compressional velocity) on the derivatives of the ray velocity for compressional waves. For this, we repeat the computations, assuming $\mathbf{m_x} = 0$ and $\mathbf{m_{xx}} = 0$, except $\nabla_\mathbf{x} v_P$ and $\nabla_\mathbf{x} \nabla_\mathbf{x} v_P$ which are taken into account. Obviously, this does not affect the directional derivatives, $\nabla_\mathbf{r} v_{\mathrm{ray}}$ and $\nabla_\mathbf{r} \nabla_\mathbf{r} v_{\mathrm{ray}}$, and the derivatives wrt the medium properties, $\nabla_\mathbf{m} v_{\mathrm{ray}}$ and $\nabla_\mathbf{m} \nabla_\mathbf{m} v_{\mathrm{ray}}$. We re-compute the spatial derivatives, $\nabla_\mathbf{x} v_{\mathrm{ray}}$ and $\nabla_\mathbf{x} \nabla_\mathbf{x} v_{\mathrm{ray}}$ and the mixed Hessian $\nabla_\mathbf{x} \nabla_\mathbf{r} v_{\mathrm{ray}}$, and compare the vector and the two tensors obtained for three cases: 1) accounting for the spatial derivatives of all model parameters, 2) accounting for the spatial derivatives of the medium properties and ignoring those of the symmetry axis direction angles, and 3) accounting only for the spatial derivatives of the axial compressional velocity. For cases 2 and 3, we use notations $\tilde{\nabla}_\mathbf{x}$ and $\hat{\nabla}_\mathbf{x}$, respectively. Case 3 is referred as FAI media – Factorized Anisotropic Inhomogeneous media (Červený, 2000). To compare the spatial gradients, we compute the ratio of their magnitudes and the angle between their directions; to compare the spatial Hessians, we compute the ratio of their corresponding eigenvalues. The non-symmetric mixed Hessian, $\nabla_\mathbf{x} \nabla_\mathbf{r} v_{\mathrm{ray}}$, has a single zero eigenvalue, and the corresponding eigenvector is the ray direction $\mathbf{r}$ (Koren and Ravve, 2021); this property holds even if we ignore the spatial derivatives of the model, so we compare the two other eigenvalues, which prove to be real for the given example and for qP waves.

Accounting for the spatial derivatives of the medium properties, $\mathbf{c_x}$ and $\mathbf{c_{xx}}$ and ignoring those of the symmetry axis direction angles, $\mathbf{a_x}$ and $\mathbf{a_{xx}}$, we obtained the following results,

$$\tilde{\nabla}_\mathbf{x} v_{\mathrm{ray}} = \begin{bmatrix} +2.9563106 \cdot 10^{-1} & +2.1489018 \cdot 10^{-1} & -1.0785298 \cdot 10^{-2} \end{bmatrix} \mathrm{s}^{-1} \quad , \quad (15.57)$$

$$\tilde{\nabla}_{\mathbf{x}}\tilde{\nabla}_{\mathbf{x}}v_{\text{ray}} = \begin{bmatrix} +1.0219286\cdot 10^{-1} & +1.2847003\cdot 10^{-1} & -3.4110536\cdot 10^{-1} \\ +1.2847003\cdot 10^{-1} & +2.1885227\cdot 10^{-1} & -3.7785968\cdot 10^{-1} \\ -3.4110536\cdot 10^{-1} & -3.7785968\cdot 10^{-1} & -1.1226839\cdot 10^{-1} \end{bmatrix} (\text{km}\cdot\text{s})^{-1} \ , \quad (15.58)$$

$$\tilde{\nabla}_{\mathbf{x}}\nabla_{\mathbf{r}}v_{\text{ray}} = \begin{bmatrix} +2.4759031\cdot 10^{-4} & +1.0839432\cdot 10^{-2} & -6.6150747\cdot 10^{-3} \\ +1.8013387\cdot 10^{-4} & +7.8862081\cdot 10^{-3} & -4.8127851\cdot 10^{-3} \\ -2.0652537\cdot 10^{-5} & -9.0416207\cdot 10^{-4} & +5.5179088\cdot 10^{-4} \end{bmatrix} \text{s}^{-1} \ . \quad (15.59)$$

Accounting for the spatial derivatives of the axial compressional velocity only, $\nabla_{\mathbf{x}}v_P$ and $\nabla_{\mathbf{xx}}v_P$, we obtained,

$$\hat{\nabla}_{\mathbf{x}}v_{\text{ray}} = \begin{bmatrix} +2.9521042\cdot 10^{-1} & +2.1457100\cdot 10^{-1} & -1.0798671\cdot 10^{-1} \end{bmatrix} \text{s}^{-1} \ , \quad (15.60)$$

$$\hat{\nabla}_{\mathbf{x}}\hat{\nabla}_{\mathbf{x}}v_{\text{ray}} = \begin{bmatrix} +1.0167580\cdot 10^{-1} & +1.2902308\cdot 10^{-1} & -3.4078923\cdot 10^{-1} \\ +1.2902308\cdot 10^{-1} & +2.1842767\cdot 10^{-1} & -3.7760288\cdot 10^{-1} \\ -3.4078923\cdot 10^{-1} & -3.7760288\cdot 10^{-1} & -1.1184338\cdot 10^{-1} \end{bmatrix} (\text{km}\cdot\text{s})^{-1} \ , \quad (15.61)$$

$$\hat{\nabla}_{\mathbf{x}}\nabla_{\mathbf{r}}v_{\text{ray}} = \begin{bmatrix} +1.4160325\cdot 10^{-4} & +6.1993488\cdot 10^{-3} & -3.7833308\cdot 10^{-3} \\ +1.0292302\cdot 10^{-4} & +4.5059400\cdot 10^{-3} & -2.7498794\cdot 10^{-3} \\ -5.1797862\cdot 10^{-5} & -2.2676953\cdot 10^{-3} & +1.3839262\cdot 10^{-3} \end{bmatrix} \text{s}^{-1} \ . \quad (15.62)$$

The magnitudes of the ray velocity gradients computed with complete and incomplete data are related as,

$$\frac{\sqrt{\tilde{\nabla}_{\mathbf{x}}v_{\text{ray}}\cdot\tilde{\nabla}_{\mathbf{x}}v_{\text{ray}}}}{\sqrt{\nabla_{\mathbf{x}}v_{\text{ray}}\cdot\nabla_{\mathbf{x}}v_{\text{ray}}}} = 1.0000857 \ , \quad \frac{\sqrt{\hat{\nabla}_{\mathbf{x}}v_{\text{ray}}\cdot\hat{\nabla}_{\mathbf{x}}v_{\text{ray}}}}{\sqrt{\nabla_{\mathbf{x}}v_{\text{ray}}\cdot\nabla_{\mathbf{x}}v_{\text{ray}}}} = 9.9885642\cdot 10^{-1} \ . \quad (15.63)$$

The angles between the corresponding gradient directions are,

$$\angle \nabla_{\mathbf{x}} v_{\text{ray}}, \tilde{\nabla}_{\mathbf{x}} v_{\text{ray}} = \arccos \frac{\nabla_{\mathbf{x}} v_{\text{ray}} \cdot \tilde{\nabla}_{\mathbf{x}} v_{\text{ray}}}{\sqrt{\nabla_{\mathbf{x}} v_{\text{ray}} \cdot \nabla_{\mathbf{x}} v_{\text{ray}}} \sqrt{\tilde{\nabla}_{\mathbf{x}} v_{\text{ray}} \cdot \tilde{\nabla}_{\mathbf{x}} v_{\text{ray}}}} = 1.2224727 \cdot 10^{-2} \text{ rad} = 0.70042524^{\text{o}} ,$$

$$\angle \nabla_{\mathbf{x}} v_{\text{ray}}, \hat{\nabla}_{\mathbf{x}} v_{\text{ray}} = \arccos \frac{\nabla_{\mathbf{x}} v_{\text{ray}} \cdot \hat{\nabla}_{\mathbf{x}} v_{\text{ray}}}{\sqrt{\nabla_{\mathbf{x}} v_{\text{ray}} \cdot \nabla_{\mathbf{x}} v_{\text{ray}}} \sqrt{\hat{\nabla}_{\mathbf{x}} v_{\text{ray}} \cdot \hat{\nabla}_{\mathbf{x}} v_{\text{ray}}}} = 1.2727741 \cdot 10^{-2} \text{ rad} = 0.72924585^{\text{o}} .$$

(15.64)

We compute the ratios of the eigenvalues (major, intermediate and minor) of the spatial Hessians,

$$\frac{\lambda_{\text{major}} \tilde{\nabla}_{\mathbf{x}} \tilde{\nabla}_{\mathbf{x}} v_{\text{ray}}}{\lambda_{\text{major}} \nabla_{\mathbf{x}} \nabla_{\mathbf{x}} v_{\text{ray}}} = 0.99791752 , \quad \frac{\lambda_{\text{major}} \hat{\nabla}_{\mathbf{x}} \hat{\nabla}_{\mathbf{x}} v_{\text{ray}}}{\lambda_{\text{major}} \nabla_{\mathbf{x}} \nabla_{\mathbf{x}} v_{\text{ray}}} = 0.99763379 ,$$

$$\frac{\lambda_{\text{interm}} \tilde{\nabla}_{\mathbf{x}} \tilde{\nabla}_{\mathbf{x}} v_{\text{ray}}}{\lambda_{\text{interm}} \nabla_{\mathbf{x}} \nabla_{\mathbf{x}} v_{\text{ray}}} = 0.73369807 , \quad \frac{\lambda_{\text{interm}} \hat{\nabla}_{\mathbf{x}} \hat{\nabla}_{\mathbf{x}} v_{\text{ray}}}{\lambda_{\text{interm}} \nabla_{\mathbf{x}} \nabla_{\mathbf{x}} v_{\text{ray}}} = 0.70452683 , \quad (15.65)$$

$$\frac{\lambda_{\text{minor}} \tilde{\nabla}_{\mathbf{x}} \tilde{\nabla}_{\mathbf{x}} v_{\text{ray}}}{\lambda_{\text{minor}} \nabla_{\mathbf{x}} \nabla_{\mathbf{x}} v_{\text{ray}}} = 1.00004069 , \quad \frac{\lambda_{\text{minor}} \hat{\nabla}_{\mathbf{x}} \hat{\nabla}_{\mathbf{x}} v_{\text{ray}}}{\lambda_{\text{minor}} \nabla_{\mathbf{x}} \nabla_{\mathbf{x}} v_{\text{ray}}} = 0.99852940 .$$

Note that the obtained "errors", associated with the incomplete ray velocity derivatives, highly depend on the magnitudes of the spatial derivatives of the model parameters. Of course, at ray points within transition zones between layers, the gradients and Hessians of the model parameters are higher and hence the "errors".

The mixed Hessians, $\tilde{\nabla}_{\mathbf{x}} \nabla_{\mathbf{r}} v_{\text{ray}}$ and $\hat{\nabla}_{\mathbf{x}} \nabla_{\mathbf{r}} v_{\text{ray}}$, computed with the incomplete spatial model derivatives, have very special eigensystems. As mentioned, the mixed Hessian, $\nabla_{\mathbf{x}} \nabla_{\mathbf{r}} v_{\text{ray}}$, computed with all spatial derivatives of the model accounted for, has a zero eigenvalue, and the corresponding eigenvector is the ray direction, $\mathbf{r}$. Even if all other eigenvalues are real, the eigenvectors are not necessarily mutually orthogonal because this matrix is not symmetric. In this specific case, the mixed Hessian computed with the "complete" data has a pair of complex-conjugate eigenvalues and a zero eigenvalue,

$$\lambda \left( \nabla_{\mathbf{x}} \nabla_{\mathbf{r}} v_{\text{ray}} \right) = \left[ +6.7286747 \cdot 10^{-4} \pm 3.2135256 \cdot 10^{-2} i \quad 0 \right] , \quad (15.66)$$

while the "incomplete" mixed Hessians, $\tilde{\nabla}_\mathbf{x} \nabla_\mathbf{r} v_{\text{ray}}$ and $\hat{\nabla}_\mathbf{x} \nabla_\mathbf{r} v_{\text{ray}}$, have two zero eigenvalues,

$$\begin{aligned} \lambda\left(\tilde{\nabla}_\mathbf{x} \nabla_\mathbf{r} v_{\text{ray}}\right) &= \begin{bmatrix} +8.6855893 \cdot 10^{-3} & 0 & 0 \end{bmatrix}, \\ \lambda\left(\hat{\nabla}_\mathbf{x} \nabla_\mathbf{r} v_{\text{ray}}\right) &= \begin{bmatrix} +6.0314695 \cdot 10^{-3} & 0 & 0 \end{bmatrix}. \end{aligned} \quad (15.67)$$

The eigenvectors corresponding to the two zero eigenvalues may be reduced to a form, where one of them is the ray direction, $\mathbf{v}_2 = \mathbf{r}$, and the other is (up to a constant factor) the cross product of the ray and the axis of symmetry directions, $\mathbf{v}_3 = \mathbf{k} \times \mathbf{r}$. Thus, the two eigenvectors are mutually orthogonal, but they are (generally) not orthogonal to the first eigenvector corresponding to the single nonzero eigenvalue. In the case when only the spatial derivatives of parameter $v_P$ are taken into account, the eigenvector corresponding to the nonzero eigenvalue of $\hat{\nabla}_\mathbf{x} \nabla_\mathbf{r} v_{\text{ray}}$ is (up to a constant factor) the spatial gradient of the axial compressional velocity, $\mathbf{v}_1 = \nabla_\mathbf{x} v_P$ (listed in the first row of Table 1).

Summarizing the above, we conclude that when the spatial derivatives of the medium symmetry axis direction angles are ignored, the accuracy of the spatial ray velocity gradient, $\nabla_\mathbf{x} v_{\text{ray}}$, and the Hessian, $\nabla_\mathbf{x} \nabla_\mathbf{x} v_{\text{ray}}$, is still adequate for some practical applications. The accuracy further decreases when the gradients and Hessians of all polar anisotropy parameters are ignored (except those of the axial compressional velocity $v_P$). However, even in this case, the accuracy may still meet the requirements, provided the ignored change of anisotropy is mild. However, we will show that for rapidly varying (in space) strong polar anisotropic media with a large negative anellipticity (referred to below as model 2), ignoring the derivatives of the symmetry axis' angles leads to a low accuracy of the ray velocity derivatives. When all spatial derivatives of the model parameters are discarded, except those of the axial compressional velocity, the results for such media may be plainly wrong, especially when the ray crosses transition regions, where the model parameters change rapidly. The eigensystem of the mixed Hessian, $\nabla_\mathbf{x}\nabla_\mathbf{r} v_{\text{ray}}$ (whose accuracy is less crucial for the ray bending method), undergoes a principal change.

## 16. MODEL 1. ACOUSTIC APPROXIMATION FOR qP WAVES

When using the acoustic approximation (AC) for qP waves in polar anisotropic media, the slowness inversion equation reduces to the fourth-degree polynomial, presented in Appendix A (equation A5). The reference Hamiltonian for qP waves in equation 2.15 reduces to,

$$H_P^{\bar{\tau}}(\mathbf{x},\mathbf{p}) = +2\left[(\mathbf{p}\cdot\mathbf{p}) - (\mathbf{k}\cdot\mathbf{p})^2\right]\left[\varepsilon - (\varepsilon - \delta)(\mathbf{k}\cdot\mathbf{p})^2 v_P^2\right]v_P^2 + (\mathbf{p}\cdot\mathbf{p})v_P^2 - 1 = 0 \quad , \quad (16.1)$$

and all its derivatives simplify accordingly. We show that although the AC can still be applied with high accuracy for computing the ray velocity magnitude, the accuracy of the ray velocity derivatives degrades, in particular for the second derivatives.

In this section we skip the intermediate results, related to the derivatives of the reference and arclength-related Hamiltonians (as the procedure is similar to the case of actual (elastic) quasi-compressional waves described in detail in the previous section), and present only the required gradients and Hessians of the ray velocity magnitude.

- Spatial gradient and Hessian of the ray velocity magnitude,

$$\nabla_\mathbf{x} v_{\text{ray}} = \begin{bmatrix} +2.9828150\cdot 10^{-1} & +2.1272276\cdot 10^{-1} & -1.0470393\cdot 10^{-1} \end{bmatrix} \text{ s}^{-1} \quad , \quad (16.2)$$

$$\nabla_{\mathbf{x}}\nabla_{\mathbf{x}}v_{\text{ray}} = \begin{bmatrix} +1.0734337\cdot 10^{-1} & +1.2111443\cdot 10^{-1} & -3.3940997\cdot 10^{-1} \\ +1.2111443\cdot 10^{-1} & +2.1853003\cdot 10^{-1} & -3.8349522\cdot 10^{-1} \\ -3.3940997\cdot 10^{-1} & -3.8349522\cdot 10^{-1} & -1.0638498\cdot 10^{-1} \end{bmatrix} (\text{km}\cdot\text{s})^{-1} \ . \quad (16.3)$$

- Directional gradient and Hessian of the ray velocity magnitude,

$$\nabla_{\mathbf{r}}v_{\text{ray}} = \begin{bmatrix} +1.6786248\cdot 10^{-3} & +7.3489703\cdot 10^{-2} & -4.4849203\cdot 10^{-2} \end{bmatrix} \text{km/s} \ , \quad (16.4)$$

$$\nabla_{\mathbf{r}}\nabla_{\mathbf{r}}v_{\text{ray}} = \begin{bmatrix} +5.1967675\cdot 10^{-1} & -1.2998389\cdot 10^{-1} & -1.5796249\cdot 10^{-1} \\ -1.2998389\cdot 10^{-1} & +4.2003550\cdot 10^{-1} & -2.8539068\cdot 10^{-1} \\ -1.5796249\cdot 10^{-1} & -2.8539068\cdot 10^{-1} & +2.9837903\cdot 10^{-1} \end{bmatrix} \text{km/s} \ . \quad (16.5)$$

- The mixed Hessian of the ray velocity magnitude,

$$\nabla_{\mathbf{x}}\nabla_{\mathbf{r}}v_{\text{ray}} = \begin{bmatrix} +1.9586285\cdot 10^{-2} & +2.3569268\cdot 10^{-2} & -2.2955389\cdot 10^{-2} \\ -4.0982388\cdot 10^{-2} & +3.1068515\cdot 10^{-3} & +1.6577964\cdot 10^{-2} \\ +3.0304186\cdot 10^{-2} & +1.2612576\cdot 10^{-2} & -2.1204429\cdot 10^{-2} \end{bmatrix} \text{s}^{-1} \ . \quad (16.6)$$

- The gradient of the ray velocity wrt the model parameters, $\nabla_{\mathbf{m}}v_{\text{ray}}$, includes the medium part: the derivatives wrt the material components, $\mathbf{c} = \{v_P, \delta, \varepsilon\}$ (where the shear factor $f$ has been eliminated), and the geometric part: the derivatives wrt the symmetry axis direction angles (see equation 14.45). The material part is,

$$\nabla_{\mathbf{c}}v_{\text{ray}} = \begin{bmatrix} +1.0017304 & +4.8127717\cdot 10^{-2} & +8.2008573\cdot 10^{-4} \end{bmatrix} \ , \quad (16.7)$$

and the geometric part is,

$$\nabla_{\mathbf{a}}v_{\text{ray}} = \begin{bmatrix} -6.9160695\cdot 10^{-2} & -2.5650553\cdot 10^{-2} \end{bmatrix} \ . \quad (16.8)$$

- The Hessian of the ray velocity wrt the model parameters, $\nabla_{\mathbf{m}}\nabla_{\mathbf{m}}v_{\text{ray}}$, includes the material , the geometric, and the mixed blocks (see equation 14.49). The material block reads,

$$\nabla_\mathbf{c}\nabla_\mathbf{c} v_{\text{ray}} = \begin{bmatrix} +6.6613381 \cdot 10^{-16} & +1.3750776 \cdot 10^{-2} & +2.3431021 \cdot 10^{-4} \\ +1.3750776 \cdot 10^{-2} & -1.5130509 \cdot 10^{-1} & -3.8439256 \cdot 10^{-3} \\ +2.3431021 \cdot 10^{-4} & -3.8439256 \cdot 10^{-3} & -1.3308136 \cdot 10^{-4} \end{bmatrix} , \qquad (16.9)$$

the mixed block reads,

$$\nabla_\mathbf{a}\nabla_\mathbf{c} v_{\text{ray}} = \begin{bmatrix} -1.9760199 \cdot 10^{-2} & -5.2444022 \cdot 10^{-1} & -1.8167098 \cdot 10^{-2} \\ -7.3287295 \cdot 10^{-3} & -1.9450617 \cdot 10^{-1} & -6.7378751 \cdot 10^{-3} \end{bmatrix} , \qquad (16.10)$$

and the geometric block reads,

$$\nabla_\mathbf{a}\nabla_\mathbf{a} v_{\text{ray}} = \begin{bmatrix} +6.2502835 \cdot 10^{-1} & -3.4561564 \cdot 10^{-2} \\ -3.4561564 \cdot 10^{-2} & +1.8321319 \cdot 10^{-1} \end{bmatrix} . \qquad (16.11)$$

<u>Validation</u>

The relative error for all gradient and Hessian components has been computed twice: comparing the numerical solution $N$ for the AC with its analytical solution $A$ (error $E$), and comparing the exact analytical solution for compressional wave $P$ with the analytical solution for the AC (error $\tilde{E}$),

$$E = \frac{N - A}{A} \quad , \quad \tilde{E} = \frac{A - P}{P} \quad . \qquad (16.12)$$

- Accuracy of the spatial gradient and Hessian of the ray velocity magnitude,

$$E(\nabla_\mathbf{x} v_{\text{ray}}) = \begin{bmatrix} +4.69 \cdot 10^{-11} & +1.34 \cdot 10^{-11} & +2.53 \cdot 10^{-11} \end{bmatrix} , \qquad (16.13)$$

$$\tilde{E}(\nabla_\mathbf{x} v_{\text{ray}}) = \begin{bmatrix} -6.05 \cdot 10^{-6} & +3.72 \cdot 10^{-5} & +2.48 \cdot 10^{-5} \end{bmatrix} , \qquad (16.14)$$

$$E\left(\nabla_\mathbf{x}\nabla_\mathbf{x} v_{\text{ray}}\right) = \begin{bmatrix} -1.56\cdot 10^{-6} & +3.65\cdot 10^{-7} & +1.67\cdot 10^{-7} \\ +3.65\cdot 10^{-7} & -5.57\cdot 10^{-7} & +1.09\cdot 10^{-7} \\ +1.67\cdot 10^{-7} & +1.09\cdot 10^{-7} & +3.02\cdot 10^{-6} \end{bmatrix} \quad , \tag{16.15}$$

$$\tilde{E}\left(\nabla_\mathbf{x}\nabla_\mathbf{x} v_{\text{ray}}\right) = \begin{bmatrix} -3.92\cdot 10^{-5} & +1.35\cdot 10^{-4} & -9.10\cdot 10^{-6} \\ +1.35\cdot 10^{-4} & -4.81\cdot 10^{-5} & -3.36\cdot 10^{-5} \\ -9.10\cdot 10^{-6} & -3.36\cdot 10^{-5} & +6.63\cdot 10^{-5} \end{bmatrix} \quad . \tag{16.16}$$

- Accuracy of the directional gradient and Hessian of the ray velocity magnitude,

$$E\left(\nabla_\mathbf{r} v_{\text{ray}}\right) = \begin{bmatrix} +2.29\cdot 10^{-8} & +1.50\cdot 10^{-9} & +1.86\cdot 10^{-9} \end{bmatrix} \quad , \tag{16.17}$$

$$\tilde{E}\left(\nabla_\mathbf{r} v_{\text{ray}}\right) = \begin{bmatrix} -1.86\cdot 10^{-3} & -1.86\cdot 10^{-3} & -1.86\cdot 10^{-3} \end{bmatrix} \quad , \tag{16.18}$$

$$E\left(\nabla_\mathbf{r}\nabla_\mathbf{r} v_{\text{ray}}\right) = \begin{bmatrix} -3.47\cdot 10^{-5} & -5.80\cdot 10^{-5} & -2.27\cdot 10^{-5} \\ -5.80\cdot 10^{-5} & -3.44\cdot 10^{-5} & -1.85\cdot 10^{-5} \\ -2.27\cdot 10^{-5} & -1.85\cdot 10^{-5} & -1.60\cdot 10^{-5} \end{bmatrix} \quad , \tag{16.19}$$

$$\tilde{E}\left(\nabla_\mathbf{r}\nabla_\mathbf{r} v_{\text{ray}}\right) = \begin{bmatrix} -1.86\cdot 10^{-3} & -1.59\cdot 10^{-3} & -1.99\cdot 10^{-3} \\ -1.59\cdot 10^{-3} & -5.45\cdot 10^{-3} & -5.08\cdot 10^{-3} \\ -1.99\cdot 10^{-3} & -5.08\cdot 10^{-3} & -3.74\cdot 10^{-3} \end{bmatrix} \quad . \tag{16.20}$$

- Accuracy of the mixed Hessian of the ray velocity magnitude,

$$E\left(\nabla_\mathbf{x}\nabla_\mathbf{r} v_{\text{ray}}\right) = \begin{bmatrix} +1.98\cdot 10^{-6} & -1.20\cdot 10^{-5} & -6.60\cdot 10^{-6} \\ +1.24\cdot 10^{-5} & +1.00\cdot 10^{-4} & +2.54\cdot 10^{-6} \\ -9.58\cdot 10^{-6} & +2.72\cdot 10^{-5} & +3.56\cdot 10^{-6} \end{bmatrix} \quad , \tag{16.21}$$

$$\tilde{E}\left(\nabla_\mathbf{x}\nabla_\mathbf{r} v_{\text{ray}}\right) = \begin{bmatrix} -1.81\cdot 10^{-3} & -1.98\cdot 10^{-4} & -8.19\cdot 10^{-4} \\ -1.94\cdot 10^{-3} & \mathbf{+4.90\cdot 10^{-2}} & -7.37\cdot 10^{-3} \\ -1.85\cdot 10^{-3} & -1.05\cdot 10^{-3} & -1.57\cdot 10^{-3} \end{bmatrix} \quad . \tag{16.22}$$

A large error component is marked with a bold number. Although the relative error of this component is large, the component itself is small (as compared to the other components; see equation 16.6), so that its absolute error is approximately the same as that of the other terms.

- Accuracy of the material part of the ray velocity gradient wrt the model parameters,

$$E(\nabla_{\mathbf{c}} v_{\text{ray}}) = \begin{bmatrix} +3.75 \cdot 10^{-14} & +1.56 \cdot 10^{-8} & +1.63 \cdot 10^{-9} \end{bmatrix} \quad , \tag{16.23}$$

$$\tilde{E}(\nabla_{\mathbf{c}} v_{\text{ray}}) = \begin{bmatrix} -1.65 \cdot 10^{-6} & +4.12 \cdot 10^{-4} & \mathbf{-4.49 \cdot 10^{-2}} \end{bmatrix} \quad . \tag{16.24}$$

- Accuracy of the geometric part of the ray velocity gradient wrt the model parameters,

$$E(\nabla_{\mathbf{c}} v_{\text{ray}}) = \begin{bmatrix} +2.56 \cdot 10^{-7} & +1.29 \cdot 10^{-7} \end{bmatrix} \quad , \tag{16.25}$$

$$\tilde{E}(\nabla_{\mathbf{c}} v_{\text{ray}}) = \begin{bmatrix} -1.86 \cdot 10^{-3} & -1.86 \cdot 10^{-3} \end{bmatrix} \quad . \tag{16.26}$$

We note that the relative discrepancies between the ray velocity derivatives for the acoustic approximation and elastic models wrt the zenith and azimuth angles of the medium symmetry axis are identical,

$$\frac{[\partial v_{\text{ray}} / \partial \theta_{\text{ax}}]_{AC} - [\partial v_{\text{ray}} / \partial \theta_{\text{ax}}]_{P}}{[\partial v_{\text{ray}} / \partial \theta_{\text{ax}}]_{P}} = \frac{[\partial v_{\text{ray}} / \partial \psi_{\text{ax}}]_{AC} - [\partial v_{\text{ray}} / \partial \psi_{\text{ax}}]_{P}}{[\partial v_{\text{ray}} / \partial \psi_{\text{ax}}]_{P}} \quad , \tag{16.27}$$

where subscripts AC and P denote the acoustic and elastic models for compressional waves, respectively. This equation can be also arranged as,

$$\frac{[\partial v_{\text{ray}} / \partial \theta_{\text{ax}}]_{AC}}{[\partial v_{\text{ray}} / \partial \theta_{\text{ax}}]_{P}} = \frac{[\partial v_{\text{ray}} / \partial \psi_{\text{ax}}]_{AC}}{[\partial v_{\text{ray}} / \partial \psi_{\text{ax}}]_{P}} \quad . \tag{16.28}$$

- Accuracy of the material part of the ray velocity Hessian wrt the model parameters,

$$E\left(\nabla_{\mathbf{c}}\nabla_{\mathbf{c}}v_{\text{ray}}\right) = \begin{bmatrix} - & -7.58 \cdot 10^{-9} & +1.62 \cdot 10^{-6} \\ -7.58 \cdot 10^{-9} & +1.28 \cdot 10^{-6} & +1.70 \cdot 10^{-6} \\ +1.62 \cdot 10^{-6} & +1.70 \cdot 10^{-6} & +1.75 \cdot 10^{-4} \end{bmatrix}, \quad (16.29)$$

$$\tilde{E}\left(\nabla_{\mathbf{c}}\nabla_{\mathbf{c}}v_{\text{ray}}\right) = \begin{bmatrix} - & +4.12 \cdot 10^{-4} & \mathbf{-4.49 \cdot 10^{-2}} \\ +4.12 \cdot 10^{-4} & -4.36 \cdot 10^{-3} & \mathbf{+3.20 \cdot 10^{-2}} \\ \mathbf{-4.49 \cdot 10^{-2}} & \mathbf{+3.20 \cdot 10^{-2}} & \mathbf{-1.49 \cdot 10^{-2}} \end{bmatrix}. \quad (16.30)$$

- Accuracy of the mixed part of the ray velocity Hessian wrt the model parameters,

$$E\left(\nabla_{\mathbf{a}}\nabla_{\mathbf{c}}v_{\text{ray}}\right) = \begin{bmatrix} +2.70 \cdot 10^{-7} & -5.20 \cdot 10^{-7} & +7.84 \cdot 10^{-6} \\ +1.41 \cdot 10^{-7} & -3.74 \cdot 10^{-7} & +5.92 \cdot 10^{-6} \end{bmatrix}, \quad (16.31)$$

$$\tilde{E}\left(\nabla_{\mathbf{a}}\nabla_{\mathbf{c}}v_{\text{ray}}\right) = \begin{bmatrix} -1.86 \cdot 10^{-3} & +8.48 \cdot 10^{-4} & \mathbf{-4.49 \cdot 10^{-2}} \\ -1.86 \cdot 10^{-3} & +8.48 \cdot 10^{-4} & \mathbf{-4.49 \cdot 10^{-2}} \end{bmatrix}. \quad (16.32)$$

The first row of the mixed Hessian matrix in equation 16.30 is related to the zenith angle $\theta_{\text{ax}}$, while the second row – to the azimuth angle $\psi_{\text{ax}}$. The columns are related to the material properties $\mathbf{c} = \{v_P, \delta, \varepsilon\}$. As we see, the two lines are identical.

- Accuracy of the geometric part of the ray velocity Hessian wrt the model parameters,

$$E\left(\nabla_{\mathbf{a}}\nabla_{\mathbf{a}}v_{\text{ray}}\right) = \begin{bmatrix} +8.60 \cdot 10^{-8} & +9.93 \cdot 10^{-7} \\ +9.93 \cdot 10^{-7} & +3.75 \cdot 10^{-8} \end{bmatrix}, \quad (16.33)$$

$$\tilde{E}\left(\nabla_{\mathbf{a}}\nabla_{\mathbf{a}}v_{\text{ray}}\right) = \begin{bmatrix} -4.00 \cdot 10^{-3} & \mathbf{+1.27 \times 10^{-2}} \\ \mathbf{+1.27 \cdot 10^{-2}} & -2.86 \cdot 10^{-3} \end{bmatrix}. \quad (16.34)$$

As we see, the numerical errors $|E|$ of the finite difference approximations are small for the gradients and acceptable for the Hessians, which indicates that the analytical formulae for the AC derivatives of the ray velocity are correct. However, the discrepancies $|\tilde{E}|$ of the gradients and, in particular, the Hessians computed with the elastic and acoustic models are large, and for some components unacceptable (bold numbers). The reason is inherent to the nature of the differential operator. The AC for the ray velocity in itself is accurate; still, it is an approximation. The accuracy of the derivatives of an approximated function is much worse than the accuracy of the function itself, and the accuracy of the second derivatives is even worse. While the accuracy of the AC slowness inversion along with the computed phase and ray velocities is very high, using the AC for computing the qP ray velocity derivatives is questionable. The reduced computational complexity of the AC is accompanied by a declined accuracy of its results when the derivatives of the approximated values are involved.

## 17. MODEL 1. RAY VELOCITY DERIVATIVES FOR qSV WAVES

For the given model properties $\mathbf{m}$ and the ray direction $\mathbf{r}$, there is no shear wave triplication: the qSV solution is unique, with the ray velocity, $v_{\text{ray}} = 1.6513176 \text{ km/s}$ (Table 3). In this section, we list the derivatives of the ray velocity magnitude for this wave type and validate them numerically with the finite-difference approximations.

- The spatial gradient and Hessian of the ray velocity magnitude are,

$$\nabla_{\mathbf{x}} v_{\text{ray}} = \begin{bmatrix} -1.5524655 \cdot 10^{-1} & -2.3999453 \cdot 10^{-1} & -3.1525198 \cdot 10^{-1} \end{bmatrix} \text{ s}^{-1} \quad , \tag{17.1}$$

$$\nabla_{\mathbf{x}} \nabla_{\mathbf{x}} v_{\text{ray}} = \begin{bmatrix} -1.5863467 \cdot 10^{-1} & -1.3552001 \cdot 10^{-1} & -5.1585529 \cdot 10^{-1} \\ -1.3552001 \cdot 10^{-1} & +2.4620300 \cdot 10^{-1} & -1.8682279 \cdot 10^{-1} \\ -5.1585529 \cdot 10^{-1} & -1.8682279 \cdot 10^{-1} & -1.9252759 \cdot 10^{-1} \end{bmatrix} (\text{km} \cdot \text{s})^{-1} \quad . \tag{17.2}$$

- The directional gradient and Hessian of the ray velocity magnitude are,

$$\nabla_{\mathbf{r}} v_{\text{ray}} = \begin{bmatrix} +2.6384053 \cdot 10^{-3} & +1.1550861 \cdot 10^{-1} & -7.0492451 \cdot 10^{-2} \end{bmatrix} \text{ km/s} \quad , \quad (17.3)$$

$$\nabla_{\mathbf{r}} \nabla_{\mathbf{r}} v_{\text{ray}} = \begin{bmatrix} +8.1678783 \cdot 10^{-1} & -2.0528496 \cdot 10^{-1} & -2.4768156 \cdot 10^{-1} \\ -2.0528496 \cdot 10^{-1} & +6.1726235 \cdot 10^{-1} & -4.2236495 \cdot 10^{-1} \\ -2.4768156 \cdot 10^{-1} & -4.2236495 \cdot 10^{-1} & +4.5299123 \cdot 10^{-1} \end{bmatrix} \text{ km/s} \quad . \quad (17.4)$$

- The mixed Hessian of the ray velocity magnitude is,

$$\nabla_{\mathbf{x}} \nabla_{\mathbf{r}} v_{\text{ray}} = \begin{bmatrix} +3.0303315 \cdot 10^{-2} & +1.5954905 \cdot 10^{-2} & -2.3209435 \cdot 10^{-2} \\ -6.4841113 \cdot 10^{-2} & -1.3784012 \cdot 10^{-2} & +3.7448908 \cdot 10^{-2} \\ +4.7343762 \cdot 10^{-2} & +7.2449732 \cdot 10^{-3} & -2.5651677 \cdot 10^{-2} \end{bmatrix} \text{ s}^{-1} \quad . \quad (17.5)$$

- The ray velocity gradients wrt the material properties, $\mathbf{c} = \{v_P, f, \delta, \varepsilon\}$, and wrt the geometric properties, $\mathbf{a} = \{\theta_{\text{ax}}, \psi_{\text{ax}}\}$, are,

$$\nabla_{\mathbf{c}} v_{\text{ray}} = \begin{bmatrix} +4.7180503 \cdot 10^{-1} & -3.7340715 & -2.7785934 \cdot 10^{-2} & +2.774608 \cdot 10^{-2} \end{bmatrix} \quad , \quad (17.6)$$

where the first component is unitless, and the others have the units of velocity, and

$$\nabla_{\mathbf{a}} v_{\text{ray}} = \begin{bmatrix} -1.0870443 \cdot 10^{-1} & -4.0316667 \cdot 10^{-2} \end{bmatrix} \frac{\text{km}}{\text{s} \cdot \text{radian}} \quad . \quad (17.7)$$

- The ray velocity Hessians wrt the material, mixed, and geometric properties of the model, are,

$$\nabla_{\mathbf{c}} \nabla_{\mathbf{c}} v_{\text{ray}} = \begin{bmatrix} +4.4408921 \cdot 10^{-16} & -1.0668776 & -7.9388383 \cdot 10^{-3} & +7.9274516 \cdot 10^{-3} \\ -1.0668776 & -8.5424260 & +8.2246372 \cdot 10^{-2} & -8.1867757 \cdot 10^{-2} \\ -7.9388383 \cdot 10^{-3} & +8.2246372 \cdot 10^{-2} & -2.1318676 \cdot 10^{-1} & +2.1283892 \cdot 10^{-1} \\ +7.9274516 \cdot 10^{-3} & -8.1867757 \cdot 10^{-2} & +2.1283892 \cdot 10^{-1} & -2.1249209 \cdot 10^{-1} \end{bmatrix} \quad ,$$

(17.8)

$$\nabla_{\mathbf{a}}\nabla_{\mathbf{c}} v_{\text{ray}} = \begin{bmatrix} -3.1058408 \cdot 10^{-2} & +3.0683985 \cdot 10^{-2} & +3.1807813 \cdot 10^{-1} & -3.1716496 \cdot 10^{-1} \\ -1.1519048 \cdot 10^{-2} & +1.1380181 \cdot 10^{-2} & +1.1796990 \cdot 10^{-1} & -1.1763122 \cdot 10^{-1} \end{bmatrix},$$

(17.9)

$$\nabla_{\mathbf{a}}\nabla_{\mathbf{a}} v_{\text{ray}} = \begin{bmatrix} +9.4437246 \cdot 10^{-1} & -6.8425839 \cdot 10^{-2} \\ -6.8425839 \cdot 10^{-2} & +2.8273764 \cdot 10^{-1} \end{bmatrix}. \tag{17.10}$$

Validation

The relative errors of the finite-difference approximations are listed below.

- The relative errors of the spatial gradient and Hessian of the ray velocity magnitude are,

$$E(\nabla_{\mathbf{x}} v_{\text{ray}}) = \begin{bmatrix} +1.59 \cdot 10^{-9} & -6.43 \cdot 10^{-10} & +2.29 \cdot 10^{-10} \end{bmatrix}, \tag{17.11}$$

$$E(\nabla_{\mathbf{x}}\nabla_{\mathbf{x}} v_{\text{ray}}) = \begin{bmatrix} +1.36 \cdot 10^{-6} & -1.87 \cdot 10^{-8} & +2.39 \cdot 10^{-8} \\ -1.87 \cdot 10^{-8} & +5.86 \cdot 10^{-7} & +2.71 \cdot 10^{-7} \\ +2.39 \cdot 10^{-8} & +2.71 \cdot 10^{-7} & +2.27 \cdot 10^{-7} \end{bmatrix}. \tag{17.12}$$

- The relative errors of the directional gradient and Hessian of the ray velocity magnitude are,

$$E(\nabla_{\mathbf{r}} v_{\text{ray}}) = \begin{bmatrix} +4.50 \cdot 10^{-9} & -2.11 \cdot 10^{-10} & -1.32 \cdot 10^{-10} \end{bmatrix}, \tag{17.13}$$

$$E(\nabla_{\mathbf{r}}\nabla_{\mathbf{r}} v_{\text{ray}}) = \begin{bmatrix} +1.55 \cdot 10^{-6} & +3.00 \cdot 10^{-5} & -1.26 \cdot 10^{-5} \\ +3.00 \cdot 10^{-5} & +5.50 \cdot 10^{-6} & -1.73 \cdot 10^{-6} \\ -1.26 \cdot 10^{-5} & -1.73 \cdot 10^{-6} & -4.07 \cdot 10^{-6} \end{bmatrix}. \tag{17.14}$$

- The relative errors of the mixed Hessian of the ray velocity magnitude are,

$$E(\nabla_{\mathbf{x}}\nabla_{\mathbf{r}}v_{\text{ray}})=\begin{bmatrix}+1.14\cdot10^{-5} & -2.33\cdot10^{-5} & -2.90\cdot10^{-6}\\ +5.16\cdot10^{-6} & -1.79\cdot10^{-5} & +7.64\cdot10^{-8}\\ +4.23\cdot10^{-6} & -1.99\cdot10^{-5} & +1.47\cdot10^{-7}\end{bmatrix} . \qquad (17.15)$$

- The relative errors of the ray velocity gradients wrt the material and geometric properties of the model are,

$$E(\nabla_{\mathbf{c}}v_{\text{ray}})=\begin{bmatrix}+2.19\cdot10^{-13} & +1.59\cdot10^{-6} & +2.52\cdot10^{-9} & +8.57\cdot10^{-9}\end{bmatrix}, \qquad (17.16)$$

$$E(\nabla_{\mathbf{a}}v_{\text{ray}})=\begin{bmatrix}-5.49\cdot10^{-11} & -7.45\cdot10^{-10}\end{bmatrix}. \qquad (17.17)$$

- The relative errors of the ray velocity Hessians wrt the material, mixed, and geometric properties of the model are,

$$E(\nabla_{\mathbf{c}}\nabla_{\mathbf{c}}v_{\text{ray}})=\begin{bmatrix}- & +1.93\cdot10^{-8} & +2.22\cdot10^{-7} & +1.30\cdot10^{-6}\\ +1.93\cdot10^{-8} & +3.39\cdot10^{-8} & +1.94\cdot10^{-6} & -3.11\cdot10^{-7}\\ +2.22\cdot10^{-7} & +1.94\cdot10^{-6} & -2.90\cdot10^{-5} & -3.10\cdot10^{-7}\\ +1.30\cdot10^{-6} & -3.11\cdot10^{-7} & -3.10\cdot10^{-7} & -1.16\cdot10^{-7}\end{bmatrix}, \qquad (17.18)$$

$$E(\nabla_{\mathbf{a}}\nabla_{\mathbf{c}}v_{\text{ray}})=\begin{bmatrix}-1.10\cdot10^{-7} & +2.72\cdot10^{-7} & +4.01\cdot10^{-7} & -2.98\cdot10^{-9}\\ -1.24\cdot10^{-7} & +6.35\cdot10^{-7} & +1.07\cdot10^{-6} & -2.20\cdot10^{-7}\end{bmatrix}, \qquad (17.19)$$

$$E(\nabla_{\mathbf{a}}\nabla_{\mathbf{a}}v_{\text{ray}})=\begin{bmatrix}+1.68\cdot10^{-8} & -2.54\cdot10^{-7}\\ -2.54\cdot10^{-7} & +6.04\cdot10^{-8}\end{bmatrix}. \qquad (17.20)$$

## 18. MODEL 1. RAY VELOCITY DERIVATIVES FOR SH WAVES

The material parameters required for computing the ray velocity magnitude and its spatial derivatives for SH waves, are the axial shear velocity $v_S$ and the Thomsen $\gamma$ parameter. However, the spatial gradients and Hessians in Tables 2 and 3, respectively, are given for the axial compressional velocity $v_P$ and the shear velocity factor $f$ (rather than $v_S$). For consistency reason, we continue with these input spatial material derivatives, and apply the relation for the axial shear velocity, $v_S = v_P \sqrt{1-f}$, for computing its spatial gradient and Hessian, $\nabla_{\mathbf{x}} v_S$ and $\nabla_{\mathbf{x}} \nabla_{\mathbf{x}} v_S$,

$$\nabla_{\mathbf{x}} v_S = \sqrt{1-f}\, \nabla_{\mathbf{x}} v_P - \frac{v_P \nabla_{\mathbf{x}} f}{2\sqrt{1-f}} \quad ,$$

$$\nabla_{\mathbf{x}} \nabla_{\mathbf{x}} v_S = \sqrt{1-f}\, \nabla_{\mathbf{x}} \nabla_{\mathbf{x}} v_P - \frac{v_P \nabla_{\mathbf{x}} \nabla_{\mathbf{x}} f}{2\sqrt{1-f}} - \frac{v_P \nabla_{\mathbf{x}} f \otimes \nabla_{\mathbf{x}} f}{4(1-f)^{3/2}} - \frac{\nabla_{\mathbf{x}} v_P \otimes \nabla_{\mathbf{x}} f + \nabla_{\mathbf{x}} f \otimes v_P}{2\sqrt{1-f}} \quad . \tag{18.1}$$

We then use a normalization factor of $v_S = v_P \sqrt{1-f} = 1.6416455\,\text{km/s}$ (the values are taken from Table 1), resulting in,

$$\overline{\nabla_{\mathbf{x}} v_S} = \begin{bmatrix} -9.6972727 \cdot 10^{-2} & -1.4372727 \cdot 10^{-1} & -1.9442272 \cdot 10^{-1} \end{bmatrix} \text{km}^{-1} \quad , \tag{18.2}$$

$$\overline{\nabla_{\mathbf{x}} \nabla_{\mathbf{x}} v_S} = \begin{bmatrix} -1.0027850 \cdot 10^{-1} & -7.5292607 \cdot 10^{-2} & -3.1444089 \cdot 10^{-1} \\ -7.5292607 \cdot 10^{-2} & +1.4962626 \cdot 10^{-1} & -1.0770544 \cdot 10^{-1} \\ -3.1444089 \cdot 10^{-1} & -1.0770544 \cdot 10^{-1} & -1.2198415 \cdot 10^{-1} \end{bmatrix} \text{km}^{-2} \quad . \tag{18.3}$$

Unlike the case of qP-qSV waves, the slowness inversion for SH waves is performed analytically: The slowness vector components are computed with equation A9; the magnitudes of the phase and ray velocities are computed with equation A11 (see results in Table 3), $v_{\text{ray}} = 1.6439470\,\text{km/s}$.

In this section, we provide the ray velocity derivatives, computed analytically for SH wave, and validate them numerically with the finite-difference approximations.

- The spatial gradient and Hessian of the ray velocity magnitude are,

$$\nabla_{\mathbf{x}} v_{\text{ray}} = \begin{bmatrix} -1.5820924 \cdot 10^{-1} & -2.3702272 \cdot 10^{-1} & -3.1853528 \cdot 10^{-1} \end{bmatrix} \text{ s}^{-1} \quad , \tag{18.4}$$

$$\nabla_{\mathbf{x}} \nabla_{\mathbf{x}} v_{\text{ray}} = \begin{bmatrix} -1.6322424 \cdot 10^{-1} & -1.2672607 \cdot 10^{-1} & -5.1671733 \cdot 10^{-1} \\ -1.2672607 \cdot 10^{-1} & +2.4635332 \cdot 10^{-1} & -1.7928789 \cdot 10^{-1} \\ -5.1671733 \cdot 10^{-1} & -1.7928789 \cdot 10^{-1} & -1.9901780 \cdot 10^{-1} \end{bmatrix} (\text{km} \cdot \text{s})^{-1} \quad . \tag{18.5}$$

- The directional gradient and Hessian of the ray velocity magnitude are,

$$\nabla_{\mathbf{r}} v_{\text{ray}} = \begin{bmatrix} +6.2489860 \cdot 10^{-4} & +2.7357878 \cdot 10^{-2} & -1.6695931 \cdot 10^{-2} \end{bmatrix} \text{ km/s} \quad , \tag{18.6}$$

$$\nabla_{\mathbf{r}} \nabla_{\mathbf{r}} v_{\text{ray}} = \begin{bmatrix} +1.9345225 \cdot 10^{-1} & -4.8689821 \cdot 10^{-2} & -5.8620741 \cdot 10^{-2} \\ -4.8689821 \cdot 10^{-2} & +1.4319010 \cdot 10^{-1} & -9.8200986 \cdot 10^{-2} \\ -5.8620741 \cdot 10^{-2} & -9.8200986 \cdot 10^{-2} & +1.0616984 \cdot 10^{-1} \end{bmatrix} \text{ km/s} \quad . \tag{18.7}$$

- The mixed Hessian of the ray velocity magnitude is,

$$\nabla_{\mathbf{x}} \nabla_{\mathbf{r}} v_{\text{ray}} = \begin{bmatrix} +7.1878668 \cdot 10^{-3} & +4.2435921 \cdot 10^{-3} & -5.7806953 \cdot 10^{-3} \\ -1.5387345 \cdot 10^{-2} & -4.5744269 \cdot 10^{-3} & +9.6689613 \cdot 10^{-3} \\ +1.1130627 \cdot 10^{-2} & -1.9004638 \cdot 10^{-3} & -3.8685039 \cdot 10^{-3} \end{bmatrix} \text{ s}^{-1} \quad . \tag{18.8}$$

- The ray velocity gradient wrt the model parameters, $\mathbf{m} = \{v_S, \gamma, \theta_{\text{ax}}, \psi_{\text{ax}}\}$, is,

$$\nabla_{\mathbf{m}} v_{\text{ray}} = \begin{bmatrix} +1.0014020 & +2.4853022 \cdot 10^{-2} & -2.5746326 \cdot 10^{-2} & -9.5488849 \cdot 10^{-3} \end{bmatrix} \quad . \tag{18.9}$$

- The ray velocity Hessian wrt the model parameters is,

$$\nabla_{\mathbf{m}} \nabla_{\mathbf{m}} v_{\text{ray}} = \begin{bmatrix} -2.220446 \cdot 10^{-16} & +1.5139092 \cdot 10^{-2} & -1.5683244 \cdot 10^{-2} & -5.8166546 \cdot 10^{-3} \\ +1.5139092 \cdot 10^{-2} & -8.4572901 \cdot 10^{-2} & -2.7860655 \cdot 10^{-1} & -1.0333054 \cdot 10^{-1} \\ -1.5683244 \cdot 10^{-2} & -2.7860655 \cdot 10^{-1} & +2.2100896 \cdot 10^{-1} & -1.7194085 \cdot 10^{-2} \\ -5.8166546 \cdot 10^{-3} & -1.0333054 \cdot 10^{-1} & -1.7194085 \cdot 10^{-2} & +6.6599291 \cdot 10^{-2} \end{bmatrix} .$$

(18.10)

Validation

The relative errors of the finite-difference approximations are listed below.

- The relative errors of the spatial gradient and Hessian of the ray velocity magnitude are,

$$E(\nabla_{\mathbf{x}} v_{\text{ray}}) = \begin{bmatrix} -2.88 \cdot 10^{-11} & +1.64 \cdot 10^{-12} & +1.50 \cdot 10^{-11} \end{bmatrix} \quad , \qquad (18.11)$$

$$E(\nabla_{\mathbf{x}} \nabla_{\mathbf{x}} v_{\text{ray}}) = \begin{bmatrix} -3.00 \cdot 10^{-7} & -4.79 \cdot 10^{-10} & +1.19 \cdot 10^{-8} \\ -4.79 \cdot 10^{-10} & +1.50 \cdot 10^{-7} & +2.24 \cdot 10^{-8} \\ +1.19 \cdot 10^{-8} & +2.24 \cdot 10^{-8} & -3.27 \cdot 10^{-7} \end{bmatrix} \quad . \qquad (18.12)$$

- The relative errors of the directional gradient and Hessian of the ray velocity magnitude are,

$$E(\nabla_{\mathbf{r}} v_{\text{ray}}) = \begin{bmatrix} -5.37 \cdot 10^{-9} & +3.76 \cdot 10^{-10} & +2.79 \cdot 10^{-10} \end{bmatrix} \quad , \qquad (18.13)$$

$$E(\nabla_{\mathbf{r}} \nabla_{\mathbf{r}} v_{\text{ray}}) = \begin{bmatrix} +2.76 \cdot 10^{-5} & +5.85 \cdot 10^{-5} & +1.18 \cdot 10^{-5} \\ +5.85 \cdot 10^{-5} & +4.17 \cdot 10^{-5} & +2.34 \cdot 10^{-5} \\ +1.18 \cdot 10^{-5} & +2.34 \cdot 10^{-5} & +1.59 \cdot 10^{-5} \end{bmatrix} \quad . \qquad (18.14)$$

- The relative errors of the mixed Hessian of the ray velocity magnitude are,

$$E(\nabla_{\mathbf{x}} \nabla_{\mathbf{r}} v_{\text{ray}}) = \begin{bmatrix} +1.17 \cdot 10^{-5} & +3.68 \cdot 10^{-6} & +8.17 \cdot 10^{-6} \\ +8.47 \cdot 10^{-6} & +4.50 \cdot 10^{-6} & +7.34 \cdot 10^{-6} \\ -7.04 \cdot 10^{-7} & +5.39 \cdot 10^{-7} & -1.68 \cdot 10^{-5} \end{bmatrix} \quad . \qquad (18.15)$$

- The relative errors of the ray velocity gradient wrt the model parameters are,

$$E(\nabla_{\mathbf{m}} v_{\text{ray}}) = \begin{bmatrix} -7.01 \cdot 10^{-13} & -1.72 \cdot 10^{-9} & -1.15 \cdot 10^{-9} & -1.76 \cdot 10^{-9} \end{bmatrix} \quad . \qquad (18.16)$$

- The relative errors of the ray velocity Hessian wrt the model parameters are,

$$E\left(\nabla_{\mathbf{m}}\nabla_{\mathbf{m}}v_{\text{ray}}\right) = \begin{bmatrix} - & +5.61\cdot 10^{-7} & -4.97\cdot 10^{-8} & +2.67\cdot 10^{-7} \\ +5.61\cdot 10^{-7} & -6.07\cdot 10^{-5} & -1.24\cdot 10^{-7} & -6.26\cdot 10^{-7} \\ -4.97\cdot 10^{-8} & -1.24\cdot 10^{-7} & +4.27\cdot 10^{-7} & -1.75\cdot 10^{-6} \\ +2.67\cdot 10^{-7} & -6.26\cdot 10^{-7} & -1.75\cdot 10^{-6} & +8.72\cdot 10^{-7} \end{bmatrix} . \quad (18.17)$$

### 19. MODEL 1. TTI AS A PARTICULAR CASE OF GENERAL ANISOTROPY

In this section we show that the same results obtained in this part (Part II) for the gradients and Hessians of the ray velocity (for the given inhomogeneous polar anisotropic model parameters and their spatial gradients and Hessians, presented in Tables 1 and 2, and the given ray direction), can be obtained by applying the theory presented in Part I for general anisotropy. For this, we first have to convert (transform) the polar anisotropic model parameters and their spatial gradients and Hessians, into "equivalent model parameters" described with the full set of the twenty-one elastic properties and their spatial gradients and Hessians. Recall that the input model parameters are given by the array $\mathbf{m}(\mathbf{x}) = \mathbf{c} \,\|\, \mathbf{a}$: The five material parameters $\mathbf{c}(\mathbf{x}) = \{v_P, f, \delta, \varepsilon, \gamma\}$ (considered as a set of scalar invariants) and the two polar angles $\mathbf{a}(\mathbf{x}) = \{\theta_{\text{ax}}, \psi_{\text{ax}}\}$, as well as their spatial gradients and Hessians; all of them are given in the global frame $\mathbf{x}$. The converted twenty-one elastic parameters are in the global frame as well. In this section, we consider quasi-compressional waves only (to be compared with the derivatives obtained in Section 14).

As mentioned, Tables 1 and 2 list the model parameters and their gradients and Hessians in the global frame. Recall that we store the relative gradients and Hessians, whose units are $\text{km}^{-1}$ and $\text{km}^{-2}$, respectively, for all model parameters $m_i$ (five material parameters and two angles).

The conversion of the set $\mathbf{m} = \mathbf{c} \| \mathbf{a}$ into the twenty-one stiffness tensor components can be done in two stages:

- First, converting set $\mathbf{c}$ into the five crystal stiffness components,

$$\{v_S \quad f \quad \delta \quad \varepsilon \quad \gamma\} \rightarrow \{\hat{C}_{11} \quad \hat{C}_{13} \quad \hat{C}_{33} \quad \hat{C}_{44} \quad \hat{C}_{66}\} \quad \text{or} \quad \mathbf{c}(\mathbf{x}) \rightarrow \hat{\mathbf{C}}(\mathbf{x}) \quad, \quad (19.1)$$

where the "hat" symbol emphasizes that these are the crystal stiffness components. We reemphasize that in this section, we use the global frame $\mathbf{x}$ only.

- Second, converting the five crystal stiffness components, $\hat{\mathbf{C}}(\mathbf{x})$, along with the two polar angles, $\mathbf{a} = \{\theta_{ax} \quad \psi_{ax}\}$, into the twenty-one stiffness components of the general anisotropic matrix $\mathbf{C}(\mathbf{x})$,

$$\left\{ \underbrace{\hat{\mathbf{C}}(\mathbf{x})}_{5 \text{ items}} \quad \underbrace{\mathbf{a}(\mathbf{x})}_{2 \text{ items}} \right\} \rightarrow \underbrace{\mathbf{C}(\mathbf{x})}_{21 \text{ items}} \quad . \quad (19.2)$$

At each of the two stages, we also compute the corresponding spatial gradients and Hessians: at stage one $\nabla_{\mathbf{x}} \hat{\mathbf{C}}(\mathbf{x})$ and $\nabla_{\mathbf{x}} \nabla_{\mathbf{x}} \hat{\mathbf{C}}(\mathbf{x})$, and at stage two $\nabla_{\mathbf{x}} \mathbf{C}(\mathbf{x})$ and $\nabla_{\mathbf{x}} \nabla_{\mathbf{x}} \mathbf{C}(\mathbf{x})$.

Remark: The crystal stiffness $\hat{\mathbf{C}}(\mathbf{x})$ can be interpreted in two ways: a) A stiffness tensor in the crystal frame that relates the strains to the stresses; b) Another set of the medium invariants – re-parameterization of the normalized set of parameters $\mathbf{c}(\mathbf{x})$ into $\hat{\mathbf{C}}(\mathbf{x})$, where all the components have the units of velocity squared. We will use definition b).

To implement the first stage of the conversion, we apply equation set 2.7b, and compute the crystal stiffness components,

$$\frac{\hat{C}_{11}(\mathbf{x})}{v_P^2(\mathbf{x})} = 1 + 2\varepsilon(\mathbf{x}) , \quad \frac{\hat{C}_{13}(\mathbf{x})}{v_P^2(\mathbf{x})} = \sqrt{f(\mathbf{x})[f(\mathbf{x}) + 2\delta(\mathbf{x})]} - [1 - f(\mathbf{x})] ,$$
$$\frac{\hat{C}_{33}(\mathbf{x})}{v_P^2(\mathbf{x})} = 1 , \quad \frac{\hat{C}_{44}(\mathbf{x})}{v_P^2(\mathbf{x})} = 1 - f(\mathbf{x}) , \quad \frac{\hat{C}_{66}(\mathbf{x})}{v_P^2(\mathbf{x})} = [1 - f(\mathbf{x})][1 + 2\gamma(\mathbf{x})] .$$
(19.3)

The equations of this set make it possible to compute the spatial gradients and Hessians of the crystal stiffness tensor components, $\nabla_\mathbf{x}\hat{\mathbf{C}}(\mathbf{x})$ and $\nabla_\mathbf{x}\nabla_\mathbf{x}\hat{\mathbf{C}}(\mathbf{x})$. The spatial gradients are,

$$\nabla_\mathbf{x}\hat{C}_{11} = 2v_P^2 \nabla_\mathbf{x}\varepsilon + 2(1 + 2\varepsilon)v_P \nabla_\mathbf{x}v_P$$
$$\nabla_\mathbf{x}\hat{C}_{13} = v_P^2 \frac{f\nabla_\mathbf{x}\delta + \delta\nabla_\mathbf{x}f + f\nabla_\mathbf{x}f}{\sqrt{f(f + 2\delta)}} + v_P^2 \nabla_\mathbf{x}f + 2\left[\sqrt{f(f + 2\delta)} - 2(1 - f)\right]v_P \nabla_\mathbf{x}v_P$$
$$\nabla_\mathbf{x}\hat{C}_{33} = 2v_P \nabla_\mathbf{x}v_P$$
$$\nabla_\mathbf{x}\hat{C}_{44} = 2(1 - f)v_P \nabla_\mathbf{x}v_P - v_P^2 \nabla_\mathbf{x}f$$
$$\nabla_\mathbf{x}\hat{C}_{66} = 2(1 - f)(1 + 2\gamma)v_P \nabla_\mathbf{x}v_P + 2(1 - f)v_P^2 \nabla_\mathbf{x}\gamma - (1 + 2\gamma)v_P^2 \nabla_\mathbf{x}f$$
(19.4)

The spatial Hessians are,

$$\nabla_\mathbf{x}\nabla_\mathbf{x}\hat{C}_{11} = 4v_P \left(\nabla_\mathbf{x}v_P \otimes \nabla_\mathbf{x}\varepsilon + \nabla_\mathbf{x}\varepsilon \otimes \nabla_\mathbf{x}v_P\right)$$
$$+ 2(1 + 2\varepsilon)\nabla_\mathbf{x}v_P \otimes \nabla_\mathbf{x}v_P + 2v_P^2 \nabla_\mathbf{x}\nabla_\mathbf{x}\varepsilon + 2(1 + 2\varepsilon)\nabla_\mathbf{x}\nabla_\mathbf{x}v_P ,$$
(19.5a)

$$\nabla_{\mathbf{x}}\nabla_{\mathbf{x}}\hat{C}_{13} =$$

$$+\frac{\delta v_P^2}{\sqrt{f}\left(f+2\delta\right)^{3/2}}\left(\nabla_{\mathbf{x}}f\otimes\nabla_{\mathbf{x}}\delta+\nabla_{\mathbf{x}}\delta\otimes\nabla_{\mathbf{x}}f\right)-2\left(1-f-\sqrt{f}\sqrt{f+2\delta}\right)\nabla_{\mathbf{x}}v_P\otimes\nabla_{\mathbf{x}}v_P$$

$$+\frac{2\sqrt{f}}{\sqrt{f+2\delta}}v_P\left(\nabla_{\mathbf{x}}v_P\otimes\nabla_{\mathbf{x}}\delta+\nabla_{\mathbf{x}}\delta\otimes\nabla_{\mathbf{x}}v_P\right)-\frac{\delta^2 v_P^2}{f^{3/2}\left(f+2\delta\right)^{3/2}}\nabla_{\mathbf{x}}f\otimes\nabla_{\mathbf{x}}f$$

$$+2\left[\frac{f+\delta}{\sqrt{f}\sqrt{f+2\delta}}+1\right]v_P\left(\nabla_{\mathbf{x}}v_P\otimes\nabla_{\mathbf{x}}f+\nabla_{\mathbf{x}}f\otimes\nabla_{\mathbf{x}}v_P\right)-\frac{\sqrt{f}v_P^2}{\left(f+2\delta\right)^{3/2}}\nabla_{\mathbf{x}}\delta\otimes\nabla_{\mathbf{x}}\delta$$

$$-2\left(1-f-\sqrt{f}\sqrt{f+2\delta}\right)v_P\nabla_{\mathbf{x}}\nabla_{\mathbf{x}}v_P+\left[\frac{f+\delta}{\sqrt{f}\sqrt{f+2\delta}}+1\right]v_P^2\nabla_{\mathbf{x}}\nabla_{\mathbf{x}}f+\frac{\sqrt{f}v_P^2}{\sqrt{f+2\delta}}\nabla_{\mathbf{x}}\nabla_{\mathbf{x}}\delta\ ,$$

(19.5b)

$$\nabla_{\mathbf{x}}\nabla_{\mathbf{x}}\hat{C}_{33} = 2\nabla_{\mathbf{x}}v_P\otimes\nabla_{\mathbf{x}}v_P+2v_P\nabla_{\mathbf{x}}\nabla_{\mathbf{x}}v_P \quad , \tag{19.5c}$$

$$\nabla_{\mathbf{x}}\nabla_{\mathbf{x}}\hat{C}_{44} = -2v_P\left(\nabla_{\mathbf{x}}v_P\otimes\nabla_{\mathbf{x}}f+\nabla_{\mathbf{x}}f\otimes\nabla_{\mathbf{x}}v_P\right)-v_P^2\nabla_{\mathbf{x}}\nabla_{\mathbf{x}}f$$
$$+2\left(1-f\right)\nabla_{\mathbf{x}}v_P\otimes\nabla_{\mathbf{x}}v_P+2\left(1-f\right)\nabla_{\mathbf{x}}\nabla_{\mathbf{x}}v_P \quad ,$$

(19.5d)

$$\nabla_{\mathbf{x}}\nabla_{\mathbf{x}}\hat{C}_{66} = +2\left(1-f\right)\left(1+2\gamma\right)\nabla_{\mathbf{x}}v_P\otimes\nabla_{\mathbf{x}}v_P$$
$$-2v_P^2\left(\nabla_{\mathbf{x}}f\otimes\nabla_{\mathbf{x}}\gamma+\nabla_{\mathbf{x}}\gamma\otimes\nabla_{\mathbf{x}}f\right)+2\left(1-f\right)\left(1+2\gamma\right)v_P\nabla_{\mathbf{x}}\nabla_{\mathbf{x}}v_P$$
$$+4\left(1-f\right)v_P\left(\nabla_{\mathbf{x}}v_P\otimes\nabla_{\mathbf{x}}\gamma+\nabla_{\mathbf{x}}\gamma\otimes\nabla_{\mathbf{x}}v_P\right)-\left(1+2\gamma\right)v_P^2\nabla_{\mathbf{x}}\nabla_{\mathbf{x}}f$$
$$-2\left(1+2\gamma\right)v_P\left(\nabla_{\mathbf{x}}v_P\otimes\nabla_{\mathbf{x}}f+\nabla_{\mathbf{x}}f\otimes\nabla_{\mathbf{x}}v_P\right)+2\left(1-f\right)v_P^2\nabla_{\mathbf{x}}\nabla_{\mathbf{x}}\gamma \quad .$$

(19.5e)

Note that the spatial gradients and Hessians of the polar anisotropic model parameters and the crystal stiffness tensor components in equation sets 19.4 and 19.5 are absolute values (rather than relative). The computed relative gradients are presented in Table 4, and the relative Hessians in Tables 5 and 6, where the normalization factors are the reciprocals of the (non-zero) crystal stiffness components, $\hat{C}_{ij}^{-1}$.

Table 4. "Crystal" stiffness components of the polar anisotropy model 1 and their spatial gradients

| # | $\hat{C}_{ij}$ | Value, $(km/s)^2$ | Relative gradient, $km^{-1}$ (in global frame) | | |
|---|---|---|---|---|---|
| | | | $x_1$ | $x_2$ | $x_3$ |
| 1 | $\hat{C}_{11}$ | +18.375 | $+1.4296667 \cdot 10^{-1}$ | $+8.64 \cdot 10^{-2}$ | $-8.1733333 \cdot 10^{-2}$ |
| 2 | $\hat{C}_{13}$ | +8.0151716 | $+4.2532494 \cdot 10^{-1}$ | $+4.0870528 \cdot 10^{-1}$ | $+1.6340855 \cdot 10^{-1}$ |
| 3 | $\hat{C}_{33}$ | +12.25 | $+1.684 \cdot 10^{-1}$ | $+1.224 \cdot 10^{-1}$ | $-6.16 \cdot 10^{-2}$ |
| 4 | $\hat{C}_{44}$ | +2.695 | $-1.9394545 \cdot 10^{-1}$ | $-2.8745455 \cdot 10^{-1}$ | $-3.8884545 \cdot 10^{-1}$ |
| 5 | $\hat{C}_{66}$ | +3.1262 | $-1.784558 \cdot 10^{-1}$ | $-2.8290282 \cdot 10^{-1}$ | $-3.9479028 \cdot 10^{-1}$ |

Table 5. Spatial Hessians of "crystal" stiffness tensor (diagonal components), model 1

| Component | | Relative Hessian, $km^{-2}$ (in global frame) | | |
|---|---|---|---|---|
| # | $m_i$ | $x_1 x_1$ | $x_2 x_2$ | $x_3 x_3$ |
| 1 | $\hat{C}_{11}$ | $+5.4246667 \cdot 10^{-2}$ | $+1.1137808 \cdot 10^{-1}$ | $-6.9688960 \cdot 10^{-2}$ |
| 2 | $\hat{C}_{13}$ | $+2.6082298 \cdot 10^{-1}$ | $+5.0164808 \cdot 10^{-3}$ | $-2.2370352 \cdot 10^{-3}$ |
| 3 | $\hat{C}_{33}$ | $+7.2179280 \cdot 10^{-2}$ | $+1.3209088 \cdot 10^{-1}$ | $-6.1902720 \cdot 10^{-2}$ |
| 4 | $\hat{C}_{44}$ | $-1.8174958 \cdot 10^{-1}$ | $+3.4056758 \cdot 10^{-5}$ | $-1.683679 \cdot 10^{-1}$ |
| 5 | $\hat{C}_{66}$ | $-1.8286132 \cdot 10^{-1}$ | $+3.4930247 \cdot 10^{-1}$ | $-1.7513776 \cdot 10^{-1}$ |

Table 6. Spatial Hessians of "crystal" stiffness tensor (off-diagonal components), model 1

| Component | | Relative Hessian, km$^{-2}$ (in global frame) | | |
|---|---|---|---|---|
| # | $m_i$ | $x_1 x_2$ | $x_2 x_3$ | $x_1 x_3$ |
| 1 | $\hat{C}_{11}$ | $+5.0763973 \cdot 10^{-2}$ | $-2.0131664 \cdot 10^{-1}$ | $-1.9244381 \cdot 10^{-1}$ |
| 2 | $\hat{C}_{13}$ | $+2.0995847 \cdot 10^{-1}$ | $-2.6707485 \cdot 10^{-1}$ | $+5.5608440 \cdot 10^{-2}$ |
| 3 | $\hat{C}_{33}$ | $+8.3906080 \cdot 10^{-2}$ | $-2.1916992 \cdot 10^{-1}$ | $-1.9958672 \cdot 10^{-1}$ |
| 4 | $\hat{C}_{44}$ | $-1.2270996 \cdot 10^{-1}$ | $-1.5952318 \cdot 10^{-1}$ | $-1.6836790 \cdot 10^{-1}$ |
| 5 | $\hat{C}_{66}$ | $-1.3385222 \cdot 10^{-1}$ | $-1.6387388 \cdot 10^{-1}$ | $-5.9465138 \cdot 10^{-1}$ |

Given the $3 \times 3$ rotation matrix $\mathbf{A}_{\text{rot}}$ (equation C.4), we apply equations C8 – C12 to build the $6 \times 6$ matrix $\mathbf{R}_{\text{rot}}$ for the global-to-local (crystal) rotation of the stiffness tensor in the Kelvin-matrix form (Appendix C). However, the rotation that we need now is local-to-global, therefore, we apply the first equation of set C13,

$$\mathbf{C}_{\text{glb}} = \mathbf{R}_{\text{rot}}^T \mathbf{C}_{\text{loc}} \mathbf{R}_{\text{rot}} \quad \text{where} \quad \mathbf{C}_{\text{glb}} \equiv \mathbf{C} \quad \text{and} \quad \mathbf{C}_{\text{loc}} \equiv \hat{\mathbf{C}} \quad . \qquad (19.6)$$

For polar anisotropy, the only difference between the Voigt and the less commonly used Kelvin forms is the factor 2 for the shear components $C_{44}$ and $C_{66}$ in the lower right block of the Kelvin form. Expanding the rotation in equation 19.6, we obtain the transform between the crystal and the global stiffness components in an analytic form. This form is needed to further compute analytically the gradients and Hessians of the global stiffness components vs. the gradients and Hessians of the crystal components and the axis zenith and azimuth. The crystal-to-global stiffness conversion, $\hat{\mathbf{C}} \rightarrow \mathbf{C}$ resulting from the Bond algorithm, reads,

$$C_{11} = \cos^2 \psi_{ax} \left[ \hat{C}_{11} + \hat{C}_{13} + \left( \hat{C}_{11} - \hat{C}_{13} \right) \cos 2\theta_{ax} \right] \sin^2 \psi_{ax} + \hat{C}_{11} \sin^4 \psi_{ax} + \hat{C}_{44} \sin^2 \theta_{ax} \sin^2 2\psi_{ax}$$
$$+ \cos^4 \psi_{ax} \left( \hat{C}_{11} \cos^4 \theta_{ax} + 2\hat{C}_{13} \cos^2 \theta_{ax} \sin^2 \theta_{ax} + \hat{C}_{33} \sin^4 \theta_{ax} + \hat{C}_{44} \sin^2 2\theta_{ax} \right) \; ,$$

(19.7a)

$$C_{12} = \left( \hat{C}_{11} - 2\hat{C}_{66} \right) \cos^2 \theta_{ax} \sin^4 \psi_{ax} + \left( \hat{C}_{13} \sin^4 \psi_{ax} - \hat{C}_{44} \sin^2 2\psi_{ax} \right) \sin^2 \theta_{ax}$$
$$+ \cos^4 \psi_{ax} \left[ \left( \hat{C}_{11} - 2\hat{C}_{66} \right) \cos^2 \theta_{ax} + \hat{C}_{13} \sin^2 \theta_{ax} \right] + \cos^2 \psi_{ax} \sin^2 \psi_{ax} \times$$
$$\left[ \hat{C}_{11} + \hat{C}_{11} \cos^4 \theta_{ax} + \hat{C}_{33} \sin^4 \theta_{ax} + \cos^2 \theta_{ax} \left( 2\hat{C}_{13} \sin^2 \theta_{ax} - 4\hat{C}_{66} \right) \right] + \hat{C}_{44} \sin^2 2\theta_{ax} \sin^2 2\psi_{ax} / 4$$

(19.7b)

$$C_{13} = \cos^2 \psi_{ax} \left[ \hat{C}_{11} + 6\hat{C}_{13} + \hat{C}_{33} - 4\hat{C}_{44} \left( \hat{C}_{11} - 2\hat{C}_{13} + \hat{C}_{33} - 4\hat{C}_{44} \right) \cos 4\theta_{ax} \right] / 8$$
$$+ \sin^2 \psi_{ax} \left[ \hat{C}_{13} \cos^2 \theta_{ax} + (\hat{C}_{11} - 2\hat{C}_{66}) \sin^2 \theta_{ax} \right] \; ,$$

(19.7c)

$$C_{14} = -\cos \theta_{ax} \sin \theta_{ax} \sin \psi_{ax} \left\{ \cos^2 \psi_{ax} \left[ \hat{C}_{11} - \hat{C}_{33} + 4\hat{C}_{44} - 4\hat{C}_{66} \right. \right.$$
$$\left. + \left( \hat{C}_{11} - 2\hat{C}_{13} + \hat{C}_{33} - 4\hat{C}_{44} \right) \cos 2\theta_{ax} \right] / 2 + \left( \hat{C}_{11} - \hat{C}_{13} - 2\hat{C}_{66} \right) \sin^2 \psi_{ax} \right\} \; ,$$

(19.7d)

$$C_{15} = -\cos \theta_{ax} \sin \theta_{ax} \cos \psi_{ax} \left\{ \cos^2 \psi_{ax} \times \right.$$
$$\left. \left[ \hat{C}_{11} - \hat{C}_{33} + \left( \hat{C}_{11} - 2\hat{C}_{13} + \hat{C}_{33} - 4\hat{C}_{44} \right) \cos 2\theta_{ax} \right] / 2 + \left( \hat{C}_{11} - \hat{C}_{13} - 2\hat{C}_{44} \right) \sin^2 \psi_{ax} \right\} \; ,$$

(19.7e)

$$C_{16} = -\sin^2 \theta_{ax} \cos \psi_{ax} \sin \psi_{ax} \left\{ \cos^2 \psi_{ax} \left[ \hat{C}_{11} - \hat{C}_{33} \right. \right.$$
$$\left. + \left( \hat{C}_{11} - 2\hat{C}_{13} + \hat{C}_{33} - 4\hat{C}_{44} \right) \cos 2\theta_{ax} \right] + 2\left( \hat{C}_{11} - \hat{C}_{13} - 2\hat{C}_{44} \right) \sin^2 \psi_{ax} \right\} / 2 \; ,$$

(19.7f)

$$C_{22} = \hat{C}_{11} \cos^4 \psi_{ax} + \cos^2 \psi_{ax} \left[ \hat{C}_{11} + \hat{C}_{13} + \left( \hat{C}_{11} - \hat{C}_{13} \right) \cos 2\theta_{ax} \right] \sin^2 \psi_{ax}$$
$$+ \hat{C}_{11} \cos^4 \theta_{ax} \sin^4 \psi_{ax} + 2\hat{C}_{13} \cos^2 \theta_{ax} \sin^2 \theta_{ax} \sin^4 \psi_{ax}$$
$$+ \hat{C}_{44} \sin^2 \theta_{ax} \sin^2 2\psi_{ax} + \hat{C}_{33} \sin^4 \theta_{ax} \sin^4 \psi_{ax} + \hat{C}_{44} \sin^2 2\theta_{ax} \sin^4 \psi_{ax} \; ,$$

(19.7g)

$$C_{23} = \left[\hat{C}_{11} + 6\hat{C}_{13} + \hat{C}_{33} - 4\hat{C}_{44} - (C_{11} - 2C_{13} + C_{33} - 4C_{44})\cos 4\theta_{ax}\right]\sin^2\psi_{ax}/8$$
$$+ \cos^2\psi_{ax}\left[\hat{C}_{13}\cos^2\theta_{ax} + (\hat{C}_{11} - 2\hat{C}_{66})\sin^2\theta_{ax}\right] , \tag{19.7h}$$

$$C_{24} = \cos\theta_{ax}\sin\theta_{ax}\sin\psi_{ax}\left\{(\hat{C}_{13} - \hat{C}_{11} + 2\hat{C}_{44})\cos^2\psi_{ax}\right.$$
$$\left. - \left[\hat{C}_{11} - \hat{C}_{33} + (\hat{C}_{11} - 2\hat{C}_{13} + \hat{C}_{33} - 4\hat{C}_{44})\cos 2\theta_{ax}\right]\sin^2\psi_{ax}/2\right\} , \tag{19.7i}$$

$$C_{25} = \cos\theta_{ax}\sin\theta_{ax}\cos\psi_{ax}\left\{(\hat{C}_{13} - \hat{C}_{11} + 2\hat{C}_{66})\cos^2\psi_{ax}\right.$$
$$\left. - \left[\hat{C}_{11} - \hat{C}_{33} + 4\hat{C}_{44} - 4\hat{C}_{66} + (\hat{C}_{11} - 2\hat{C}_{13} + \hat{C}_{33} - 4\hat{C}_{44})\cos 2\theta_{ax}\right]\sin^2\psi_{ax}/2\right\} , \tag{19.7j}$$

$$C_{26} = -\sin^2\theta_{ax}\cos\psi_{ax}\sin\psi_{ax}\left\{2(\hat{C}_{11} - \hat{C}_{13} - 2C_{44})\cos^2\psi_{ax}\right.$$
$$\left. + \left[\hat{C}_{11} - \hat{C}_{33} + (\hat{C}_{11} - 2\hat{C}_{13} + \hat{C}_{33} - 4\hat{C}_{44})\cos 2\theta_{ax}\right]\sin^2\psi_{ax}\right\}/2 , \tag{19.7k}$$

$$C_{33} = \hat{C}_{33}\cos^4\theta_{ax} + 2\hat{C}_{13}\cos^2\theta_{ax}\sin^2\theta_{ax} + \hat{C}_{11}\sin^4\theta_{ax} + \hat{C}_{44}\sin^2 2\theta_{ax} , \tag{19.7l}$$

(Note: the global stiffness component $C_{33}$ is independent of the axial azimuth, $\psi_{ax}$),

$$C_{34} = \left[\hat{C}_{33} - \hat{C}_{11} + (\hat{C}_{11} - 2\hat{C}_{13} + \hat{C}_{33} - 4\hat{C}_{44})\cos 2\theta_{ax}\right]\sin 2\theta_{ax}\sin\psi_{ax}/4 , \tag{19.7m}$$

$$C_{35} = \sin 2\theta_{ax}\cos\psi_{ax}\left[\hat{C}_{33} - \hat{C}_{11} + (\hat{C}_{11} - 2\hat{C}_{13} + \hat{C}_{33} - 4\hat{C}_{44})\cos 2\theta_{ax}\right]/4 , \tag{19.7n}$$

$$C_{36} = \sin^2\theta_{ax}\sin 2\psi_{ax}\left[\hat{C}_{33} - \hat{C}_{11} - 4\hat{C}_{44} + 4\hat{C}_{66} + (\hat{C}_{11} - 2\hat{C}_{13} + \hat{C}_{33} - 4\hat{C}_{44})\cos 2\theta_{ax}\right]/4 , \tag{19.7o}$$

$$C_{44} = \left[\hat{C}_{11} - 2\hat{C}_{13} + \hat{C}_{33} + 4\hat{C}_{44} - (\hat{C}_{11} - 2\hat{C}_{13} + \hat{C}_{33} - 4\hat{C}_{44})\cos 4\theta_{ax}\right]\sin^2\psi_{ax}/8$$
$$+ (\hat{C}_{44}\cos^2\theta_{ax} + \hat{C}_{66}\sin^2\theta_{ax})\cos^2\psi_{ax} , \tag{19.7p}$$

$$C_{45} = \sin^2 \theta_{ax} \sin 2\psi_{ax} \left[ \hat{C}_{11} - 2\hat{C}_{13} + \hat{C}_{33} - 2(\hat{C}_{44} + \hat{C}_{66}) \right.$$
$$\left. + (\hat{C}_{11} - 2\hat{C}_{13} + \hat{C}_{33} - 4\hat{C}_{44}) \cos 2\theta_{ax} \right]/4 \;,$$
(19.7q)

$$C_{46} = \cos \theta_{ax} \sin \theta_{ax} \cos \psi_{ax} \left\{ 2C_{44} \cos 2\psi_{ax} - 2C_{66} + 2\sin^2 \psi_{ax} \times \right.$$
$$\left. \left[ 2C_{44} \cos 2\theta_{ax} + (C_{11} - 2C_{13} + C_{33}) \sin^2 \theta_{ax} \right] \right\}/2 \;,$$
(19.7r)

$$C_{55} = \cos^2 \psi_{ax} \left[ \hat{C}_{11} - 2\hat{C}_{13} + \hat{C}_{33} + 4\hat{C}_{44} - (\hat{C}_{11} - 2\hat{C}_{13} + \hat{C}_{33} - 4\hat{C}_{44}) \cos 4\theta_{ax} \right]/8$$
$$+ (\hat{C}_{44} \cos^2 \theta_{ax} + \hat{C}_{66} \sin^2 \theta_{ax}) \sin^2 \psi_{ax} \;,$$
(19.7s)

$$C_{56} = \cos \theta_{ax} \sin \theta_{ax} \sin \psi_{ax} \left\{ \cos^2 \psi_{ax} \left[ \hat{C}_{11} - 2\hat{C}_{13} + \hat{C}_{33} - 2(\hat{C}_{44} + \hat{C}_{66}) \right. \right.$$
$$\left. \left. - (\hat{C}_{11} - 2\hat{C}_{13} + \hat{C}_{33} - 4C_{44}) \cos 2\theta_{ax} \right] + 2(\hat{C}_{44} - \hat{C}_{66}) \sin^2 \psi_{ax} \right\}/2 \;,$$
(19.7t)

$$C_{66} = \left[ 3\hat{C}_{11} - 6\hat{C}_{13} + 3\hat{C}_{33} + 20\hat{C}_{44} + 32\hat{C}_{66} \right.$$
$$-4(\hat{C}_{11} - 2\hat{C}_{13} + \hat{C}_{33} + 4\hat{C}_{44} - 8\hat{C}_{66}) \cos 2\theta_{ax} + (\hat{C}_{11} - 2\hat{C}_{13} + \hat{C}_{33} - 4\hat{C}_{44}) \times$$
$$\left. (4\cos 2\theta_{ax} - 3)\cos 4\psi_{ax} + 2(\hat{C}_{11} - 2\hat{C}_{13} + \hat{C}_{33} - 4\hat{C}_{44}) \cos 4\theta_{ax} \sin^2 2\psi_{ax} \right]/64 \;.$$
(19.7u)

Equation set 19.7 defines the twenty-one components of the global stiffness tensor that depend in a linear way on the five polar anisotropy components of the "crystal" stiffness tensor, while the coefficients of these linear functions depend on the polar angles of the symmetry axis.

The gradient and the Hessian of the fourth-order stiffness tensor, $\nabla_{\mathbf{x}} \tilde{\mathbf{C}}$ and $\nabla_{\mathbf{x}} \nabla_{\mathbf{x}} \tilde{\mathbf{C}}$ (where the tilde emphasizes that this is the fourth-order stiffness tensor rather than its Voigt or Kelvin matrix representation), are tensors of the fifth and sixth orders, respectively. Working with these tensors is inconvenient and unnecessary. For the further analysis, we arrange the global stiffness components as an array of length 21, and its gradient and Hessian will represent 2D and 3D arrays, respectively (representing collections of vectors and second-order tensors) of dimensions $21 \times 3$ and $21 \times 3 \times 3$. Furthermore, due to the symmetry of the Hessians, the 3D array can be reduced to a 2D $21 \times 6$ array, with the Hessians of each component represented by vectors of length six.

With this in mind, the "crystal to global" transform can be arranged as,

$$\underbrace{C_i(\mathbf{x})}_{21 \times 1} = \sum_{j=1}^{5} \underbrace{A_{ij}\left[\theta_{\mathrm{ax}}(\mathbf{x}), \psi_{\mathrm{ax}}(\mathbf{x})\right]}_{21 \times 5} \underbrace{\hat{C}_j(\mathbf{x})}_{5 \times 1} \quad , \tag{19.8}$$

where $\mathbf{A}(\theta_{\mathrm{ax}}, \psi_{\mathrm{ax}})$ is a conversion matrix of dimensions $21 \times 5$ whose components are defined in equation set 19.7. The order (sequence) of the crystal frame components is,

$$\hat{\mathbf{C}} = \left\{\hat{C}_1, \hat{C}_2, \hat{C}_3, \hat{C}_4, \hat{C}_5\right\} = \left\{\hat{C}_{11}, \hat{C}_{13}, \hat{C}_{33}, \hat{C}_{44}, \hat{C}_{66}\right\} \quad , \tag{19.9}$$

and the order of the global tensor components is given in Table 1. The local (crystal) and global stiffness components and the axis direction angles are spatially dependent.

The components of matrix $\mathbf{A}(\theta_{\mathrm{ax}}, \psi_{\mathrm{ax}})$ follow from equation set 19.7,

$$\begin{aligned}
A_{1,1} &= \left(\cos^2\theta_{\mathrm{ax}}\cos^2\psi_{\mathrm{ax}} + \sin^2\psi_{\mathrm{ax}}\right)^2 \quad , \\
A_{1,2} &= 2\left(\cos^2\theta_{\mathrm{ax}}\cos^2\psi_{\mathrm{ax}} + \sin^2\psi_{\mathrm{ax}}\right)\sin^2\theta_{\mathrm{ax}}\cos^2\psi_{\mathrm{ax}} \quad , \\
A_{1,3} &= \sin^4\theta_{\mathrm{ax}}\cos^4\psi_{\mathrm{ax}} \quad , \\
A_{1,4} &= 4\left(\cos^2\theta_{\mathrm{ax}}\cos^2\psi_{\mathrm{ax}} + \sin^2\psi_{\mathrm{ax}}\right)\sin^2\theta_{\mathrm{ax}}\cos^2\psi_{\mathrm{ax}} \quad , \\
A_{1,5} &= 0 \quad ,
\end{aligned} \tag{19.10a}$$

$$A_{2,1} = \left(\cos^2\theta_{ax}\cos^2\psi_{ax} + \sin^2\psi_{ax}\right)\left(\cos^2\theta_{ax}\sin^2\psi_{ax} + \cos^2\psi_{ax}\right) \quad,$$
$$A_{2,2} = \left(\cos^4\psi_{ax} + 2\cos^2\theta_{ax}\cos^2\psi_{ax}\sin^2\psi_{ax} + \sin^4\psi_{ax}\right)\sin^2\theta_{ax} \quad,$$
$$A_{2,3} = +\sin^4\theta_{ax}\cos^2\psi_{ax}\sin^2\psi_{ax} \quad,$$
$$A_{2,4} = -\sin^4\theta_{ax}\sin^2 2\psi_{ax} \quad,$$
$$A_{2,5} = -2\cos^2\theta_{ax} \quad,$$
(19.10b)

$$A_{3,1} = \left(\cos^2\theta_{ax}\cos^2\psi_{ax} + \sin^2\psi_{ax}\right)\sin^2\theta_{ax} \quad,$$
$$A_{3,2} = \left(3 + \cos 4\theta_{ax}\right)\cos^2\psi_{ax}/4 + \cos^2\theta_{ax}\sin^2\psi_{ax} \quad,$$
$$A_{3,3} = +\cos^2\theta_{ax}\sin^2\theta_{ax}\cos^2\psi_{ax} \quad,$$
$$A_{3,4} = -4\cos^2\theta_{ax}\sin^2\theta_{ax}\cos^2\psi_{ax} \quad,$$
$$A_{3,5} = -2\sin^2\theta_{ax}\sin^2\psi_{ax} \quad,$$
(19.10c)

$$A_{4,1} = -\left(\cos^2\theta_{ax}\cos^2\psi_{ax} + \sin^2\psi_{ax}\right)\cos\theta_{ax}\sin\theta_{ax}\sin\psi_{ax} \quad,$$
$$A_{4,2} = +\left(\cos 2\theta_{ax}\cos^2\psi_{ax} + \sin^2\psi_{ax}\right)\cos\theta_{ax}\sin\theta_{ax}\sin\psi_{ax} \quad,$$
$$A_{4,3} = \cos\theta_{ax}\sin^3\theta_{ax}\cos^2\psi_{ax}\sin\psi_{ax} \quad,$$
$$A_{4,4} = -4\cos\theta_{ax}\sin^3\theta_{ax}\cos^2\psi_{ax}\sin\psi_{ax} \quad,$$
$$A_{4,5} = \sin 2\theta_{ax}\sin\psi_{ax} \quad,$$
(19.10d)

$$A_{5,1} = -\left(\cos^2\theta_{ax}\cos^2\psi_{ax} + \sin^2\psi_{ax}\right)\cos\theta_{ax}\sin\theta_{ax}\cos\psi_{ax} \quad,$$
$$A_{5,2} = +\left(\cos 2\theta_{ax}\cos^2\psi_{ax} + \sin^2\psi_{ax}\right)\cos\theta_{ax}\sin\theta_{ax}\cos\psi_{ax} \quad,$$
$$A_{5,3} = \cos\theta_{ax}\sin^3\theta_{ax}\cos^3\psi_{ax} \quad,$$
$$A_{5,4} = +\left(\cos 2\theta_{ax}\cos^2\psi_{ax} + \sin^2\psi_{ax}\right)\sin 2\theta_{ax}\cos\psi_{ax} \quad,$$
$$A_{5,5} = 0 \quad,$$
(19.10e)

$$A_{6,1} = -\left(\cos^2\theta_{ax}\cos^2\psi_{ax} + \sin^2\psi_{ax}\right)\sin^2\theta_{ax}\cos\psi_{ax}\sin\psi_{ax} \quad,$$
$$A_{6,2} = +\left(\cos 2\theta_{ax}\cos^2\psi_{ax} + \sin^2\psi_{ax}\right)\sin^2\theta_{ax}\cos\psi_{ax}\sin\psi_{ax} \quad,$$
$$A_{6,3} = \sin^4\theta_{ax}\cos^3\psi_{ax}\sin\psi_{ax} \quad,$$
$$A_{6,4} = +\left(\cos 2\theta_{ax}\cos^2\psi_{ax} + \sin^2\psi_{ax}\right)\sin^2\theta_{ax}\sin 2\psi_{ax} \quad,$$
$$A_{6,5} = 0 \quad,$$
(19.10f)

$$A_{7,1} = +\left(\cos^2\theta_{ax}\sin^2\psi_{ax} + \cos^2\psi_{ax}\right)^2 \quad,$$
$$A_{7,2} = 2\left(\cos^2\theta_{ax}\sin^2\psi_{ax} + \cos^2\psi_{ax}\right)\sin^2\theta_{ax}\sin^2\psi_{ax} \quad,$$
$$A_{7,3} = \sin^4\theta_{ax}\sin^4\psi_{ax} \quad,$$
$$A_{7,4} = 4\left(\cos^2\theta_{ax}\sin^2\psi_{ax} + \cos^2\psi_{ax}\right)\sin^2\theta_{ax}\sin^2\psi_{ax} \quad,$$
$$A_{7,5} = 0 \quad,$$
(19.10g)

$$A_{8,1} = \left(\cos^2\theta_{ax}\sin^2\psi_{ax} + \cos^2\psi_{ax}\right)\sin^2\theta_{ax} \quad,$$
$$A_{8,2} = (3+\cos 4\theta_{ax})\sin^2\psi_{ax}/4 + \cos^2\theta_{ax}\cos^2\psi_{ax} \quad,$$
$$A_{8,3} = \cos^2\theta_{ax}\sin^2\theta_{ax}\sin^2\psi_{ax} \quad,$$
$$A_{8,4} = -\sin^2 2\theta_{ax}\sin^2\psi_{ax} \quad,$$
$$A_{8,5} = -2\sin^2\theta_{ax}\cos^2\psi_{ax} \quad,$$
(19.10h)

$$A_{9,1} = -\left(\cos^2\theta_{ax}\sin^2\psi_{ax} + \cos^2\psi_{ax}\right)\cos\theta_{ax}\sin\theta_{ax}\sin\psi_{ax} \quad,$$
$$A_{9,2} = +\left(\cos 2\theta_{ax}\sin^2\psi_{ax} + \cos^2\psi_{ax}\right)\cos\theta_{ax}\sin\theta_{ax}\sin\psi_{ax} \quad,$$
$$A_{9,3} = \cos\theta_{ax}\sin^3\theta_{ax}\sin^3\psi_{ax} \quad,$$
$$A_{9,4} = 2\left(\cos 2\theta_{ax}\sin^2\psi_{ax} + \cos^2\psi_{ax}\right)\cos\theta_{ax}\sin\theta_{ax}\sin\psi_{ax} \quad,$$
$$A_{9,5} = 0 \quad,$$
(19.10i)

$$A_{10,1} = -\left(\cos^2\theta_{ax}\sin^2\psi_{ax} + \cos^2\psi_{ax}\right)\cos\theta_{ax}\sin\theta_{ax}\cos\psi_{ax} \quad ,$$
$$A_{10,2} = +\left(\cos 2\theta_{ax}\sin^2\psi_{ax} + \cos^2\psi_{ax}\right)\cos\theta_{ax}\sin\theta_{ax}\cos\psi_{ax} \quad ,$$
$$A_{10,3} = \cos\theta_{ax}\sin^3\theta_{ax}\cos\psi_{ax}\sin^2\psi_{ax} \quad ,$$
$$A_{10,4} = -4\cos\theta_{ax}\sin^3\theta_{ax}\cos\psi_{ax}\sin^2\psi_{ax} \quad ,$$
$$A_{10,5} = \sin 2\theta_{ax}\cos\psi_{ax} \quad ,$$
(19.10j)

$$A_{11,1} = -\left(\cos^2\theta_{ax}\sin^2\psi_{ax} + \cos^2\psi_{ax}\right)\sin^2\theta_{ax}\cos\psi_{ax}\sin\psi_{ax} \quad ,$$
$$A_{11,2} = +\left(\cos 2\theta_{ax}\sin^2\psi_{ax} + \cos^2\psi_{ax}\right)\sin^2\theta_{ax}\cos\psi_{ax}\sin\psi_{ax} \quad ,$$
$$A_{11,3} = \sin^4\theta_{ax}\cos\psi_{ax}\sin^3\psi_{ax} \quad ,$$
$$A_{11,4} = 2\left(\cos 2\theta_{ax}\sin^2\psi_{ax} + \cos^2\psi_{ax}\right)\sin^2\theta_{ax}\cos\psi_{ax}\sin\psi_{ax} \quad ,$$
$$A_{11,5} = 0 \quad ,$$
(19.10k)

$$A_{12,1} = \sin^4\theta_{ax} \quad ,$$
$$A_{12,2} = 2\cos^2\theta_{ax}\sin^2\theta_{ax} \quad ,$$
$$A_{12,3} = \cos^4\theta_{ax} \quad ,$$
$$A_{12,4} = \sin^2 2\theta_{ax} \quad ,$$
$$A_{12,5} = 0 \quad ,$$
(19.10l)

$$A_{13,1} = -\cos\theta_{ax}\sin^3\theta_{ax}\sin\psi_{ax} \quad ,$$
$$A_{13,2} = -\sin 4\theta_{ax}\sin\psi_{ax}/4 \quad ,$$
$$A_{13,3} = +\cos^3\theta_{ax}\sin\theta_{ax}\sin\psi_{ax} \quad ,$$
$$A_{13,4} = -\sin 4\theta_{ax}\sin\psi_{ax}/2 \quad ,$$
$$A_{13,5} = 0 \quad ,$$
(19.10m)

$$A_{14,1} = -\cos\theta_{ax}\sin^3\theta_{ax}\cos\psi_{ax} \quad ,$$
$$A_{14,2} = -\sin 4\theta_{ax}\cos\psi_{ax}/4 \quad ,$$
$$A_{13,3} = +\cos^3\theta_{ax}\sin\theta_{ax}\cos\psi_{ax} \quad ,$$
$$A_{14,3} = -\sin 4\theta_{ax}\cos\psi_{ax}/2 \quad ,$$
$$A_{14,3} = 0 \quad ,$$
(19.10n)

$$A_{15,1} = -\sin^4\theta_{ax}\cos\psi_{ax}\sin\psi_{ax} \quad,$$
$$A_{15,2} = -\cos 2\theta_{ax}\sin^2\theta_{ax}\sin 2\psi_{ax}/2 \quad,$$
$$A_{15,3} = +\cos^2\theta_{ax}\sin^2\theta_{ax}\cos\psi_{ax}\sin\psi \quad, \quad (19.10\text{o})$$
$$A_{15,4} = -4\cos^2\theta_{ax}\sin^2\theta_{ax}\cos\psi_{ax}\sin\psi \quad,$$
$$A_{15,5} = \sin^2\theta_{ax}\sin 2\psi_{ax} \quad,$$

$$A_{16,1} = +\cos^2\theta_{ax}\sin^2\theta_{ax}\sin^2\psi_{ax} \quad,$$
$$A_{16,2} = -2\cos^2\theta_{ax}\sin^2\theta_{ax}\sin^2\psi_{ax} \quad,$$
$$A_{16,3} = +\cos^2\theta_{ax}\sin^2\theta_{ax}\sin^2\psi \quad, \quad (19.10\text{p})$$
$$A_{16,4} = \cos^2\theta_{ax}\cos^2\psi_{ax} + \cos^2 2\theta_{ax}\sin^2\psi_{ax} \quad,$$
$$A_{16,5} = \sin^2\theta_{ax}\cos^2\psi_{ax} \quad,$$

$$A_{17,1} = +\cos^2\theta_{ax}\sin^2\theta_{ax}\cos\psi_{ax}\sin\psi_{ax} \quad,$$
$$A_{17,2} = -2\cos^2\theta_{ax}\sin^2\theta_{ax}\cos\psi_{ax}\sin\psi_{ax} \quad,$$
$$A_{17,3} = +\cos^2\theta_{ax}\sin^2\theta_{ax}\cos\psi_{ax}\sin\psi_{ax} \quad, \quad (19.10\text{q})$$
$$A_{17,4} = -(1+2\cos 2\theta_{ax})\sin^2\theta_{ax}\sin 2\psi_{ax}/2 \quad,$$
$$A_{17,5} = -\sin^2\theta_{ax}\,s\cos\psi_{ax}\,in\psi_{ax} \quad,$$

$$A_{18,1} = +\cos\theta_{ax}\sin^3\theta_{ax}\cos\psi_{ax}\sin^2\psi_{ax} \quad,$$
$$A_{18,2} = -2\cos\theta_{ax}\sin^3\theta_{ax}\cos\psi_{ax}\sin^2\psi_{ax} \quad,$$
$$A_{18,3} = +\cos\theta_{ax}\sin^3\theta_{ax}\cos\psi_{ax}\sin^2\psi_{ax} \quad, \quad (19.10\text{r})$$
$$A_{18,4} = +\left(2\cos 2\theta_{ax}\sin^2\psi_{ax} + \cos 2\psi_{ax}\right)\cos\theta_{ax}\cos\psi_{ax}\sin\theta_{ax} \quad,$$
$$A_{18,5} = -\cos\theta_{ax}\sin\theta_{ax}\cos\psi_{ax} \quad,$$

$$A_{19,1} = +\cos^2\theta_{ax}\sin^2\theta_{ax}\cos^2\psi_{ax} \quad,$$
$$A_{19,2} = -2\cos^2\theta_{ax}\sin^2\theta_{ax}\cos^2\psi_{ax} \quad,$$
$$A_{19,3} = +\cos^2\theta_{ax}\sin^2\theta_{ax}\cos^2\psi_{ax} \quad, \quad (19.10\text{s})$$
$$A_{19,4} = \cos^2 2\theta_{ax}\cos^2\psi_{ax} + \cos^2\theta_{ax}\sin^2\psi_{ax} \quad,$$
$$A_{19,5} = \sin^2\theta_{ax}\sin^2\psi_{ax} \quad,$$

$$\begin{aligned}
A_{20,1} &= +\cos\theta_{ax}\sin^3\theta_{ax}\cos^2\psi_{ax}\sin\psi_{ax} \quad, \\
A_{20,2} &= -2\cos\theta_{ax}\sin^3\theta_{ax}\cos^2\psi_{ax}\sin\psi_{ax} \quad, \\
A_{20,3} &= +\cos\theta_{ax}\sin^3\theta_{ax}\cos^2\psi_{ax}\sin\psi_{ax} \quad, \\
A_{20,4} &= +\cos^3\theta_{ax}\sin\theta_{ax}\sin\psi_{ax} - \cos\theta_{ax}\sin^3\theta_{ax}\sin 3\psi_{ax} \quad, \\
A_{20,5} &= -\cos\theta_{ax}\sin\theta_{ax}\sin\psi_{ax} \quad,
\end{aligned}$$ (19.10t)

$$\begin{aligned}
A_{21,1} &= +\sin^4\theta_{ax}\cos^2\psi_{ax}\sin^2\psi_{ax} \quad, \\
A_{21,2} &= -2\sin^4\theta_{ax}\cos^2\psi_{ax}\sin^2\psi_{ax} \quad, \\
A_{21,3} &= +\sin^4\theta_{ax}\cos^2\psi_{ax}\sin^2\psi_{ax} \quad, \\
A_{21,4} &= \left(5 + 3\cos 4\psi_{ax} - 8\cos 2\theta_{ax}\cos^2 2\psi_{ax} - 2\cos 4\theta_{ax}\sin^2 2\psi_{ax}\right)/16 \quad, \\
A_{21,5} &= +\cos^2\theta_{ax} \quad.
\end{aligned}$$ (19.10u)

Thus, the components of matrix $\mathbf{A}(21\times 5)$ are the known trigonometric functions of the two polar angles, $\theta_{ax}$ and $\psi_{ax}$. We arrange equation 19.8 as,

$$\mathbf{C} = \mathbf{A}\hat{\mathbf{C}} \quad, \tag{19.11}$$

where $\hat{\mathbf{C}}$ and $\mathbf{C}$ are arrays of lengths 5 and 21, respectively. The gradient of the global stiffness is an array $21\times 3$, representing a collection of vectors,

$$\nabla_{\mathbf{x}}\mathbf{C} = \mathbf{A}\nabla_{\mathbf{x}}\hat{\mathbf{C}} + \left(\frac{\partial\mathbf{A}}{\partial\theta_{ax}}\cdot\hat{\mathbf{C}}\right)\otimes\nabla_{\mathbf{x}}\theta_{ax} + \left(\frac{\partial\mathbf{A}}{\partial\psi_{ax}}\cdot\hat{\mathbf{C}}\right)\otimes\nabla_{\mathbf{x}}\psi_{ax} \quad. \tag{19.12}$$

The Hessian of the global stiffness is an array $21\times 3\times 3$, representing a collection of symmetric second-order tensors,

$$\begin{aligned}
\nabla_\mathbf{x} \nabla_\mathbf{x} \mathbf{C} &= \mathbf{A} \cdot \nabla_\mathbf{x} \nabla_\mathbf{x} \hat{\mathbf{C}} \\
&+ \frac{\partial \mathbf{A}}{\partial \theta_{ax}} \cdot \left( \nabla_\mathbf{x} \hat{\mathbf{C}} \otimes \nabla_\mathbf{x} \theta_{ax} \right) + \frac{\partial \mathbf{A}}{\partial \theta_{ax}} \cdot \left[ \nabla_\mathbf{x} \theta_{ax} \otimes \nabla_\mathbf{x} \hat{\mathbf{C}}^T \right]^{T\{3,1,2\}} \\
&+ \frac{\partial \mathbf{A}}{\partial \psi_{ax}} \cdot \left( \nabla_\mathbf{x} \hat{\mathbf{C}} \otimes \nabla_\mathbf{x} \psi_{ax} \right) + \frac{\partial \mathbf{A}}{\partial \psi_{ax}} \cdot \left[ \nabla_\mathbf{x} \psi_{ax} \otimes \nabla_\mathbf{x} \hat{\mathbf{C}}^T \right]^{T\{3,1,2\}} \\
&+ \left( \frac{\partial^2 \mathbf{A}}{\partial \theta_{ax}^2} \cdot \hat{\mathbf{C}} \right) \otimes \left( \nabla_\mathbf{x} \theta_{ax} \otimes \nabla_\mathbf{x} \theta_{ax} \right) + \left( \frac{\partial^2 \mathbf{A}}{\partial \psi_{ax}^2} \cdot \hat{\mathbf{C}} \right) \otimes \left( \nabla_\mathbf{x} \psi_{ax} \otimes \nabla_\mathbf{x} \psi_{ax} \right) \\
&+ \left( \frac{\partial^2 \mathbf{A}}{\partial \theta_{ax} \partial \psi_{ax}} \cdot \hat{\mathbf{C}} \right) \otimes \left( \nabla_\mathbf{x} \theta_{ax} \otimes \nabla_\mathbf{x} \psi_{ax} + \nabla_\mathbf{x} \psi_{ax} \otimes \nabla_\mathbf{x} \theta_{ax} \right) \\
&+ \frac{\partial \mathbf{A}}{\partial \theta_{ax}} \cdot \left( \hat{\mathbf{C}} \otimes \nabla_\mathbf{x} \nabla_\mathbf{x} \theta_{ax} \right) + \frac{\partial \mathbf{A}}{\partial \psi_{ax}} \cdot \left( \hat{\mathbf{C}} \otimes \nabla_\mathbf{x} \nabla_\mathbf{x} \psi_{ax} \right) \quad .
\end{aligned} \qquad (19.13)$$

Multi-dimensional arrays in this relationship obey tensor multiplication rules, even when they are not true physical tensors. Notation $T\{3,1,2\}$ means that the 3D array is transposed: indices 1, 2, 3 of the original become 3, 1, 2 of the result.

We apply equations 19.11, 19.12, and 19.13 to compute the global stiffness components, their gradients, and their Hessians, respectively. The results are listed in Tables 7, 8, 9, where the units of the (density-normalized) stiffness components are $(\text{km}/\text{s})^2$, while the gradients and Hessians are presented in the "relative form", with the units $\text{km}^{-1}$ and $\text{km}^{-2}$, respectively.

Table 7. Stiffness components and their spatial gradients for polar anisotropy of model 1 presented as general (triclinic) anisotropy

| Component | | Value, $(\text{km}/\text{s})^2$ | Relative gradient, $\text{km}^{-1}$ (in global frame) | | |
|---|---|---|---|---|---|
| # | $C_{ij}$ | | $x_1$ | $x_2$ | $x_3$ |
| 1 | $C_{11}$ | $+1.7192147 \cdot 10^{+1}$ | $+1.6132821 \cdot 10^{-1}$ | $+6.8401473 \cdot 10^{-2}$ | $-5.8359797 \cdot 10^{-2}$ |
| 2 | $C_{12}$ | $+1.1155347 \cdot 10^{+1}$ | $+3.4058675 \cdot 10^{-1}$ | $+2.8068990 \cdot 10^{-1}$ | $+1.1054008 \cdot 10^{-1}$ |
| 3 | $C_{13}$ | $+8.8862242$ | $+3.7543581 \cdot 10^{-1}$ | $+3.8147322 \cdot 10^{-1}$ | $+1.1895402 \cdot 10^{-1}$ |
| 4 | $C_{14}$ | $-1.1116386$ | $+8.3840966 \cdot 10^{-2}$ | $-2.9826347 \cdot 10^{-2}$ | $-6.1125491 * 10^{-2}$ |
| 5 | $C_{15}$ | $-1.3756936$ | $+1.0783425 \cdot 10^{-2}$ | $+9.7174204 * 10^{-2}$ | $-2.2091868 \cdot 10^{-1}$ |
| 6 | $C_{16}$ | $-5.6162454 \cdot 10^{-1}$ | $-7.5124920 \cdot 10^{-2}$ | $+1.5729301 \cdot 10^{-1}$ | $-3.2000103 \cdot 10^{-1}$ |
| 7 | $C_{22}$ | $+1.7192147 \cdot 10^{+1}$ | $+1.5679206 \cdot 10^{-1}$ | $+9.0959106 \cdot 10^{-2}$ | $-6.9217838 \cdot 10^{-2}$ |
| 8 | $C_{23}$ | $+8.8862242$ | $+3.7421849 \cdot 10^{-1}$ | $+3.8752678 \cdot 10^{-1}$ | $+1.1604017 \cdot 10^{-1}$ |
| 9 | $C_{24}$ | $-1.3756936$ | $+3.8129651 \cdot 10^{-2}$ | $-3.8814496 \cdot 10^{-2}$ | $-1.5546098 \cdot 10^{-1}$ |
| 10 | $C_{25}$ | $-1.1116386$ | $+4.0007736 \cdot 10^{-2}$ | $+1.8814976 \cdot 10^{-1}$ | $-1.6604757 \cdot 10^{-1}$ |
| 11 | $C_{26}$ | $-5.6162454 \cdot 10^{-1}$ | $-8.2493292 \cdot 10^{-2}$ | $+1.9393483 \cdot 10^{-1}$ | $-3.3763845 \cdot 10^{-1}$ |
| 12 | $C_{33}$ | $+1.3066002 \cdot 10^{+1}$ | $+1.5671391 \cdot 10^{-1}$ | $+1.3942024 \cdot 10^{-1}$ | $-7.9255362 \cdot 10^{-2}$ |
| 13 | $C_{34}$ | $-6.4569645 \cdot 10^{-1}$ | $+3.3866432 \cdot 10^{-2}$ | $+1.5052517 \cdot 10^{-1}$ | $-2.2022270 \cdot 10^{-1}$ |
| 14 | $C_{35}$ | $-6.4569645 \cdot 10^{-1}$ | $-8.4816728 \cdot 10^{-4}$ | $+3.2315568 \cdot 10^{-1}$ | $-3.0331782 \cdot 10^{-1}$ |
| 15 | $C_{36}$ | $-1.5580447 \cdot 10^{-1}$ | $+1.6880325 \cdot 10^{-2}$ | $+6.7253962 \cdot 10^{-1}$ | $-2.3600675 \cdot 10^{-1}$ |
| 16 | $C_{44}$ | $+3.1065241$ | $-1.8198193 \cdot 10^{-1}$ | $-2.6104261 \cdot 10^{-1}$ | $-3.7720224 \cdot 10^{-1}$ |
| 17 | $C_{45}$ | $+3.0372408 \cdot 10^{-1}$ | $-9.1176383 \cdot 10^{-2}$ | $+6.5647434 \cdot 10^{-2}$ | $-2.6268503 \cdot 10^{-1}$ |
| 18 | $C_{46}$ | $+1.3971923 \cdot 10^{-2}$ | $-6.5595707 \cdot 10^{-1}$ | $+2.0529577$ | $+7.5963197 \cdot 10^{-1}$ |
| 19 | $C_{55}$ | $+3.1065241$ | $-1.8877000 \cdot 10^{-1}$ | $-2.2728652 \cdot 10^{-1}$ | $-3.9345062 \cdot 10^{-1}$ |
| 20 | $C_{56}$ | $+1.3971923 \cdot 10^{-2}$ | $-1.3467419$ | $+5.4881278$ | $-8.9387572 \cdot 10^{-1}$ |
| 21 | $C_{66}$ | $+3.0780040$ | $-1.7915491 \cdot 10^{-1}$ | $-2.7946435 \cdot 10^{-1}$ | $-3.9009910 \cdot 10^{-1}$ |

Table 8. Hessians of polar anisotropy stiffness components presented

as general (triclinic) anisotropy: diagonal elements, model 1

| Component | | Relative Hessian, $km^{-1}$ (in global frame) | | |
|---|---|---|---|---|
| # | $C_{ij}$ | $x_1 x_1$ | $x_2 x_2$ | $x_3 x_3$ |
| 1 | $C_{11}$ | $+8.0031180 \cdot 10^{-2}$ | $+1.0611903 \cdot 10^{-1}$ | $-8.1321886 \cdot 10^{-2}$ |
| 2 | $C_{12}$ | $+2.0949607 \cdot 10^{-1}$ | $-2.4155167 \cdot 10^{-2}$ | $-1.0416276 \cdot 10^{-2}$ |
| 3 | $C_{13}$ | $+2.1474649 \cdot 10^{-1}$ | $+1.1477856 \cdot 10^{-2}$ | $-1.1417243 \cdot 10^{-2}$ |
| 4 | $C_{14}$ | $+2.2296092 \cdot 10^{-2}$ | $-9.8673368 \cdot 10^{-3}$ | $-1.7823804 \cdot 10^{-1}$ |
| 5 | $C_{15}$ | $-1.2216634 \cdot 10^{-1}$ | $+1.9622265 \cdot 10^{-2}$ | $-4.1599009 \cdot 10^{-2}$ |
| 6 | $C_{16}$ | $-2.1598952 \cdot 10^{-1}$ | $+1.5096846 \cdot 10^{-1}$ | $-1.1577661 \cdot 10^{-1}$ |
| 7 | $C_{22}$ | $+6.8122918 \cdot 10^{-2}$ | $+1.0558535 \cdot 10^{-1}$ | $-5.7042334 \cdot 10^{-2}$ |
| 8 | $C_{23}$ | $+2.1131782 \cdot 10^{-1}$ | $+1.7350969 \cdot 10^{-2}$ | $-5.4424859 \cdot 10^{-3}$ |
| 9 | $C_{24}$ | $-4.1180796 \cdot 10^{-2}$ | $+7.4960508 \cdot 10^{-2}$ | $-1.599478 \cdot 10^{-1}$ |
| 10 | $C_{25}$ | $-1.0003696 \cdot 10^{-1}$ | $-4.6391242 \cdot 10^{-2}$ | $+2.6360959 \cdot 10^{-2}$ |
| 11 | $C_{26}$ | $-2.3178415 \cdot 10^{-1}$ | $+1.6239578 \cdot 10^{-1}$ | $-6.6727899 \cdot 10^{-2}$ |
| 12 | $C_{33}$ | $+5.5628244 \cdot 10^{-2}$ | $+1.5185850 \cdot 10^{-1}$ | $-6.5153372 \cdot 10^{-2}$ |
| 13 | $C_{34}$ | $-6.1359155 \cdot 10^{-2}$ | $+2.2082765 \cdot 10^{-1}$ | $-1.9513772 \cdot 10^{-1}$ |
| 14 | $C_{35}$ | $-1.5910934 \cdot 10^{-1}$ | $+2.3788261 \cdot 10^{-1}$ | $-2.0472561 \cdot 10^{-2}$ |
| 15 | $C_{36}$ | $-1.6834503 \cdot 10^{-1}$ | $+3.6460005 \cdot 10^{-1}$ | $-6.2930942 \cdot 10^{-2}$ |
| 16 | $C_{44}$ | $-1.8284592 \cdot 10^{-1}$ | $+3.1201024 \cdot 10^{-1}$ | $-1.7254692 \cdot 10^{-1}$ |
| 17 | $C_{45}$ | $-2.1794757 \cdot 10^{-1}$ | $-7.0079453 \cdot 10^{-2}$ | $-6.9714547 \cdot 10^{-2}$ |
| 18 | $C_{46}$ | $-4.0074558 \cdot 10^{-1}$ | $-1.7685679$ | $-4.9305893 \cdot 10^{-1}$ |
| 19 | $C_{55}$ | $-2.0049799 \cdot 10^{-1}$ | $+3.0378755 \cdot 10^{-1}$ | $-1.3836326 \cdot 10^{-1}$ |
| 20 | $C_{56}$ | $-2.0786628$ | $-2.9371336$ | $-2.9371336$ |
| 21 | $C_{66}$ | $-1.8145087 \cdot 10^{-1}$ | $+3.4384906 \cdot 10^{-1}$ | $-1.7343105 \cdot 10^{-1}$ |

Table 9. Hessians of polar anisotropy stiffness components presented as general (triclinic) anisotropy: off-diagonal elements, model 1

| Component | | Relative Hessian, $\text{km}^{-1}$ (in global frame) | | |
|---|---|---|---|---|
| # | $C_{ij}$ | $x_1 x_2$ | $x_2 x_3$ | $x_1 x_3$ |
| 1 | $C_{11}$ | $+1.4046489 \cdot 10^{-2}$ | $-1.3362002 \cdot 10^{-1}$ | $-8.9484845 \cdot 10^{-2}$ |
| 2 | $C_{12}$ | $+1.1143068 \cdot 10^{-1}$ | $+4.4173639 \cdot 10^{-2}$ | $-3.4899066 \cdot 10^{-2}$ |
| 3 | $C_{13}$ | $+1.5315024 \cdot 10^{-1}$ | $-2.6299025 \cdot 10^{-2}$ | $-9.7131122 \cdot 10^{-3}$ |
| 4 | $C_{14}$ | $+3.0689364 \cdot 10^{-1}$ | $+1.0652913 \cdot 10^{-1}$ | $-2.8941059 \cdot 10^{-3}$ |
| 5 | $C_{15}$ | $+6.7231472 \cdot 10^{-2}$ | $+1.7425876 \cdot 10^{-1}$ | $-1.570753 \cdot 10^{-1}$ |
| 6 | $C_{16}$ | $-5.7735958 \cdot 10^{-2}$ | $+9.7091642 \cdot 10^{-2}$ | $-1.1270939 \cdot 10^{-1}$ |
| 7 | $C_{22}$ | $+4.2858194 \cdot 10^{-2}$ | $-1.1096614 \cdot 10^{-1}$ | $-1.2506246 \cdot 10^{-1}$ |
| 8 | $C_{23}$ | $+1.3141530 \cdot 10^{-1}$ | $-8.8555138 \cdot 10^{-3}$ | $+4.2471284 \cdot 10^{-2}$ |
| 9 | $C_{24}$ | $+1.2492918 \cdot 10^{-1}$ | $-7.3212741 \cdot 10^{-2}$ | $-1.3150462 \cdot 10^{-1}$ |
| 10 | $C_{25}$ | $+2.7641566$ | $+3.2199762$ | $+6.4081316 \cdot 10^{-1}$ |
| 11 | $C_{26}$ | $-1.1490330 \cdot 10^{-1}$ | $+6.6302601 \cdot 10^{-2}$ | $-3.2183826 \cdot 10^{-2}$ |
| 12 | $C_{33}$ | $+6.7305519 \cdot 10^{-2}$ | $-9.9799979 \cdot 10^{-2}$ | $-1.2714903 \cdot 10^{-1}$ |
| 13 | $C_{34}$ | $-7.8583562 \cdot 10^{-1}$ | $+5.1162243 \cdot 10^{-2}$ | $+8.5146917 \cdot 10^{-2}$ |
| 14 | $C_{35}$ | $-1.1825760$ | $-5.3959285 \cdot 10^{-1}$ | $+4.8718245 \cdot 10^{-1}$ |
| 15 | $C_{36}$ | $-1.9031972 \cdot 10^{-1}$ | $-3.7261288 \cdot 10^{-1}$ | $-4.0171345 \cdot 10^{-1}$ |
| 16 | $C_{44}$ | $-1.7419612 \cdot 10^{-1}$ | $-6.8933524 \cdot 10^{-1}$ | $-5.9262649 \cdot 10^{-1}$ |
| 17 | $C_{45}$ | $+1.9227142 \cdot 10^{-1}$ | $+3.5832830 \cdot 10^{-1}$ | $-1.2957257 \cdot 10^{-1}$ |
| 18 | $C_{46}$ | $+9.1433584 \cdot 10^{-2}$ | $+3.0765028 \cdot 10^{-1}$ | $+4.1552149 \cdot 10^{-2}$ |
| 19 | $C_{55}$ | $+5.3613438 \cdot 10^{-2}$ | $-3.4570020 \cdot 10^{-2}$ | $-3.8139665 \cdot 10^{-1}$ |
| 20 | $C_{56}$ | $+3.9373203 \cdot 10^{-1}$ | $-1.8517986 \cdot 10^{-1}$ | $-9.8498727 \cdot 10^{-1}$ |
| 21 | $C_{66}$ | $-1.8525098 \cdot 10^{-2}$ | $-3.6276162 \cdot 10^{-1}$ | $-3.1987921 \cdot 10^{-1}$ |

Now we have all the necessary data to compute the ray velocity and its derivatives for a general triclinic medium applying the workflow and technique of Part I.

The gradients and Hessians of the reference Hamiltonian

- The slowness gradient and Hessian are,

$$H_{\mathbf{p}}^{\bar{\tau}}(P) = [+1.5292672 \quad +2.0390230 \quad +3.3983716] \text{ km/s} \quad , \quad (19.14)$$

$$H_{\mathbf{pp}}^{\bar{\tau}}(P) = \begin{bmatrix} +1.3156487 \cdot 10^{+1} & -6.5837978 & -1.0570487 \cdot 10^{+1} \\ -6.5837978 & +9.1268467 & -1.4589380 \cdot 10^{+1} \\ -1.0570487 \cdot 10^{+1} & -1.4589380 \cdot 10^{+1} & -5.1864173 \end{bmatrix} (\text{km/s})^2 \quad . \quad (19.15)$$

- The spatial gradient and Hessian are,

$$H_{\mathbf{x}}^{\bar{\tau}}(P) = [+1.0307910 \cdot 10^{-1} \quad +7.3508827 \cdot 10^{-2} \quad -3.6182110 \cdot 10^{-2}] \text{ km}^{-1} \quad , \quad (19.16)$$

$$H_{\mathbf{xx}}^{\bar{\tau}}(P) = \begin{bmatrix} +6.7942998 \cdot 10^{-2} & +7.2816551 \cdot 10^{-2} & -1.0171649 \cdot 10^{-1} \\ +7.2816551 \cdot 10^{-2} & +1.0451836 \cdot 10^{-1} & -1.2496685 \cdot 10^{-1} \\ -1.0171649 \cdot 10^{-1} & -1.0498619 \cdot 10^{-1} & -5.1176013 \cdot 10^{-2} \end{bmatrix} \text{ km}^{-2} \quad . \quad (19.17)$$

- The mixed Hessian is,

$$H_{\mathbf{xp}}^{\bar{\tau}}(P) = \begin{bmatrix} +3.0292332 \cdot 10^{-1} & +3.7083701 \cdot 10^{-1} & +5.9581842 \cdot 10^{-1} \\ +2.7168949 \cdot 10^{-1} & +4.2522278 \cdot 10^{-2} & +7.7610092 \cdot 10^{-1} \\ +3.3735610 \cdot 10^{-1} & +4.1993330 \cdot 10^{-1} & +6.1264755 \cdot 10^{-1} \end{bmatrix} \text{ km}^{-2} \quad . \quad (19.18)$$

The gradients and Hessians of the arclength-related Hamiltonian

- The slowness gradient and Hessian are,

$$H_{\mathbf{p}}(P) = [0.36 \quad 0.48 \quad 0.80] = \mathbf{r} \quad , \quad (19.19)$$

$$H_{\mathbf{pp}}(P) = \begin{bmatrix} +4.2632858 & +4.6285195 \cdot 10^{-2} & +7.4954193 \cdot 10^{-2} \\ +4.6285195 \cdot 10^{-2} & +4.3317688 & +7.5048987 \cdot 10^{-2} \\ +7.4954193 \cdot 10^{-2} & +7.5048987 \cdot 10^{-2} & +4.4128055 \end{bmatrix} \text{ km/s} \quad . \quad (19.20)$$

- The inverse of this matrix is,

$$H_{\mathbf{pp}}^{-1}(P) = \begin{bmatrix} +2.3465669 \cdot 10^{-1} & -2.4389839 \cdot 10^{-3} & -3.9443069 \cdot 10^{-3} \\ -2.4389839 \cdot 10^{-3} & +2.3094597 \cdot 10^{-1} & -3.8862917 \cdot 10^{-3} \\ -3.9443069 \cdot 10^{-3} & -3.8862917 \cdot 10^{-3} & +2.2674629 \cdot 10^{-1} \end{bmatrix} \text{ s/km} \quad . \quad (19.21)$$

- The spatial gradient and Hessian are,

$$H_{\mathbf{x}}(P) = \begin{bmatrix} +2.4265528 \cdot 10^{-2} & +1.7304482^{-2} & -8.5175169 \cdot 10^{-3} \end{bmatrix} \text{ s/km}^2 \quad , \quad (19.22)$$

$$H_{\mathbf{xx}}(P) = \begin{bmatrix} +7.2692223 \cdot 10^{-3} & +8.7592149 \cdot 10^{-3} & -2.7058305 \cdot 10^{-2} \\ +8.7592149 \cdot 10^{-3} & +1.7086149 \cdot 10^{-2} & -3.0880151 \cdot 10^{-2} \\ -2.7058305 \cdot 10^{-2} & -3.0880151 \cdot 10^{-2} & -8.7863832 \cdot 10^{-3} \end{bmatrix} \text{ s/km}^3 \quad . \quad (19.23)$$

The mixed Hessian is,

$$H_{\mathbf{xp}}(P) = \begin{bmatrix} +4.5890609 \cdot 10^{-2} & +5.6187094 \cdot 10^{-2} & +8.1874695 \cdot 10^{-2} \\ +1.3781332 \cdot 10^{-2} & +3.5182881 \cdot 10^{-2} & +6.9843915 \cdot 10^{-2} \\ -3.2898342 \cdot 10^{-3} & -1.2395824 \cdot 10^{-2} & -3.8903298 \cdot 10^{-2} \end{bmatrix} \text{ s/km}^3 \quad . \quad (19.24)$$

Finally, we compute the two gradients, $\nabla_{\mathbf{x}} v_{\text{ray}}(\mathbf{x},\mathbf{r})$, $\nabla_{\mathbf{r}} v_{\text{ray}}(\mathbf{x},\mathbf{r})$, and the three Hessians, $\nabla_{\mathbf{x}} \nabla_{\mathbf{x}} v_{\text{ray}}(\mathbf{x},\mathbf{r})$, $\nabla_{\mathbf{r}} \nabla_{\mathbf{r}} v_{\text{ray}}(\mathbf{x},\mathbf{r})$, $\nabla_{\mathbf{x}} \nabla_{\mathbf{r}} v_{\text{ray}}(\mathbf{x},\mathbf{r})$ of the ray velocity magnitude. As expected, the resulted ray velocity derivatives fully coincide with the derivatives listed in equations 14.37-14.41 for compressional waves, computed directly for polar anisotropic media (without its conversion to general anisotropy). We define the discrepancies $E$ between the two methods as,

$$E = \frac{T - P}{P} \quad , \quad (19.25)$$

where $T$ and $P$ stand for the numbers related to triclinic and polar anisotropies, respectively. These discrepancies are listed below.

- For the inverted slowness vector,

$$E(\mathbf{p}) = \begin{bmatrix} -2.98 \cdot 10^{-15} & -8.48 \cdot 10^{-16} & +1.32 \cdot 10^{-15} \end{bmatrix} \quad . \tag{19.26}$$

- For the ray velocity magnitude,

$$E(v_{\text{ray}}) = 3.89 \cdot 10^{-16} \quad . \tag{19.27}$$

- For the spatial gradient and Hessian,

$$E(\nabla_\mathbf{x} v_{\text{ray}}) = \begin{bmatrix} +4.28 \cdot 10^{-15} & +1.36 \cdot 10^{-14} & -2.19 \cdot 10^{-14} \end{bmatrix} \quad , \tag{19.28}$$

$$E(\nabla_\mathbf{x} \nabla_\mathbf{x} v_{\text{ray}}) = \begin{bmatrix} -4.78 \cdot 10^{-15} & +4.01 \cdot 10^{-15} & -9.65 \cdot 10^{-15} \\ +4.01 \cdot 10^{-15} & -1.61 \cdot 10^{-14} & +2.89 \cdot 10^{-16} \\ -9.65 \cdot 10^{-15} & +2.89 \cdot 10^{-16} & -5.61 \cdot 10^{-15} \end{bmatrix} \quad . \tag{19.29}$$

- For the directional gradient,

$$E(\nabla_\mathbf{r} v_{\text{ray}}) = \begin{bmatrix} +2.43 \cdot 10^{-12} & +2.45 \cdot 10^{-14} & +6.50 \cdot 10^{-14} \end{bmatrix} \quad . \tag{19.30}$$

The first component of the relative error in this array essentially exceeds the other two, but the first component of the directional gradient $\nabla_\mathbf{r} v_{\text{ray}}$ is small (as compared to the other components; see equation 14.39), thus, the corresponding absolute error is not particularly large.

- For the directional Hessian,

$$E(\nabla_\mathbf{r} \nabla_\mathbf{r} v_{\text{ray}}) = \begin{bmatrix} -5.71 \cdot 10^{-14} & +1.47 \cdot 10^{-13} & +7.07 \cdot 10^{-14} \\ +1.47 \cdot 10^{-13} & -6.16 \cdot 10^{-14} & -2.48 \cdot 10^{-14} \\ +7.07 \cdot 10^{-14} & -2.48 \cdot 10^{-14} & +1.10 \cdot 10^{-13} \end{bmatrix} \quad . \tag{19.31}$$

- For the mixed Hessian,

$$E(\nabla_\mathbf{x}\nabla_\mathbf{r} v_{\text{ray}}) = \begin{bmatrix} +7.55 \cdot 10^{-14} & +4.14 \cdot 10^{-14} & +5.60 \cdot 10^{-14} \\ -4.66 \cdot 10^{-14} & +5.23 \cdot 10^{-13} & -1.09 \cdot 10^{-13} \\ -5.23 \cdot 10^{-14} & -41.9 \cdot 10^{-14} & -5.00 \cdot 10^{-14} \end{bmatrix} \quad . \tag{19.32}$$

As expected, the discrepancies between the results of the two approaches are "almost" zero, where the worst relative discrepancy does not exceed $E_{\max} = 2.5 \cdot 10^{-12}$, and most of relative discrepancies are much smaller than that. Note that both methods are "almost" analytic and exact. We say "almost", because computing the ray velocity derivatives includes the slowness vector inversion, which is a numerical procedure. In addition to the round-off errors (which are the main source of the tiny discrepancies in equations 19.28-19.32), the governing equations for this inversion are formulated differently for polar and triclinic anisotropies. Still, the inverted slowness vectors with the two methods are pretty close: the relative discrepancy for the inverted slowness components does not exceed $3 \cdot 10^{-15}$ (equation 19.26).

## 20. TTI MODEL 2: PARAMETER SETTING AND SLOWNESS INVERSION

Next, we consider another model representing an extreme (but still realistic) case of a large negative anellipticity. Unlike compacted sediment shale/sand rocks, this model, suggested by Grechka (2013), is characterized with high positive $\delta$ and negative $\varepsilon$ values,

$$v_P = 3 \text{ km/s}, \quad f = 0.75, \quad \delta = 0.3, \quad \varepsilon = -0.15 \quad . \tag{20.1}$$

These values represent polar anisotropic materials in unconsolidated shallow sand rocks (also used by Ravve and Koren, 2021b), and for the selected ray direction, the slowness inversion yields a shear wave triplication (i.e., the same ray velocity direction corresponds to three qSV slowness vectors with their corresponding ray velocity magnitudes; obviously, the triplication makes the problem more challenging).

Remark: The medium properties suggested by Grechka (2013) do not include the Thomsen parameter $\gamma$ because this parameter is only related to SH shear waves. Since we also consider SH waves in this study, we assigned the value for $\gamma$ imposing the stability condition: The stiffness tensor should be positive definite (positive semidefinite in the case of the acoustic approximation) (Slawinski, 2015; Adamus, 2020). To find the eigenvalues, we arrange the polar anisotropic stiffness tensor in the Kelvin matrix form (rather than the commonly used Voight matrix form),

$$\mathbf{C} = \begin{bmatrix} C_{11} & C_{12} & C_{13} & & & \\ C_{12} & C_{11} & C_{13} & & & \\ C_{13} & C_{13} & C_{33} & & & \\ & & & 2C_{44} & & \\ & & & & 2C_{44} & \\ & & & & & 2C_{66} \end{bmatrix} \quad , \quad C_{12} = C_{11} - 2C_{66} \quad . \tag{20.2}$$

This polar anisotropy stiffness matrix has two double eigenvalues, $2C_{44}$ and $2C_{66}$, and two simple eigenvalues (e.g., Bona et al, 2007), which are the roots of a quadratic equation. If the smaller root of this equation is positive, then of course, the larger root is also positive, and the material is stable. Thus, we impose the condition $\lambda_{\min} > 0$, where,

$$\frac{2\lambda_{\min}}{v_P^2} = A - \sqrt{A^2 - 16B(1-f) + 4\left[3 - A - 4(1-f-\delta)f\right]} \quad ,$$

$$\text{where} \quad A \equiv 1 + 4\varepsilon + 2f - 4\gamma(1-f) \text{ and } B \equiv \sqrt{f(f+2\delta)} \quad . \tag{20.3}$$

Solving equation 20.2 for $\gamma$, we find its upper limit value, $\gamma < \gamma_{\text{crit}}$, where,

$$\gamma_{\text{crit}} = B + f + \delta + \sigma - 1/2, \qquad \sigma = \frac{\varepsilon - \delta}{1 - f}. \qquad (20.4)$$

For the polar anisotropic parameters $\mathbf{c} = \{v_P, f, \delta, \varepsilon\}$ given in equation 20.1 and Table 10, the stability condition leads to a large negative critical $\gamma$ parameter (equation 20.4), $\gamma_{\text{crit}} = -0.24376941$, which means that a material with the given values of $f, \delta, \varepsilon$ and a value of $\gamma > \gamma_{\text{crit}}$, does not exist; therefore, we assigned the value $\gamma = -0.25$. Equation 20.4 cannot be applied for the acoustic approximation (AC) due to the factor $1 - f$ in the denominator (but equation 20.3 still can). Note that parameter $\gamma$ is irrelevant for the AC. For $f = 1$, equation 20.3 yields $\varepsilon \geq \delta$, i.e., only polar anisotropic (transversely isotropic) media with a non-negative anellipticity parameters, $\eta = (\varepsilon - \delta)/(1 + 2\delta)$, are stable under the AC conditions. However, the AC can be considered an abstract model, introduced to decrease the number of medium parameters and to simplify the governing equations of compressional waves in elastic media; thus, to the best of our understanding, it can be (carefully) applied to polar anisotropy with negative anellipticities as well; we will show that even in this case it provides a reasonable accuracy for the quasi-compressional ray velocity; however the AC accuracy of the ray velocity derivatives (in particular the second derivatives) may be questionable for both positive and negative anellipticities $\eta$.

Table 10. Properties of the polar anisotropy and their spatial gradients, model 2

| Component | | Value | Relative gradient, km$^{-1}$ (in global frame) | | |
|---|---|---|---|---|---|
| # | $m_i$ | | $x_1$ | $x_2$ | $x_3$ |
| 1 | $v_P$ | 3 km/s | −0.0541 | +0.0385 | +0.0254 |
| 2 | $f$ | 0.75 | +0.0650 | −0.1241 | +0.0874 |
| 3 | $\delta$ | +0.30 | −0.0633 | −0.0352 | +0.0759 |
| 4 | $\varepsilon$ | −0.15 | +0.0429 | −0.1155 | +0.0924 |
| 5 | $\gamma$ | −0.25 | −0.0212 | +0.1170 | +0.0763 |
| 6 | $\theta_{ax}$ | 0.69509432 rad | −0.0405 | +0.1252 | −0.1635 |
| 7 | $\psi_{ax}$ | 1.11832143 rad | +0.0815 | −0.1544 | +0.0916 |

Table 11. Spatial Hessians of polar anisotropy properties, model 2

| Component | | Relative Hessian, km$^{-2}$ (in global frame) | | | | | |
|---|---|---|---|---|---|---|---|
| # | $m_i$ | $x_1 x_1$ | $x_1 x_2$ | $x_1 x_3$ | $x_2 x_2$ | $x_2 x_3$ | $x_3 x_3$ |
| 1 | $v_P$ | −0.0134 | +0.0256 | −0.0571 | +0.0479 | −0.0563 | +0.0211 |
| 3 | $f$ | +0.0186 | +0.0098 | +0.0692 | −0.0315 | +0.0559 | −0.0687 |
| 3 | $\delta$ | −0.0196 | +0.0473 | −0.0347 | +0.0292 | −0.0112 | −0.0812 |
| 4 | $\varepsilon$ | −0.0405 | −0.0651 | +0.0137 | −0.0321 | +0.0543 | −0.0301 |
| 5 | $\gamma$ | +0.0244 | −0.0105 | −0.0289 | +0.0913 | +0.0487 | −0.0507 |
| 6 | $\theta_{ax}$ | −0.0272 | −0.0921 | +0.0177 | +0.0599 | −0.0141 | −0.0614 |
| 7 | $\psi_{ax}$ | +0.0719 | +0.0208 | −0.0255 | +0.0414 | −0.0362 | +0.0413 |

The symmetry axis direction angles lead to the following Cartesian components of the symmetry axis direction,

$$\mathbf{k} = \begin{bmatrix} 0.280 & 0.576 & 0.768 \end{bmatrix} \quad . \tag{20.5}$$

We assume the following ray velocity direction,

$$\mathbf{r} = \begin{bmatrix} +0.5696 & +0.48 & -0.6672 \end{bmatrix} \quad , \tag{20.6}$$

and perform the slowness inversion. We obtain five solutions (also obtained in Ravve and Koren, 2021b) listed in Table 12: One for the quasi-compressional qP waves, three for the triplicated qSV shear waves ($qSV_1, qSV_2, qSV_3$) and one for the SH waves. In addition, we obtain a solution for the AC waves, $qP_{ac}$, by setting $f = 1$.

Table 12. Results of the slowness vector inversion, model 2

| Mode | $p_1$, s/km | $p_2$, s/km | $p_3$, s/km | $v_{phs}$, km/s | $v_{ray}$, km/s |
|---|---|---|---|---|---|
| $qP_{ac}$ | +0.23331742 | +0.20380847 | −0.25011372 | 2.5115500 | 2.5150808 |
| qP | +0.23355822 | +0.20428276 | −0.24952117 | 2.5114714 | 2.5152739 |
| $qSV_1$ | +0.39826410 | +0.35649389 | −0.39921519 | 1.4989371 | 1.5052881 |
| $qSV_2$ | +1.0331849 | +1.2499129 | +0.012121941 | 0.61663811 | 0.84719014 |
| $qSV_3$ | +0.45355759 | +0.057425633 | −1.5780707 | 0.60865734 | 0.74693784 |
| SH | +0.41700904 | +0.35443873 | −0.47870791 | 1.3753134 | 1.3754158 |

## 21. MODEL 2. RAY VELOCITY DERIVATIVES FOR qP WAVES

The workflow and formulae used in this section are described in detail in section 15 for the parameters of model 1. Thus, in this section we only list the final results and discuss their accuracy.

Gradients and Hessians of the ray velocity magnitude

After computing all necessary derivatives of the reference Hamiltonian and then the arclength-related Hamiltonian, we can establish the gradients and Hessians of the ray velocity magnitude. The computation of the compressional ray velocity yields $v_{\text{ray}} = 2.5152739 \text{ km/s}$. The spatial gradient and Hessian of the ray velocity magnitude are, respectively,

$$\nabla_{\mathbf{x}} v_{\text{ray}} = \begin{bmatrix} -1.5304772 \cdot 10^{-1} & +1.4064440 \cdot 10^{-1} & +3.2022975 \cdot 10^{-2} \end{bmatrix} \text{ s}^{-1} \quad , \quad (21.1)$$

$$\nabla_{\mathbf{x}} \nabla_{\mathbf{x}} v_{\text{ray}} = \begin{bmatrix} -1.9082789 \cdot 10^{-3} & +9.6134592 \cdot 10^{-2} & -1.4220633 \cdot 10^{-1} \\ +9.6134592 \cdot 10^{-2} & +1.6118634 \cdot 10^{-1} & -1.9719449 \cdot 10^{-1} \\ -1.4220633 \cdot 10^{-1} & -1.9719449 \cdot 10^{-1} & +1.0376433 \cdot 10^{-1} \end{bmatrix} (\text{km} \cdot \text{s})^{-1} . \quad (21.2)$$

The directional gradient and Hessian of the ray velocity magnitude are, respectively,

$$\nabla_{\mathbf{r}} v_{\text{ray}} = \begin{bmatrix} -4.4930092 \cdot 10^{-2} & -8.5084409 \cdot 10^{-2} & -9.9569389 \cdot 10^{-2} \end{bmatrix} \text{ km/s} \quad , \quad (21.3)$$

$$\nabla_{\mathbf{r}} \nabla_{\mathbf{r}} v_{\text{ray}} = \begin{bmatrix} +2.3617230 \cdot 10^{-1} & +4.3682610 \cdot 10^{-1} & +4.4854643 \cdot 10^{-1} \\ +4.3682610 \cdot 10^{-1} & +7.6261840 \cdot 10^{-1} & +7.9404761 \cdot 10^{-1} \\ +4.4854643 \cdot 10^{-1} & +7.9404761 \cdot 10^{-1} & +8.0495430 \cdot 10^{-1} \end{bmatrix} \text{ km/s} \quad , \quad (21.4)$$

and the mixed "spatial-directional" Hessian of the ray velocity magnitude is,

$$\nabla_{\mathbf{x}} \nabla_{\mathbf{r}} v_{\text{ray}} = \begin{bmatrix} -1.8312703 \cdot 10^{-2} & -4.8123132 \cdot 10^{-2} & -5.0254825 \cdot 10^{-2} \\ +5.8889453 \cdot 10^{-2} & +1.3407109 \cdot 10^{-1} & +1.4672895 \cdot 10^{-1} \\ -7.4386047 \cdot 10^{-2} & -1.4887329 \cdot 10^{-1} & -1.7060772 \cdot 10^{-1} \end{bmatrix} \text{ s}^{-1} . \quad (21.5)$$

The gradient of the ray velocity wrt the model parameters is presented as a concatenation of the material and geometric counterparts, where the material part reads,

$$\nabla_{\mathbf{c}} v_{\text{ray}} = \begin{bmatrix} +8.3842464 \cdot 10^{-1} & -1.0988204 \cdot 10^{-3} & +2.7597849 \cdot 10^{-3} & 3.5790534 \end{bmatrix} \quad , \quad (21.6)$$

and the geometric part (symmetry axis direction angles) reads,

$$\nabla_{\mathbf{a}} v_{\text{ray}} = \begin{bmatrix} -1.3194051 \cdot 10^{-1} & +2.6897635 \cdot 10^{-2} \end{bmatrix} \text{ km}/(\text{s} \cdot \text{radian}) \quad . \tag{21.7}$$

The $6 \times 6$ Hessian matrix of the ray velocity wrt the model parameters $\nabla_{\mathbf{m}} \nabla_{\mathbf{m}} v_{\text{ray}}$ consists of four blocks: two symmetric diagonal blocks ($\nabla_{\mathbf{c}} \nabla_{\mathbf{c}} v_{\text{ray}}$ and $\nabla_{\mathbf{a}} \nabla_{\mathbf{a}} v_{\text{ray}}$) of dimensions $4 \times 4$ and $2 \times 2$, and related to either material or geometric properties of the model, respectively, and two mutually transposed off-diagonal blocks $2 \times 4$ and $4 \times 2$, $\nabla_{\mathbf{a}} \nabla_{\mathbf{c}} v_{\text{ray}}$ and $\nabla_{\mathbf{c}} \nabla_{\mathbf{a}} v_{\text{ray}}$, respectively, related to both the material and geometric properties (equation 14.46). The material block reads,

$$\nabla_{\mathbf{c}} \nabla_{\mathbf{c}} v_{\text{ray}} = \begin{bmatrix} -1.1102230 \cdot 10^{-16} & -3.6627346 \cdot 10^{-4} & +9.1992831 \cdot 10^{-4} & +1.1930178 \\ -3.6627346 \cdot 10^{-4} & +3.7134583 \cdot 10^{-3} & +4.7864340 \cdot 10^{-4} & +6.1041148 \cdot 10^{-3} \\ +9.1992831 \cdot 10^{-4} & +4.7864340 \cdot 10^{-4} & -7.3454004 \cdot 10^{-3} & +2.1584586 \cdot 10^{-2} \\ +1.1930178 & +6.1041148 \cdot 10^{-3} & +2.1584586 \cdot 10^{-2} & -5.2135064 \end{bmatrix} .$$

(21.8)

The mixed block reads,

$$\nabla_{\mathbf{a}} \nabla_{\mathbf{c}} v_{\text{ray}} = \begin{bmatrix} -4.3980170 \cdot 10^{-2} & +2.7406544 \cdot 10^{-2} & -6.9154265 \cdot 10^{-2} & +1.6725630 \cdot 10^{-1} \\ +8.9658784 \cdot 10^{-3} & -5.5871486 \cdot 10^{-3} & +1.4097916 \cdot 10^{-2} & -3.4097178 \cdot 10^{-2} \end{bmatrix}$$

(21.9)

The first row of the mixed block has the units $\text{radian}^{-1}$, while the other components have the units $\text{km}/(\text{s} \cdot \text{radian})$. Finally, the geometric block reads,

$$\nabla_{\mathbf{a}} \nabla_{\mathbf{a}} v_{\text{ray}} = \begin{bmatrix} +1.6464628 & -3.0556052 \cdot 10^{-1} \\ -3.0556052 \cdot 10^{-1} & +1.2941038 \cdot 10^{-1} \end{bmatrix} \frac{\text{km}}{\text{s} \cdot \text{radian}^2} \quad . \tag{21.10}$$

Validation test

As in section 15, in order to check the correctness of all the ray velocity derivatives, obtained analytically, we compare their values to numerical values obtained by the finite differences method. The numerical directional derivatives of the ray velocity, approximating the gradient and the Hessians $\nabla_{\mathbf{r}} v_{\text{ray}}$, $\nabla_{\mathbf{r}} \nabla_{\mathbf{r}} v_{\text{ray}}$ and the mixed Hessian $\nabla_{\mathbf{x}} \nabla_{\mathbf{r}} v_{\text{ray}}$, were computed wrt the zenith and azimuth angles, $\theta_{\text{ray}}$ and $\psi_{\text{ray}}$, of the ray direction vector, and then were related analytically to the Cartesian components of this direction. The numerical approximations for the ray velocity derivatives were computed with the spatial resolution $10^{-4}$ km and angular resolution $10^{-5}$ radians for the spatial and directional gradients and Hessians of the ray velocity magnitude, including the mixed Hessian. When computing the ray velocity derivatives wrt the model parameters, $\nabla_{\mathbf{m}} v_{\text{ray}}$ and $\nabla_{\mathbf{m}} \nabla_{\mathbf{m}} v_{\text{ray}}$, the resolution was $10^{-4}$ times the absolute value of the parameter. The (unitless) relative errors of the numerical approximations of the ray velocity derivatives for qP waves are listed below.

- For the spatial gradient and Hessian,

$$E(\nabla_{\mathbf{x}} v_{\text{ray}}) = \begin{bmatrix} -3.21 \cdot 10^{-11} & +2.50 \cdot 10^{-10} & +1.98 \cdot 10^{-9} \end{bmatrix} \quad , \tag{21.11}$$

$$E(\nabla_{\mathbf{x}} \nabla_{\mathbf{x}} v_{\text{ray}}) = \begin{bmatrix} +8.80 \cdot 10^{-6} & +2.06 \cdot 10^{-7} & +2.85 \cdot 10^{-7} \\ +2.06 \cdot 10^{-7} & -2.17 \cdot 10^{-7} & -2.28 \cdot 10^{-7} \\ +2.85 \cdot 10^{-7} & -2.28 \cdot 10^{-7} & +1.31 \cdot 10^{-6} \end{bmatrix} \quad . \tag{21.12}$$

- For the directional gradient and Hessian,

$$E(\nabla_{\mathbf{r}} v_{\text{ray}}) = \begin{bmatrix} -6.60 \cdot 10^{-11} & +3.87 \cdot 10^{-10} & +2.13 \cdot 10^{-10} \end{bmatrix} \quad , \tag{21.13}$$

$$E(\nabla_{\mathbf{r}} \nabla_{\mathbf{r}} v_{\text{ray}}) = \begin{bmatrix} -2.02 \cdot 10^{-5} & -1.80 \cdot 10^{-6} & -1.03 \cdot 10^{-5} \\ -1.80 \cdot 10^{-6} & -8.55 \cdot 10^{-6} & -6.75 \cdot 10^{-6} \\ -1.03 \cdot 10^{-5} & -6.75 \cdot 10^{-6} & -9.71 \cdot 10^{-6} \end{bmatrix} \quad . \tag{21.14}$$

- For the mixed Hessian,

$$E(\nabla_{\mathbf{x}}\nabla_{\mathbf{r}}v_{\text{ray}}) = \begin{bmatrix} +1.29 \cdot 10^{-6} & +2.03 \cdot 10^{-6} & +1.80 \cdot 10^{-6} \\ -1.31 \cdot 10^{-6} & +1.13 \cdot 10^{-6} & +2.91 \cdot 10^{-7} \\ -5.94 \cdot 10^{-7} & +1.37 \cdot 10^{-6} & +6.37 \cdot 10^{-7} \end{bmatrix} \quad . \tag{21.15}$$

- For the material, $\nabla_{\mathbf{c}}v_{\text{ray}}$, and geometric, $\nabla_{\mathbf{a}}v_{\text{ray}}$, parts of the model-related gradient $\nabla_{\mathbf{m}}v_{\text{ray}}$,

$$E(\nabla_{\mathbf{c}}v_{\text{ray}}) = \begin{bmatrix} -5.14 \cdot 10^{-14} & +1.61 \cdot 10^{-6} & +1.59 \cdot 10^{-7} & +2.30 \cdot 10^{-8} \end{bmatrix} \quad , \tag{21.16}$$

$$E(\nabla_{\mathbf{a}}v_{\text{ray}}) = \begin{bmatrix} +6.20 \cdot 10^{-8} & +3.44 \cdot 10^{-6} \end{bmatrix} \quad . \tag{21.17}$$

- For the material, $\nabla_{\mathbf{c}}\nabla_{\mathbf{c}}v_{\text{ray}}$, geometric, $\nabla_{\mathbf{a}}\nabla_{\mathbf{a}}v_{\text{ray}}$, and mixed, $\nabla_{\mathbf{a}}\nabla_{\mathbf{c}}v_{\text{ray}}$, parts of the model-related Hessian $\nabla_{\mathbf{m}}\nabla_{\mathbf{m}}v_{\text{ray}}$,

$$E(\nabla_{\mathbf{c}}\nabla_{\mathbf{c}}v_{\text{ray}}) = \begin{bmatrix} - & +1.66 \cdot 10^{-6} & +1.61 \cdot 10^{-7} & +2.29 \cdot 10^{-8} \\ +1.66 \cdot 10^{-6} & +1.67 \cdot 10^{-6} & +3.92 \cdot 10^{-6} & +2.58 \cdot 10^{-6} \\ +1.61 \cdot 10^{-7} & +3.92 \cdot 10^{-6} & +1.06 \cdot 10^{-6} & +3.23 \cdot 10^{-7} \\ +2.29 \cdot 10^{-8} & +2.58 \cdot 10^{-6} & +3.23 \cdot 10^{-7} & +5.98 \cdot 10^{-8} \end{bmatrix} \quad , \tag{21.18}$$

$$E(\nabla_{\mathbf{a}}\nabla_{\mathbf{c}}v_{\text{ray}}) = \begin{bmatrix} +6.89 \cdot 10^{-8} & +1.77 \cdot 10^{-6} & +1.02 \cdot 10^{-6} & +1.83 \cdot 10^{-6} \\ +3.44 \cdot 10^{-6} & +5.15 \cdot 10^{-6} & +3.75 \cdot 10^{-6} & +2.81 \cdot 10^{-6} \end{bmatrix} \quad , \tag{21.19}$$

$$E(\nabla_{\mathbf{a}}\nabla_{\mathbf{a}}v_{\text{ray}}) = \begin{bmatrix} +2.95 \cdot 10^{-8} & -8.31 \cdot 10^{-7} \\ -8.31 \cdot 10^{-7} & +6.43 \cdot 10^{-7} \end{bmatrix} \quad . \tag{21.20}$$

As in section 15, a dash has been placed to the upper left corner of $E(\nabla_{\mathbf{c}}\nabla_{\mathbf{c}}v_{\text{ray}})$ because the relative error cannot be estimated for a parameter whose exact theoretical value is zero; however, the numerical (finite difference) estimate was $\partial^2 v_{\text{ray}} / \partial v_P^2 = 4.93 \cdot 10^{-11}$ s / km or $1.24 \cdot 10^{-10}$ (unitless) after the normalization with the ray velocity magnitude, which is well below the numerical error for the other second derivatives and indicates the correctness of the vanishing value.

Sensitivity of the ray velocity derivatives to the spatial derivatives of the model parameters

As in section 15, our next test checks the significance (the effect) of the spatial gradients and Hessians of the model parameters $f, \delta, \varepsilon, \theta_{ax}$ and $\psi_{ax}$ (all model parameters except the axial compressional velocity $v_P$) on the derivatives of the ray velocity magnitude for compressional waves. For this, we repeat the computations, assuming $\mathbf{m_x} = 0$ and $\mathbf{m_{xx}} = 0$, except $\nabla_{\mathbf{x}} v_P$ and $\nabla_{\mathbf{x}} \nabla_{\mathbf{x}} v_P$ which are taken into account. Obviously, this does not affect the directional derivatives, $\nabla_{\mathbf{r}} v_{\text{ray}}$ and $\nabla_{\mathbf{r}} \nabla_{\mathbf{r}} v_{\text{ray}}$, and the derivatives wrt the medium properties, $\nabla_{\mathbf{m}} v_{\text{ray}}$ and $\nabla_{\mathbf{m}} \nabla_{\mathbf{m}} v_{\text{ray}}$. We re-compute the spatial derivatives, $\nabla_{\mathbf{x}} v_{\text{ray}}$ and $\nabla_{\mathbf{x}} \nabla_{\mathbf{x}} v_{\text{ray}}$ and the mixed Hessian $\nabla_{\mathbf{x}} \nabla_{\mathbf{r}} v_{\text{ray}}$, and compare the vector and the two tensors obtained for three cases: 1) accounting for the spatial derivatives of all model parameters, 2) accounting for the spatial derivatives of the medium properties and ignoring those of the symmetry axis direction angles, and 3) accounting only for the spatial derivatives of the axial compressional velocity. For cases 2 and 3, we use notations $\tilde{\nabla}_{\mathbf{x}}$ and $\hat{\nabla}_{\mathbf{x}}$, respectively. As mentioned in Section 15, case 3 is referred to as FAI media – Factorized Anisotropic Inhomogeneous media (Červený, 2000). To compare the spatial gradients, we compute the ratio of their magnitudes and the angle between their directions; to compare the spatial Hessians, we compute the ratio of their corresponding eigenvalues. The non-symmetric mixed Hessian, $3 \times 3$, has a single zero eigenvalue, and the corresponding eigenvector is the ray direction **r** (Koren and Ravve, 2021); this property holds even if we ignore the spatial derivatives of the model, so we compare the two other eigenvalues, which prove to be real for the given example and for qP waves.

Accounting for the spatial derivatives of the medium properties, $\mathbf{c_x}$ and $\mathbf{c_{xx}}$ and ignoring those of the symmetry axis direction angles, $\mathbf{a_x}$ and $\mathbf{a_{xx}}$, we obtained the following results,

$$\tilde{\nabla}_{\mathbf{x}} v_{\text{ray}} = \begin{bmatrix} -1.592135 \cdot 10^{-1} & 1.5677084 \cdot 10^{-1} & +1.4273089 \cdot 10^{-2} \end{bmatrix} \text{ s}^{-1} \quad , \quad (21.21)$$

$$\tilde{\nabla}_{\mathbf{x}}\tilde{\nabla}_{\mathbf{x}} v_{\text{ray}} = \begin{bmatrix} -9.6991852 \cdot 10^{-3} & +9.5785035 \cdot 10^{-2} & -1.4942056 \cdot 10^{-1} \\ +9.5785035 \cdot 10^{-2} & +1.4092704 \cdot 10^{-1} & -1.6999229 \cdot 10^{-1} \\ -1.4942056 \cdot 10^{-1} & -1.6999229 \cdot 10^{-1} & +6.5688008 \cdot 10^{-2} \end{bmatrix} (\text{km} \cdot \text{s})^{-1} \quad , (21.22)$$

$$\tilde{\nabla}_{\mathbf{x}}\nabla_{\mathbf{r}} v_{\text{ray}} = \begin{bmatrix} +2.9663811 \cdot 10^{-3} & +5.6174554 \cdot 10^{-3} & +6.5737849 \cdot 10^{-3} \\ -1.3971873 \cdot 10^{-3} & -2.6458628 \cdot 10^{-3} & -3.096301 \cdot 10^{-3} \\ -1.855089 \cdot 10^{-3} & -3.5129942 \cdot 10^{-3} & -4.111055 \cdot 10^{-1} \end{bmatrix} \text{ s}^{-1} \quad . \quad (21.23)$$

Accounting for the spatial derivatives of the axial compressional velocity only, $\nabla_{\mathbf{x}} v_P$ and $\nabla_{\mathbf{xx}} v_P$, we obtained,

$$\hat{\nabla}_{\mathbf{x}} v_{\text{ray}} = \begin{bmatrix} -1.3607632 \cdot 10^{-1} & 9.6838046 \cdot 10^{-2} & +6.3887958 \cdot 10^{-2} \end{bmatrix} \text{ s}^{-1} \quad , \quad (21.24)$$

$$\hat{\nabla}_{\mathbf{x}}\hat{\nabla}_{\mathbf{x}} v_{\text{ray}} = \begin{bmatrix} -3.3704670 \cdot 10^{-2} & +6.4391012 \cdot 10^{-2} & -1.4362214 \cdot 10^{-1} \\ +6.4391012 \cdot 10^{-2} & +1.2048162 \cdot 10^{-1} & -1.4160992 \cdot 10^{-1} \\ -1.4362214 \cdot 10^{-1} & -1.4160992 \cdot 10^{-1} & +5.3072280 \cdot 10^{-2} \end{bmatrix} (\text{km} \cdot \text{s})^{-1} \quad , (21.25)$$

$$\hat{\nabla}_{\mathbf{x}}\nabla_{\mathbf{r}} v_{\text{ray}} = \begin{bmatrix} +2.4307180 \cdot 10^{-3} & +4.6030665 \cdot 10^{-3} & +5.3867039 \cdot 10^{-3} \\ -1.7298085 \cdot 10^{-3} & -3.2757497 \cdot 10^{-3} & -3.8334215 \cdot 10^{-3} \\ -1.1412243 \cdot 10^{-3} & -2.1611440 \cdot 10^{-3} & -2.5290625 \cdot 10^{-1} \end{bmatrix} \text{ s}^{-1} \quad . \quad (21.26)$$

The magnitudes of the ray velocity gradients computed with complete and incomplete data are related as,

$$\frac{\sqrt{\tilde{\nabla}_{\mathbf{x}} v_{\text{ray}} \cdot \tilde{\nabla}_{\mathbf{x}} v_{\text{ray}}}}{\sqrt{\nabla_{\mathbf{x}} v_{\text{ray}} \cdot \nabla_{\mathbf{x}} v_{\text{ray}}}} = 1.0646079 \quad , \quad \frac{\sqrt{\hat{\nabla}_{\mathbf{x}} v_{\text{ray}} \cdot \hat{\nabla}_{\mathbf{x}} v_{\text{ray}}}}{\sqrt{\nabla_{\mathbf{x}} v_{\text{ray}} \cdot \nabla_{\mathbf{x}} v_{\text{ray}}}} = 0.85026491 \quad . \tag{21.27}$$

The angles between the corresponding gradient directions are,

$$\angle \nabla_{\mathbf{x}} v_{\text{ray}}, \tilde{\nabla}_{\mathbf{x}} v_{\text{ray}} = \arccos \frac{\nabla_{\mathbf{x}} v_{\text{ray}} \cdot \tilde{\nabla}_{\mathbf{x}} v_{\text{ray}}}{\sqrt{\nabla_{\mathbf{x}} v_{\text{ray}} \cdot \nabla_{\mathbf{x}} v_{\text{ray}}} \sqrt{\tilde{\nabla}_{\mathbf{x}} v_{\text{ray}} \cdot \tilde{\nabla}_{\mathbf{x}} v_{\text{ray}}}} = 9.5432004 \cdot 10^{-2} \text{ rad} = 5.4678511^{\circ} \quad ,$$

$$\angle \nabla_{\mathbf{x}} v_{\text{ray}}, \hat{\nabla}_{\mathbf{x}} v_{\text{ray}} = \arccos \frac{\nabla_{\mathbf{x}} v_{\text{ray}} \cdot \hat{\nabla}_{\mathbf{x}} v_{\text{ray}}}{\sqrt{\nabla_{\mathbf{x}} v_{\text{ray}} \cdot \nabla_{\mathbf{x}} v_{\text{ray}}} \sqrt{\hat{\nabla}_{\mathbf{x}} v_{\text{ray}} \cdot \hat{\nabla}_{\mathbf{x}} v_{\text{ray}}}} = 2.4415670 \cdot 10^{-1} \text{ rad} = 13.989148^{\circ} \quad .$$

(21.28)

We compute the ratios of the eigenvalues (major, intermediate and minor) of the spatial Hessians,

$$\frac{\lambda_{\text{major}} \tilde{\nabla}_{\mathbf{x}} \tilde{\nabla}_{\mathbf{x}} v_{\text{ray}}}{\lambda_{\text{major}} \nabla_{\mathbf{x}} \nabla_{\mathbf{x}} v_{\text{ray}}} = 0.88930125 \quad , \quad \frac{\lambda_{\text{major}} \hat{\nabla}_{\mathbf{x}} \hat{\nabla}_{\mathbf{x}} v_{\text{ray}}}{\lambda_{\text{major}} \nabla_{\mathbf{x}} \nabla_{\mathbf{x}} v_{\text{ray}}} = 0.73499760 \quad ,$$

$$\frac{\lambda_{\text{interm}} \tilde{\nabla}_{\mathbf{x}} \tilde{\nabla}_{\mathbf{x}} v_{\text{ray}}}{\lambda_{\text{interm}} \nabla_{\mathbf{x}} \nabla_{\mathbf{x}} v_{\text{ray}}} = 1.01261071 \quad , \quad \frac{\lambda_{\text{interm}} \hat{\nabla}_{\mathbf{x}} \hat{\nabla}_{\mathbf{x}} v_{\text{ray}}}{\lambda_{\text{interm}} \nabla_{\mathbf{x}} \nabla_{\mathbf{x}} v_{\text{ray}}} = 0.31614778 \quad , \tag{21.29}$$

$$\frac{\lambda_{\text{minor}} \tilde{\nabla}_{\mathbf{x}} \tilde{\nabla}_{\mathbf{x}} v_{\text{ray}}}{\lambda_{\text{minor}} \nabla_{\mathbf{x}} \nabla_{\mathbf{x}} v_{\text{ray}}} = 1.19452327 \quad , \quad \frac{\lambda_{\text{minor}} \hat{\nabla}_{\mathbf{x}} \hat{\nabla}_{\mathbf{x}} v_{\text{ray}}}{\lambda_{\text{minor}} \nabla_{\mathbf{x}} \nabla_{\mathbf{x}} v_{\text{ray}}} = 1.32545392 \quad .$$

The "complete" mixed Hessian, $\nabla_{\mathbf{x}} \nabla_{\mathbf{r}} v_{\text{ray}}$, has a zero eigenvalue, and the "incomplete" mixed Hessians, $\tilde{\nabla}_{\mathbf{x}} \nabla_{\mathbf{r}} v_{\text{ray}}$ and $\hat{\nabla}_{\mathbf{x}} \nabla_{\mathbf{r}} v_{\text{ray}}$, have two zero eigenvalues,

$$\lambda(\nabla_{\mathbf{x}} \nabla_{\mathbf{r}} v_{\text{ray}}) = \begin{bmatrix} -7.2334381 \cdot 10^{-2} & +1.7485044 \cdot 10^{-2} & 0 \end{bmatrix} \quad ,$$

$$\lambda(\tilde{\nabla}_{\mathbf{x}} \nabla_{\mathbf{r}} v_{\text{ray}}) = \begin{bmatrix} -3.7905367 \cdot 10^{-3} & 0 & 0 \end{bmatrix} \quad , \tag{21.30}$$

$$\lambda(\hat{\nabla}_{\mathbf{x}} \nabla_{\mathbf{r}} v_{\text{ray}}) = \begin{bmatrix} -3.3740943 \cdot 10^{-3} & 0 & 0 \end{bmatrix} \quad .$$

As also noted in section 15 for model 1, the obtained "errors" highly depend on the magnitudes of the spatial derivatives of the model parameters. Of course, at ray points within transition zones between layers, the gradients and Hessians of the model parameters are higher and the "errors" increase.

## 22. MODEL 2. ACOUSTIC APPROXIMATION FOR COMPRESSIONAL WAVES

Following the workflow and formulae described in Section 16, we compute and list the ray velocity derivatives using the acoustic approximation (AC) for qP waves, but now for the polar anisotropic parameters of model 2. Recall that the medium with the properties listed in Table 10 has a (huge) negative anellipticity; as mentioned, such a medium is formally unstable under the assumption of AC. We show that although the AC can still be applied with high accuracy for computing the ray velocity magnitude, the accuracy of the ray velocity derivatives considerably degrades, in particular the second derivatives.

Skipping the intermediate results, related to the derivatives of the reference and arclength-related Hamiltonians, we present only the required gradients and Hessians of the ray velocity magnitude.

- Spatial gradient and Hessian of the ray velocity magnitude,

$$\nabla_{\mathbf{x}} v_{\text{ray}} = \begin{bmatrix} -1.5321642 \cdot 10^{-1} & +1.4113915 \cdot 10^{-1} & +3.1430038 \cdot 10^{-2} \end{bmatrix} \text{ s}^{-1} \quad , \quad (22.1)$$

$$\nabla_{\mathbf{x}} \nabla_{\mathbf{x}} v_{\text{ray}} = \begin{bmatrix} -2.0769204 \cdot 10^{-3} & +9.5915004 \cdot 10^{-2} & -1.4219269 \cdot 10^{-1} \\ +9.5915004 \cdot 10^{-2} & +1.6093135 \cdot 10^{-1} & -1.9662911 \cdot 10^{-1} \\ -1.4219269 \cdot 10^{-1} & -1.9662911 \cdot 10^{-1} & +1.0273439 \cdot 10^{-1} \end{bmatrix} (\text{km} \cdot \text{s})^{-1} \quad . \quad (22.2)$$

- Directional gradient and Hessian of the ray velocity magnitude,

$$\nabla_{\mathbf{r}} v_{\text{ray}} = \begin{bmatrix} -4.3289968 \cdot 10^{-2} & -8.1978497 \cdot 10^{-2} & -9.5934719 \cdot 10^{-2} \end{bmatrix} \text{ km/s} \quad , \quad (22.3)$$

$$\nabla_\mathbf{r}\nabla_\mathbf{r} v_{\text{ray}} = \begin{bmatrix} +2.2753472 \cdot 10^{-1} & +4.2084925 \cdot 10^{-1} & +4.3213646 \cdot 10^{-1} \\ +4.2084925 \cdot 10^{-1} & +7.3472115 \cdot 10^{-1} & +7.6499309 \cdot 10^{-1} \\ +4.3213646 \cdot 10^{-1} & +7.6499309 \cdot 10^{-1} & +7.7548994 \cdot 10^{-1} \end{bmatrix} \text{ km/s} \quad . \tag{22.4}$$

- The mixed Hessian of the ray velocity magnitude,

$$\nabla_\mathbf{x}\nabla_\mathbf{r} v_{\text{ray}} = \begin{bmatrix} -1.8012127 \cdot 10^{-2} & -4.7063162 \cdot 10^{-2} & -4.9235650 \cdot 10^{-2} \\ +5.7478903 \cdot 10^{-2} & +1.3057670 \cdot 10^{-1} & +1.4301079 \cdot 10^{-1} \\ -7.2197431 \cdot 10^{-2} & -1.4443638 \cdot 10^{-1} & -1.6554724 \cdot 10^{-1} \end{bmatrix} \text{ s}^{-1} \quad . \tag{22.5}$$

- The gradient of the ray velocity wrt the model parameters, $\nabla_\mathbf{m} v_{\text{ray}}$, includes the medium part: the derivatives wrt the material components, $\mathbf{c} = \{v_P, \delta, \varepsilon\}$ (where the shear factor $f$ has been eliminated), and the geometric part: the derivatives wrt the symmetry axis direction angles. The material part is,

$$\nabla_\mathbf{c} v_{\text{ray}} = \begin{bmatrix} +8.3836026 \cdot 10^{-1} & +2.8279271 \cdot 10^{-3} & +3.5800398 \end{bmatrix} \quad , \tag{22.6}$$

and the geometric part is,

$$\nabla_\mathbf{a} v_{\text{ray}} = \begin{bmatrix} -1.2712417 \cdot 10^{-1} & +2.5915767 \cdot 10^{-2} \end{bmatrix} \quad . \tag{22.7}$$

- The Hessian of the ray velocity wrt the model parameters, $\nabla_\mathbf{m}\nabla_\mathbf{m} v_{\text{ray}}$, includes the material, the geometric, and the mixed blocks. The material block reads,

$$\nabla_\mathbf{c}\nabla_\mathbf{c} v_{\text{ray}} = \begin{bmatrix} 0 & +9.4264237 \cdot 10^{-4} & +1.1933466 \\ +9.4264237 \cdot 10^{-4} & -7.0562750 \cdot 10^{-3} & +2.0156696 \cdot 10^{-2} \\ +1.1933466 & +2.0156696 \cdot 10^{-2} & -5.2065677 \end{bmatrix} \quad , \tag{22.8}$$

the mixed block reads,

$$\nabla_\mathbf{a}\nabla_\mathbf{c} v_{\text{ray}} = \begin{bmatrix} -4.2374722 \cdot 10^{-2} & -7.0813587 \cdot 10^{-2} & +1.4236922 \cdot 10^{-1} \\ +8.638589 \cdot 10^{-3} & +1.4436188 \cdot 10^{-2} & -2.9023651 \cdot 10^{-2} \end{bmatrix} \quad , \tag{22.9}$$

and the geometric block reads,

$$\nabla_\mathbf{a}\nabla_\mathbf{a} v_{\text{ray}} = \begin{bmatrix} +1.5862193 & -2.9437760 \cdot 10^{-1} \\ -2.9437760 \cdot 10^{-1} & +1.2468053 \cdot 10^{-1} \end{bmatrix} \quad . \tag{22.10}$$

Validation

As in Section 16, the relative error for all gradient and Hessian components has been computed twice: comparing the numerical solution $N$ for the AC with its analytical solution $A$ (error $E$), and comparing the exact analytical solution for compressional wave $P$ with the analytical solution for the AC (error $\tilde{E}$),

$$E = \frac{N - A}{A} \quad , \quad \tilde{E} = \frac{A - P}{P} \quad . \tag{22.11}$$

- Accuracy of the spatial gradient and Hessian of the ray velocity magnitude,

$$E(\nabla_\mathbf{x} v_{\text{ray}}) = \begin{bmatrix} -6.30 \cdot 10^{-11} & +2.49 \cdot 10^{-10} & +2.04 \cdot 10^{-9} \end{bmatrix} \quad , \tag{22.12}$$

$$\tilde{E}(\nabla_\mathbf{x} v_{\text{ray}}) = \begin{bmatrix} +1.10 \cdot 10^{-3} & +3.52 \cdot 10^{-3} & -1.85 \cdot 10^{-2} \end{bmatrix} \quad , \tag{22.13}$$

$$E(\nabla_\mathbf{x}\nabla_\mathbf{x} v_{\text{ray}}) = \begin{bmatrix} -4.47 \cdot 10^{-5} & +4.03 \cdot 10^{-7} & +1.12 \cdot 10^{-7} \\ +4.03 \cdot 10^{-7} & +6.10 \cdot 10^{-7} & -1.43 \cdot 10^{-7} \\ +1.12 \cdot 10^{-7} & -1.43 \cdot 10^{-7} & +5.52 \cdot 10^{-7} \end{bmatrix} \quad , \tag{22.14}$$

$$\tilde{E}(\nabla_\mathbf{x}\nabla_\mathbf{x} v_{\text{ray}}) = \begin{bmatrix} +8.84 \cdot 10^{-2} & -2.28 \cdot 10^{-3} & -9.59 \cdot 10^{-5} \\ -2.28 \cdot 10^{-3} & -1.58 \cdot 10^{-3} & -2.87 \cdot 10^{-3} \\ -9.59 \cdot 10^{-5} & -2.87 \cdot 10^{-3} & -9.93 \cdot 10^{-3} \end{bmatrix} \quad . \tag{22.15}$$

- Accuracy of the directional gradient and Hessian of the ray velocity magnitude,

$$E(\nabla_\mathbf{r} v_{\text{ray}}) = \begin{bmatrix} +9.05 \cdot 10^{-10} & +1.64 \cdot 10^{-10} & +4.49 \cdot 10^{-10} \end{bmatrix} \quad , \tag{22.16}$$

$$\tilde{E}(\nabla_{\mathbf{r}} v_{\text{ray}}) = \begin{bmatrix} -1.07 \cdot 10^{-2} & +2.21 \cdot 10^{-1} & +1.14 \cdot 10^{-1} \end{bmatrix} \quad , \qquad (22.17)$$

$$E(\nabla_{\mathbf{r}} \nabla_{\mathbf{r}} v_{\text{ray}}) = \begin{bmatrix} -2.70 \cdot 10^{-5} & +1.54 \cdot 10^{-5} & -1.36 \cdot 10^{-6} \\ +1.54 \cdot 10^{-5} & -8.26 \cdot 10^{-6} & +1.52 \cdot 10^{-6} \\ -1.36 \cdot 10^{-6} & +1.52 \cdot 10^{-6} & +4.29 \cdot 10^{-7} \end{bmatrix} \quad , \qquad (22.18)$$

$$\tilde{E}(\nabla_{\mathbf{r}} \nabla_{\mathbf{r}} v_{\text{ray}}) = \begin{bmatrix} -3.6573 \cdot 10^{-2} & -3.6575 \cdot 10^{-2} & -3.6585 \cdot 10^{-2} \\ -3.6575 \cdot 10^{-2} & -3.6581 \cdot 10^{-2} & -3.6590 \cdot 10^{-2} \\ -3.6585 \cdot 10^{-2} & -3.6590 \cdot 10^{-2} & -3.6604 \cdot 10^{-2} \end{bmatrix} \quad . \qquad (22.19)$$

- Accuracy of the mixed Hessian of the ray velocity magnitude,

$$E(\nabla_{\mathbf{x}} \nabla_{\mathbf{r}} v_{\text{ray}}) = \begin{bmatrix} +6.44 \cdot 10^{-6} & +2.27 \cdot 10^{-6} & +3.57 \cdot 10^{-6} \\ -2.20 \cdot 10^{-6} & -1.34 \cdot 10^{-6} & -1.64 \cdot 10^{-6} \\ +1.51 \cdot 10^{-7} & -2.01 \cdot 10^{-6} & -1.21 \cdot 10^{-6} \end{bmatrix} \quad , \qquad (22.20)$$

$$\tilde{E}(\nabla_{\mathbf{x}} \nabla_{\mathbf{r}} v_{\text{ray}}) = \begin{bmatrix} -1.64 \cdot 10^{-2} & -2.20 \cdot 10^{-2} & -2.03 \cdot 10^{-2} \\ -2.40 \cdot 10^{-2} & -2.61 \cdot 10^{-2} & -2.53 \cdot 10^{-2} \\ -2.94 \cdot 10^{-2} & -2.98 \cdot 10^{-2} & -2.97 \cdot 10^{-2} \end{bmatrix} \quad . \qquad (22.21)$$

The matrices in equations 22.19 and 22.21 look like systematic discrepancies, rather than random ones.

- Accuracy of the material part of the ray velocity gradient wrt the model parameters,

$$E(\nabla_{\mathbf{c}} v_{\text{ray}}) = \begin{bmatrix} +7.57 \cdot 10^{-13} & +7.28 \cdot 10^{-10} & +2.26 \cdot 10^{-10} \end{bmatrix} \quad , \qquad (22.22)$$

$$\tilde{E}(\nabla_{\mathbf{c}} v_{\text{ray}}) = \begin{bmatrix} -7.68 \cdot 10^{-5} & +2.47 \cdot 10^{-2} & +2.76 \cdot 10^{-4} \end{bmatrix} \quad . \qquad (22.23)$$

- Accuracy of the geometric part of the ray velocity gradient wrt the model parameters,

$$E(\nabla_{\mathbf{c}} v_{\text{ray}}) = \begin{bmatrix} +7.16 \cdot 10^{-10} & +3.45 \cdot 10^{-8} \end{bmatrix} \quad , \qquad (22.24)$$

$$\tilde{E}(\nabla_{\mathbf{c}} v_{\text{ray}}) = \begin{bmatrix} -3.65 \cdot 10^{-2} & -3.65 \cdot 10^{-2} \end{bmatrix} \quad . \qquad (22.25)$$

We re-emphasize that the relative discrepancies between the ray velocity derivatives for the acoustic approximation and elastic models wrt the zenith and azimuth angles of the medium symmetry axis are identical.

Accuracy of the material part of the ray velocity Hessian wrt the model parameters,

$$E(\nabla_\mathbf{c}\nabla_\mathbf{c} v_{\text{ray}}) = \begin{bmatrix} - & +1.17 \cdot 10^{-5} & +9.85 \cdot 10^{-9} \\ +1.17 \cdot 10^{-5} & +4.29 \cdot 10^{-5} & -2.35 \cdot 10^{-5} \\ +9.85 \cdot 10^{-9} & -2.35 \cdot 10^{-5} & -5.00 \cdot 10^{-7} \end{bmatrix} \quad , \qquad (22.26)$$

$$\tilde{E}(\nabla_\mathbf{c}\nabla_\mathbf{c} v_{\text{ray}}) = \begin{bmatrix} - & +2.47 \cdot 10^{-2} & +2.76 \cdot 10^{-4} \\ +2.47 \cdot 10^{-2} & -3.94 \cdot 10^{-2} & -6.62 \cdot 10^{-2} \\ +2.76 \cdot 10^{-4} & -6.62 \cdot 10^{-2} & -1.33 \cdot 10^{-3} \end{bmatrix} \quad . \qquad (22.27)$$

- Accuracy of the mixed part of the ray velocity Hessian wrt the model parameters,

$$E(\nabla_\mathbf{a}\nabla_\mathbf{c} v_{\text{ray}}) = \begin{bmatrix} +6.92 \cdot 10^{-8} & +1.23 \cdot 10^{-6} & -1.49 \cdot 10^{-6} \\ +5.01 \cdot 10^{-7} & -6.31 \cdot 10^{-6} & +3.61 \cdot 10^{-6} \end{bmatrix} \quad , \qquad (22.28)$$

$$\tilde{E}(\nabla_\mathbf{a}\nabla_\mathbf{c} v_{\text{ray}}) = \begin{bmatrix} -3.65 \cdot 10^{-2} & +2.40 \cdot 10^{-2} & -1.49 \cdot 10^{-1} \\ -3.65 \cdot 10^{-2} & +2.40 \cdot 10^{-2} & -1.49 \cdot 10^{-1} \end{bmatrix} \quad . \qquad (22.29)$$

The first row of the mixed Hessian matrix in equation 22.9 and its acoustic approximation error in equation 22.29 is related to the zenith angle $\theta_{\text{ax}}$, while the second row – to the azimuth angle $\psi_{\text{ax}}$. The columns are related to the material properties $\mathbf{c} = \{v_P, \delta, \varepsilon\}$. As we see, the two lines are identical.

- Accuracy of the geometric part of the ray velocity Hessian wrt the model parameters,

$$E(\nabla_\mathbf{a}\nabla_\mathbf{a} v_{\text{ray}}) = \begin{bmatrix} +7.44 \cdot 10^{-9} & -6.35 \cdot 10^{-8} \\ -6.35 \cdot 10^{-8} & -2.00 \cdot 10^{-7} \end{bmatrix} \quad , \qquad (22.30)$$

$$\tilde{E}\left(\nabla_{\mathbf{a}}\nabla_{\mathbf{a}} v_{\text{ray}}\right) = \begin{bmatrix} -3.6590 \cdot 10^{-2} & -3.6598 \cdot 10^{-2} \\ -3.6598 \cdot 10^{-2} & -3.6549 \cdot 10^{-2} \end{bmatrix} \quad . \tag{22.31}$$

The discrepancies in the last equation seem systematic, rather than random.

As we see, the numerical errors $|E|$ of the finite difference approximations are small for the gradients and acceptable for the Hessians, which indicates that the analytical formulae for the AC derivatives of the ray velocity are correct. However, the discrepancies $|\tilde{E}|$ of the gradients and, in particular, the Hessians computed with the elastic and acoustic models are large, and for some components unacceptable. The reason is inherent to the nature of the differential operator. The AC for the ray velocity in itself is accurate; still, it is an approximation. The accuracy of the derivatives of an approximated function is much worse than the accuracy of the function itself, and the accuracy of the second derivatives is even worse. While the accuracy of the AC slowness inversion along with the computed phase and ray velocities is very high, using the AC for computing the qP ray velocity derivatives is questionable. The reduced computational complexity of the AC is accompanied by a declined accuracy of its results when the derivatives of the approximated values are involved.

### 23. MODEL 2. COMPUTING RAY VELOCITY DERIVATIVES FOR qSV$_1$ WAVES

For the given model properties $\mathbf{m}$ and the ray direction $\mathbf{r}$, the shear waves have been triplicated; qSV$_1$ is the fast shear wave with the highest ray velocity, $v_{\text{ray}} = 1.5052881 \, \text{km/s}$ (Table 12). In this section, we list the derivatives of the ray velocity magnitude for this wave type and validate them numerically with the finite-difference approximations.

- The spatial gradient and Hessian of the ray velocity magnitude are,

$$\nabla_{\mathbf{x}} v_{\text{ray}} = \begin{bmatrix} -2.2218753 \cdot 10^{-1} & +3.2246069 \cdot 10^{-1} & -1.4169736 \cdot 10^{-1} \end{bmatrix} \text{ s}^{-1} \quad , \quad (23.1)$$

$$\nabla_{\mathbf{x}} \nabla_{\mathbf{x}} v_{\text{ray}} = \begin{bmatrix} -5.4139253 \cdot 10^{-2} & +2.6664177 \cdot 10^{-2} & -2.5063221 \cdot 10^{-1} \\ +2.6664177 \cdot 10^{-2} & +1.2459680 \cdot 10^{-1} & -1.9422356 \cdot 10^{-1} \\ -2.5063221 \cdot 10^{-1} & -1.9422356 \cdot 10^{-1} & +1.8219036 \cdot 10^{-1} \end{bmatrix} (\text{km} \cdot \text{s})^{-1} \quad . \quad (23.2)$$

- The directional gradient and Hessian of the ray velocity magnitude are,

$$\nabla_{\mathbf{r}} v_{\text{ray}} = \begin{bmatrix} -4.5011424 \cdot 10^{-2} & -8.5238429 \cdot 10^{-2} & -9.9749630 \cdot 10^{-2} \end{bmatrix} \text{ km/s} \quad , \quad (23.3)$$

$$\nabla_{\mathbf{r}} \nabla_{\mathbf{r}} v_{\text{ray}} = \begin{bmatrix} +2.3882582 \cdot 10^{-1} & +4.4183224 \cdot 10^{-1} & +4.5429142 \cdot 10^{-1} \\ +4.4183224 \cdot 10^{-1} & +7.7198161 \cdot 10^{-1} & +8.0482672 \cdot 10^{-1} \\ +4.5429142 \cdot 10^{-1} & +8.0482672 \cdot 10^{-1} & +8.1734351 \cdot 10^{-1} \end{bmatrix} \text{ km/s} \quad . \quad (23.4)$$

- The mixed Hessian of the ray velocity magnitude is,

$$\nabla_{\mathbf{x}} \nabla_{\mathbf{r}} v_{\text{ray}} = \begin{bmatrix} -1.3399489 \cdot 10^{-2} & -3.8843283 \cdot 10^{-2} & -3.9384179 \cdot 10^{-2} \\ +4.7186527 \cdot 10^{-2} & +1.1195001 \cdot 10^{-1} & +1.2082352 \cdot 10^{-1} \\ -6.5600793 \cdot 10^{-2} & -1.3225110 \cdot 10^{-1} & -1.5114919 \cdot 10^{-1} \end{bmatrix} \text{ s}^{-1} \quad . \quad (23.5)$$

- The ray velocity gradients wrt the material properties, $\mathbf{c} = \{v_P, f, \delta, \varepsilon\}$, and wrt the geometric properties, $\mathbf{a} = \{\theta_{\text{ax}}, \psi_{\text{ax}}\}$, are,

$$\nabla_{\mathbf{c}} v_{\text{ray}} = \begin{bmatrix} +5.0176269 \cdot 10^{-1} & -3.0139010 & -2.3708889 \cdot 10^{-3} & +7.107072 \cdot 10^{-3} \end{bmatrix} \quad , \quad (23.6)$$

where the first component is unitless, and the others have the units of velocity, and

$$\nabla_{\mathbf{a}} v_{\text{ray}} = \begin{bmatrix} -1.3217935 \cdot 10^{-1} & +2.6946326 \cdot 10^{-2} \end{bmatrix} \frac{\text{km}}{\text{s} \cdot \text{radian}} \quad . \quad (23.7)$$

- The ray velocity Hessians wrt the material, mixed, and geometric properties of the model, are,

$$\nabla_{\mathbf{c}}\nabla_{\mathbf{c}}v_{\text{ray}} = \begin{bmatrix} +6.6613381\cdot10^{-16} & -1.0046337 & -7.9029631\cdot10^{-4} & +2.3690240\cdot10^{-3} \\ -1.0046337 & -6.0136097 & +1.5177692\cdot10^{-2} & -5.5966855\cdot10^{-2} \\ -7.9029631\cdot10^{-4} & +1.5177692\cdot10^{-2} & +1.2697265\cdot10^{-2} & -2.7515334\cdot10^{-2} \\ +2.3690240\cdot10^{-3} & -5.5966855\cdot10^{-2} & -2.7515334\cdot10^{-2} & +5.0866535\cdot10^{-2} \end{bmatrix},$$

(23.8)

$$\nabla_{\mathbf{a}}\nabla_{\mathbf{c}}v_{\text{ray}} = \begin{bmatrix} -4.4059783\cdot10^{-2} & +3.4833342\cdot10^{-1} & +5.9780979\cdot10^{-2} & -1.7906146\cdot10^{-1} \\ +8.9821085\cdot10^{-3} & -7.1011892\cdot10^{-2} & -1.2187060\cdot10^{-2} & +3.6503798\cdot10^{-2} \end{bmatrix},$$

(23.9)

$$\nabla_{\mathbf{a}}\nabla_{\mathbf{a}}v_{\text{ray}} = \begin{bmatrix} +1.668639 & -3.1002695\cdot10^{-1} \\ -3.1002695\cdot10^{-1} & +1.3044241\cdot10^{-1} \end{bmatrix}.$$

(23.10)

Validation

The relative errors of the finite-difference approximations are listed below.

- The relative errors of the spatial gradient and Hessian of the ray velocity magnitude are,

$$E(\nabla_{\mathbf{x}}v_{\text{ray}}) = \begin{bmatrix} -1.37\cdot10^{-11} & +2.99\cdot10^{-10} & -8.73\cdot10^{-10} \end{bmatrix},$$

(23.11)

$$E(\nabla_{\mathbf{x}}\nabla_{\mathbf{x}}v_{\text{ray}}) = \begin{bmatrix} +2.38\cdot10^{-6} & -1.08\cdot10^{-6} & +1.26\cdot10^{-7} \\ -1.08\cdot10^{-6} & -1.49\cdot10^{-7} & -1.01\cdot10^{-7} \\ +1.26\cdot10^{-7} & -1.01\cdot10^{-7} & -1.66\cdot10^{-7} \end{bmatrix}.$$

(23.12)

- The relative errors of the directional gradient and Hessian of the ray velocity magnitude are,

$$E(\nabla_{\mathbf{r}}v_{\text{ray}}) = \begin{bmatrix} +1.28\cdot10^{-9} & -1.56\cdot10^{-10} & +3.98\cdot10^{-10} \end{bmatrix},$$

(23.13)

$$E\left(\nabla_\mathbf{r}\nabla_\mathbf{r} v_\text{ray}\right) = \begin{bmatrix} -3.86\cdot 10^{-5} & +4.93\cdot 10^{-6} & -1.39\cdot 10^{-5} \\ +4.93\cdot 10^{-6} & -1.37\cdot 10^{-5} & -7.14\cdot 10^{-6} \\ -1.39\cdot 10^{-5} & -7.14\cdot 10^{-6} & -1.16\cdot 10^{-5} \end{bmatrix} . \quad (23.14)$$

- The relative errors of the mixed Hessian of the ray velocity magnitude are,

$$E\left(\nabla_\mathbf{x}\nabla_\mathbf{r} v_\text{ray}\right) = \begin{bmatrix} +5.51\cdot 10^{-6} & -2.88\cdot 10^{-6} & -4.43\cdot 10^{-7} \\ +1.89\cdot 10^{-7} & +1.40\cdot 10^{-6} & +9.96\cdot 10^{-7} \\ +1.15\cdot 10^{-6} & -1.33\cdot 10^{-6} & -4.11\cdot 10^{-7} \end{bmatrix} . \quad (23.15)$$

- The relative errors of the ray velocity gradients wrt the material and geometric properties of the model are,

$$E\left(\nabla_\mathbf{c} v_\text{ray}\right) = \begin{bmatrix} +3.15\cdot 10^{-13} & +1.12\cdot 10^{-8} & +8.79\cdot 10^{-9} & -7.12\cdot 10^{-10} \end{bmatrix} , \quad (23.16)$$

$$E\left(\nabla_\mathbf{a} v_\text{ray}\right) = \begin{bmatrix} +5.09\cdot 10^{-9} & +3.53\cdot 10^{-8} \end{bmatrix} . \quad (23.17)$$

- The relative errors of the ray velocity Hessians wrt the material, mixed, and geometric properties of the model are,

$$E\left(\nabla_\mathbf{c}\nabla_\mathbf{c} v_\text{ray}\right) = \begin{bmatrix} - & +7.61\cdot 10^{-9} & +4.57\cdot 10^{-6} & -2.30\cdot 10^{-6} \\ +7.61\cdot 10^{-9} & +4.94\cdot 10^{-8} & +2.88\cdot 10^{-6} & -2.09\cdot 10^{-6} \\ +4.57\cdot 10^{-6} & +2.88\cdot 10^{-6} & -7.89\cdot 10^{-5} & -6.68\cdot 10^{-6} \\ -2.30\cdot 10^{-6} & -2.09\cdot 10^{-6} & -6.68\cdot 10^{-6} & -3.03\cdot 10^{-5} \end{bmatrix} , \quad (23.18)$$

$$E\left(\nabla_\mathbf{a}\nabla_\mathbf{c} v_\text{ray}\right) = \begin{bmatrix} +4.11\cdot 10^{-8} & -9.69\cdot 10^{-9} & -7.69\cdot 10^{-7} & -8.14\cdot 10^{-7} \\ -1.04\cdot 10^{-7} & -3.88\cdot 10^{-7} & +7.22\cdot 10^{-7} & +2.59\cdot 10^{-6} \end{bmatrix} , \quad (23.19)$$

$$E\left(\nabla_\mathbf{a}\nabla_\mathbf{a} v_\text{ray}\right) = \begin{bmatrix} -8.92\cdot 10^{-8} & -5.33\cdot 10^{-8} \\ -5.33\cdot 10^{-8} & -7.19\cdot 10^{-7} \end{bmatrix} . \quad (23.20)$$

## 24. MODEL 2. COMPUTING RAY VELOCITY DERIVATIVES FOR qSV$_2$ WAVES

As in the previous section, here we list the derivatives of the ray velocity magnitude for the wave type $\text{qSV}_2$, with the intermediate ray velocity $v_{\text{ray}} = 0.84719014\,\text{km/s}$ (Table 12) and validate them numerically with the finite-difference approximations.

- The spatial gradient and Hessian of the ray velocity magnitude are,

$$\nabla_{\mathbf{x}} v_{\text{ray}} = \begin{bmatrix} -4.7331235 \cdot 10^{-1} & +1.0222045 & -7.0749072 \cdot 10^{-1} \end{bmatrix}\ \text{s}^{-1}\ , \quad (24.1)$$

$$\nabla_{\mathbf{x}} \nabla_{\mathbf{x}} v_{\text{ray}} = \begin{bmatrix} -3.4929669 \cdot 10^{-1} & +7.2280948 \cdot 10^{-1} & -1.1479050 \\ +7.2280948 \cdot 10^{-1} & -1.5256972 & +9.3881815 \cdot 10^{-1} \\ -1.1479050 & +9.3881815 \cdot 10^{-1} & -4.7066773 \cdot 10^{-1} \end{bmatrix} (\text{km} \cdot \text{s})^{-1}\ . \quad (24.2)$$

- The directional gradient and Hessian of the ray velocity magnitude are,

$$\nabla_{\mathbf{r}} v_{\text{ray}} = \begin{bmatrix} -2.5898950 \cdot 10^{-1} & -4.9045012 \cdot 10^{-1} & -5.7394556 \cdot 10^{-1} \end{bmatrix}\ \text{km/s}\ , \quad (24.3)$$

$$\nabla_{\mathbf{r}} \nabla_{\mathbf{r}} v_{\text{ray}} = \begin{bmatrix} +4.8775432 \cdot 10^{-1} & +8.6362904 \cdot 10^{-1} & +6.4954631 \cdot 10^{-1} \\ +8.6362904 \cdot 10^{-1} & +1.2630759 & +9.1089538 \cdot 10^{-1} \\ +6.4954631 \cdot 10^{-1} & +9.1089538 \cdot 10^{-1} & +3.4961901 \cdot 10^{-1} \end{bmatrix}\ \text{km/s}\ . \quad (24.4)$$

- The mixed Hessian of the ray velocity magnitude is,

$$\nabla_{\mathbf{x}} \nabla_{\mathbf{r}} v_{\text{ray}} = \begin{bmatrix} +1.1907525 \cdot 10^{-1} & +1.4799723 \cdot 10^{-1} & +2.0812939 \cdot 10^{-2} \\ -2.2605512 \cdot 10^{-1} & -2.9808753 \cdot 10^{-1} & -4.0743857 \cdot 10^{-1} \\ +1.1198717 \cdot 10^{-1} & +1.6591081 \cdot 10^{-1} & +2.1496565 \cdot 10^{-1} \end{bmatrix}\ \text{s}^{-1}\ . \quad (24.5)$$

- The ray velocity gradients wrt the material and geometric properties are,

$$\nabla_{\mathbf{c}} v_{\text{ray}} = \begin{bmatrix} +2.8239671 \cdot 10^{-1} & -1.0099778 \cdot 10^{+1} & -3.6513276 & +6.2152554 \end{bmatrix}\ , \quad (24.6)$$

where the first component is unitless, and the others have the units of velocity, and

$$\nabla_{\mathbf{a}} v_{\text{ray}} = \begin{bmatrix} -7.6054167 \cdot 10^{-1} & +1.5504543 \cdot 10^{-1} \end{bmatrix}\ \frac{\text{km}}{\text{s} \cdot \text{radian}}\ . \quad (24.7)$$

- The ray velocity Hessians wrt the material, mixed, and geometric properties of the model, are,

$$\nabla_{\mathbf{c}}\nabla_{\mathbf{c}}v_{\text{ray}} = \begin{bmatrix} -1.1102230 \cdot 10^{-15} & -3.3665926 & -1.2171092 & +2.0717518 \\ -3.3665926 & -1.5011095 \cdot 10^{+2} & -5.7055458 \cdot 10^{+1} & +8.5804428 \cdot 10^{+1} \\ -1.2171092 & -5.7055458 \cdot 10^{+1} & -1.8203190 \cdot 10^{+1} & +3.2156455 \cdot 10^{+1} \\ +2.0717518 & +8.5804428 \cdot 10^{+1} & +3.2156455 \cdot 10^{+1} & -5.8866481 \cdot 10^{+1} \end{bmatrix},$$

(24.8)

$$\nabla_{\mathbf{a}}\nabla_{\mathbf{c}}v_{\text{ray}} = \begin{bmatrix} -2.5351389 \cdot 10^{-1} & +6.8007056 & +2.3349644 & -4.5807429 \\ +5.1681809 \cdot 10^{-2} & -1.3864044 & -4.7601015 \cdot 10^{-1} & +9.3383868 \cdot 10^{-1} \end{bmatrix}, \quad (24.9)$$

$$\nabla_{\mathbf{a}}\nabla_{\mathbf{a}}v_{\text{ray}} = \begin{bmatrix} +1.9571393 & -2.2553655 \cdot 10^{-1} \\ -2.2553655 \cdot 10^{-1} & +4.3286648 \cdot 10^{-1} \end{bmatrix}. \quad (24.10)$$

Validation

The relative errors of the finite-difference approximations are listed below.

- The relative errors of the spatial gradient and Hessian of the ray velocity magnitude are,

$$E(\nabla_{\mathbf{x}}v_{\text{ray}}) = \begin{bmatrix} +2.43 \cdot 10^{-9} & +1.03 \cdot 10^{-8} & +2.87 \cdot 10^{-9} \end{bmatrix}, \quad (24.11)$$

$$E(\nabla_{\mathbf{x}}\nabla_{\mathbf{x}}v_{\text{ray}}) = \begin{bmatrix} +3.54 \cdot 10^{-7} & -4.74 \cdot 10^{-8} & +3.47 \cdot 10^{-8} \\ -4.74 \cdot 10^{-8} & +3.38 \cdot 10^{-8} & +3.01 \cdot 10^{-8} \\ +3.47 \cdot 10^{-8} & +3.01 \cdot 10^{-8} & +2.73 \cdot 10^{-7} \end{bmatrix}. \quad (24.12)$$

- The relative errors of the directional gradient and Hessian of the ray velocity magnitude are,

$$E(\nabla_{\mathbf{r}}v_{\text{ray}}) = \begin{bmatrix} +4.08 \cdot 10^{-10} & +9.54 \cdot 10^{-11} & +2.16 \cdot 10^{-11} \end{bmatrix}, \quad (24.13)$$

$$E(\nabla_{\mathbf{r}}\nabla_{\mathbf{r}}v_{\text{ray}}) = \begin{bmatrix} -6.94\cdot10^{-6} & +5.86\cdot10^{-7} & -3.89\cdot10^{-6} \\ +5.86\cdot10^{-7} & -4.10\cdot10^{-6} & -3.61\cdot10^{-6} \\ -3.89\cdot10^{-6} & -3.61\cdot10^{-6} & -1.29\cdot10^{-5} \end{bmatrix} . \tag{24.14}$$

- The relative errors of the mixed Hessian of the ray velocity magnitude are,

$$E(\nabla_{\mathbf{x}}\nabla_{\mathbf{r}}v_{\text{ray}}) = \begin{bmatrix} +4.73\cdot10^{-8} & +1.62\cdot10^{-6} & +8.51\cdot10^{-7} \\ +5.47\cdot10^{-7} & -1.43\cdot10^{-7} & +1.84\cdot10^{-7} \\ -2.67\cdot10^{-7} & -1.53\cdot10^{-6} & -9.68\cdot10^{-7} \end{bmatrix} . \tag{24.15}$$

- The relative errors of the ray velocity gradients wrt the material and geometric properties of the model are,

$$E(\nabla_{\mathbf{c}}v_{\text{ray}}) = \begin{bmatrix} +4.13\cdot10^{-13} & +5.20\cdot10^{-7} & +9.31\cdot10^{-9} & +7.56\cdot10^{-9} \end{bmatrix} , \tag{24.16}$$

$$E(\nabla_{\mathbf{a}}v_{\text{ray}}) = \begin{bmatrix} +5.97\cdot10^{-9} & +6.31\cdot10^{-9} \end{bmatrix} . \tag{24.17}$$

- The relative errors of the ray velocity Hessians wrt the material, mixed, and geometric properties of the model are,

$$E(\nabla_{\mathbf{c}}\nabla_{\mathbf{c}}v_{\text{ray}}) = \begin{bmatrix} - & +5.21\cdot10^{-7} & +7.43\cdot10^{-9} & +1.04\cdot10^{-8} \\ +5.21\cdot10^{-7} & +1.09\cdot10^{-6} & +2.16\cdot10^{-6} & +2.24\cdot10^{-6} \\ +7.43\cdot10^{-9} & +2.16\cdot10^{-6} & +4.11\cdot10^{-8} & +8.97\cdot10^{-8} \\ +1.04\cdot10^{-8} & +2.24\cdot10^{-6} & +8.97\cdot10^{-8} & +1.02\cdot10^{-7} \end{bmatrix} , \tag{24.18}$$

$$E(\nabla_{\mathbf{a}}\nabla_{\mathbf{c}}v_{\text{ray}}) = \begin{bmatrix} +6.00\cdot10^{-9} & +8.99\cdot10^{-7} & +1.86\cdot10^{-8} & +2.66\cdot10^{-8} \\ -1.10\cdot10^{-8} & +8.73\cdot10^{-7} & +7.64\cdot10^{-9} & -1.55\cdot10^{-7} \end{bmatrix} , \tag{24.19}$$

$$E(\nabla_{\mathbf{a}}\nabla_{\mathbf{a}}v_{\text{ray}}) = \begin{bmatrix} -9.84\cdot10^{-8} & -2.61\cdot10^{-8} \\ -2.61\cdot10^{-8} & -1.96\cdot10^{-7} \end{bmatrix} . \tag{24.20}$$

## 25. MODEL 2. COMPUTING RAY VELOCITY DERIVATIVES FOR qSV$_3$ WAVES

As in the two previous sections, here we list the derivatives of the ray velocity magnitude for the wave type qSV$_3$ (slow shear), with the lowest ray velocity $v_{\text{ray}} = 0.74693784$ km/s (Table 12) and validate them numerically with the finite-difference approximations.

- The spatial gradient and Hessian of the ray velocity magnitude are,

$$\nabla_{\mathbf{x}} v_{\text{ray}} = \begin{bmatrix} -4.8313847 \cdot 10^{-1} & +1.0710803 & -8.0219936 \cdot 10^{-1} \end{bmatrix} \text{ s}^{-1} \quad , \qquad (25.1)$$

$$\nabla_{\mathbf{x}} \nabla_{\mathbf{x}} v_{\text{ray}} = \begin{bmatrix} -2.9134706 \cdot 10^{-1} & +4.0865666 \cdot 10^{-1} & -8.6968745 \cdot 10^{-1} \\ +4.0865666 \cdot 10^{-1} & -9.5623840 \cdot 10^{-1} & +5.0625030 \cdot 10^{-1} \\ -8.6968745 \cdot 10^{-1} & +5.0625030 \cdot 10^{-1} & -1.9745030 \cdot 10^{-1} \end{bmatrix} (\text{km} \cdot \text{s})^{-1} \quad . \qquad (25.2)$$

- The directional gradient and Hessian of the ray velocity magnitude are,

$$\nabla_{\mathbf{r}} v_{\text{ray}} = \begin{bmatrix} +1.7240870 \cdot 10^{-1} & +3.2649148 \cdot 10^{-1} & +3.8207419 \cdot 10^{-1} \end{bmatrix} \text{ km/s} \quad , \qquad (25.3)$$

$$\nabla_{\mathbf{r}} \nabla_{\mathbf{r}} v_{\text{ray}} = \begin{bmatrix} -2.5584494 \cdot 10^{-2} & -8.4846890 \cdot 10^{-3} & +2.3046031 \cdot 10^{-1} \\ -8.4846890 \cdot 10^{-3} & +2.3182771 \cdot 10^{-1} & +6.4888474 \cdot 10^{-1} \\ +2.3046031 \cdot 10^{-1} & +6.4888474 \cdot 10^{-1} & +1.2362246 \end{bmatrix} \text{ km/s} \quad . \qquad (25.4)$$

- The mixed Hessian of the ray velocity magnitude is,

$$\nabla_{\mathbf{x}} \nabla_{\mathbf{r}} v_{\text{ray}} = \begin{bmatrix} -1.2775429 \cdot 10^{-1} & -1.9033999 \cdot 10^{-1} & -2.4600125 \cdot 10^{-1} \\ +2.7575306 \cdot 10^{-1} & +4.3565832 \cdot 10^{-1} & +5.4883833 \cdot 10^{-1} \\ -2.0661035 \cdot 10^{-1} & -3.6053075 \cdot 10^{-1} & -4.3576141 \cdot 10^{-1} \end{bmatrix} \text{ s}^{-1} \quad . \qquad (25.5)$$

- The ray velocity gradients wrt the material and geometric properties are,

$$\nabla_{\mathbf{c}} v_{\text{ray}} = \begin{bmatrix} +2.4897928 \cdot 10^{-1} & -9.1541263 & -3.3229890 & +5.5803112 \end{bmatrix} \quad , \qquad (25.6)$$

where the first component is unitless, and the others have the units of velocity, and

$$\nabla_{\mathbf{a}} v_{\text{ray}} = \begin{bmatrix} +5.0629078 \cdot 10^{-1} & -1.0321337 \cdot 10^{-1} \end{bmatrix} \frac{\text{km}}{\text{s} \cdot \text{radian}} \quad . \qquad (25.7)$$

- The ray velocity Hessians wrt the material, mixed, and geometric properties of the model, are,

$$\nabla_\mathbf{c}\nabla_\mathbf{c} v_{\text{ray}} = \begin{bmatrix} +2.2204460 \cdot 10^{-15} & -3.0513754 & -1.1076630 & +1.8601037 \\ -3.0513754 & -1.2749144 \cdot 10^{+2} & -4.8166500 \cdot 10^{+1} & +7.2655075 \cdot 10^{+1} \\ -1.1076630 & -4.8166500 \cdot 10^{+1} & -1.4995226 \cdot 10^{+1} & +2.6931307 \cdot 10^{+1} \\ +1.8601037 & +7.2655075 \cdot 10^{+1} & +2.6931307 \cdot 10^{+1} & -4.9992637 \cdot 10^{+1} \end{bmatrix},$$

(25.8)

$$\nabla_\mathbf{a}\nabla_\mathbf{c} v_{\text{ray}} = \begin{bmatrix} +1.6876359 \cdot 10^{-1} & -5.0132788 & -1.7574243 & +3.3609318 \\ -3.4404457 \cdot 10^{-2} & +1.0220163 & +3.5827176 \cdot 10^{-1} & -6.8516574 \cdot 10^{-1} \end{bmatrix},$$

(25.9)

$$\nabla_\mathbf{a}\nabla_\mathbf{a} v_{\text{ray}} = \begin{bmatrix} +1.2765217 & -3.7569897 \cdot 10^{-1} \\ -3.7569897 \cdot 10^{-1} & -1.8095986 \cdot 10^{-1} \end{bmatrix}.$$

(25.10)

Validation

The relative errors of the finite-difference approximations are listed below.

- The relative errors of the spatial gradient and Hessian of the ray velocity magnitude are,

$$E(\nabla_\mathbf{x} v_{\text{ray}}) = \begin{bmatrix} +1.57 \cdot 10^{-9} & +7.11 \cdot 10^{-9} & +1.03 \cdot 10^{-9} \end{bmatrix},$$

(25.11)

$$E(\nabla_\mathbf{x}\nabla_\mathbf{x} v_{\text{ray}}) = \begin{bmatrix} +5.88 \cdot 10^{-7} & -1.77 \cdot 10^{-7} & +3.61 \cdot 10^{-8} \\ -1.77 \cdot 10^{-7} & -6.72 \cdot 10^{-8} & +6.42 \cdot 10^{-8} \\ +3.61 \cdot 10^{-8} & +6.42 \cdot 10^{-8} & +2.19 \cdot 10^{-7} \end{bmatrix}.$$

(25.12)

- The relative errors of the directional gradient and Hessian of the ray velocity magnitude are,

$$E(\nabla_{\mathbf{r}} v_{\text{ray}}) = \begin{bmatrix} +3.71 \cdot 10^{-10} & -2.19 \cdot 10^{-11} & +1.29 \cdot 10^{-10} \end{bmatrix} , \qquad (25.13)$$

$$E(\nabla_{\mathbf{r}} \nabla_{\mathbf{r}} v_{\text{ray}}) = \begin{bmatrix} -2.25 \cdot 10^{-4} & +5.78 \cdot 10^{-4} & +6.02 \cdot 10^{-6} \\ +5.78 \cdot 10^{-4} & +3.75 \cdot 10^{-5} & +3.20 \cdot 10^{-6} \\ +6.02 \cdot 10^{-6} & +3.20 \cdot 10^{-6} & +2.17 \cdot 10^{-6} \end{bmatrix} . \qquad (25.14)$$

- The relative errors of the mixed Hessian of the ray velocity magnitude are,

$$E(\nabla_{\mathbf{x}} \nabla_{\mathbf{r}} v_{\text{ray}}) = \begin{bmatrix} +5.91 \cdot 10^{-7} & +3.90 \cdot 10^{-7} & +4.79 \cdot 10^{-7} \\ -1.39 \cdot 10^{-6} & -4.90 \cdot 10^{-8} & -6.24 \cdot 10^{-7} \\ -6.03 \cdot 10^{-7} & +8.18 \cdot 10^{-8} & -1.96 \cdot 10^{-7} \end{bmatrix} . \qquad (25.15)$$

- The relative errors of the ray velocity gradients wrt the material and geometric properties of the model are,

$$E(\nabla_{\mathbf{c}} v_{\text{ray}}) = \begin{bmatrix} +3.68 \cdot 10^{-13} & +4.89 \cdot 10^{-7} & +8.85 \cdot 10^{-9} & +7.36 \cdot 10^{-9} \end{bmatrix} , \qquad (25.16)$$

$$E(\nabla_{\mathbf{a}} v_{\text{ray}}) = \begin{bmatrix} +4.96 \cdot 10^{-9} & -8.53 \cdot 10^{-9} \end{bmatrix} . \qquad (25.17)$$

- The relative errors of the ray velocity Hessians wrt the material, mixed, and geometric properties of the model are,

$$E(\nabla_{\mathbf{c}} \nabla_{\mathbf{c}} v_{\text{ray}}) = \begin{bmatrix} - & +4.89 \cdot 10^{-7} & +1.05 \cdot 10^{-8} & +5.20 \cdot 10^{-9} \\ +4.89 \cdot 10^{-7} & +1.10 \cdot 10^{-6} & +2.20 \cdot 10^{-6} & +2.28 \cdot 10^{-6} \\ +1.05 \cdot 10^{-8} & +2.20 \cdot 10^{-6} & +2.58 \cdot 10^{-8} & +7.67 \cdot 10^{-8} \\ +5.20 \cdot 10^{-9} & +2.28 \cdot 10^{-6} & +7.67 \cdot 10^{-8} & +8.14 \cdot 10^{-8} \end{bmatrix} , \qquad (25.18)$$

$$E(\nabla_{\mathbf{a}} \nabla_{\mathbf{c}} v_{\text{ray}}) = \begin{bmatrix} +4.86 \cdot 10^{-9} & +7.44 \cdot 10^{-7} & +4.08 \cdot 10^{-8} & -5.34 \cdot 10^{-9} \\ -6.37 \cdot 10^{-9} & +7.35 \cdot 10^{-7} & -7.26 \cdot 10^{-8} & +2.23 \cdot 10^{-7} \end{bmatrix} , \qquad (25.19)$$

$$E(\nabla_{\mathbf{a}} \nabla_{\mathbf{a}} v_{\text{ray}}) = \begin{bmatrix} +1.98 \cdot 10^{-7} & -1.45 \cdot 10^{-8} \\ -1.45 \cdot 10^{-8} & -4.70 \cdot 10^{-7} \end{bmatrix} . \qquad (25.20)$$

# 26. MODEL 2. COMPUTING RAY VELOCITY DERIVATIVES FOR SH WAVES

Recall that the spatial gradients and Hessians in Tables 10 and 11, respectively, are given for the axial compressional velocity $v_P$ and the shear velocity factor $f$ (rather than axial shear velocity $v_S = v_P\sqrt{1-f}$). Applying equation set 19.1, we compute the spatial gradient and Hessian of the axial shear velocity, $\nabla_\mathbf{x} v_S$ and $\nabla_\mathbf{x}\nabla_\mathbf{x} v_S$. We then use a normalization factor of $v_S = 1.5\,\text{km/s}$ (the values are taken from Table 10), resulting in,

$$\overline{\nabla_\mathbf{x} v_S} = \begin{bmatrix} -1.516\cdot 10^{-1} & +2.2465\cdot 10^{-1} & -1.057\cdot 10^{-1} \end{bmatrix} \text{km}^{-1} \quad, \tag{26.1}$$

$$\overline{\nabla_\mathbf{x}\nabla_\mathbf{x} v_S} = \begin{bmatrix} -4.0256750\cdot 10^{-2} & +1.5225160\cdot 10^{-2} & -1.6906624\cdot 10^{-1} \\ +1.5225160\cdot 10^{-2} & +7.4831727\cdot 10^{-2} & -1.1606488\cdot 10^{-1} \\ -1.6906624\cdot 10^{-1} & -1.1606488\cdot 10^{-1} & +1.0030291\cdot 10^{-1} \end{bmatrix} \text{km}^{-2} \quad. \tag{26.2}$$

As already mentioned, unlike the case of qP-qSV waves, the slowness inversion for SH waves is performed analytically: The slowness vector components are computed with equation A9; the magnitudes of the phase and ray velocities are computed with equation A11 (see results in Table 12), $v_\text{ray} = 1.0622130\,\text{km/s}$.

In this section, we provide the ray velocity derivatives, computed analytically for SH wave, and validate them numerically with the finite-difference approximations.

- The spatial gradient and Hessian of the ray velocity magnitude are,

$$\nabla_\mathbf{x} v_\text{ray} = \begin{bmatrix} -1.4799718\cdot 10^{-1} & +1.7194039\cdot 10^{-1} & -1.4747621\cdot 10^{-1} \end{bmatrix} \text{s}^{-1} \quad, \tag{26.3}$$

$$\nabla_{\mathbf{x}}\nabla_{\mathbf{x}}v_{\text{ray}} = \begin{bmatrix} -5.7271685 \cdot 10^{-2} & +3.5328847 \cdot 10^{-2} & -1.5749660 \cdot 10^{-1} \\ +3.5328847 \cdot 10^{-2} & +2.6555509 \cdot 10^{-3} & -1.6012227 \cdot 10^{-1} \\ -1.5749660 \cdot 10^{-1} & -1.6012227 \cdot 10^{-1} & -6.2688069 \cdot 10^{-3} \end{bmatrix} (\text{km} \cdot \text{s})^{-1} \quad . \quad (26.4)$$

- The directional gradient and Hessian of the ray velocity magnitude are,

$$\nabla_{\mathbf{r}}v_{\text{ray}} = \begin{bmatrix} -1.3173817 \cdot 10^{-2} & -2.4947343 \cdot 10^{-2} & -2.919444 \cdot 10^{-2} \end{bmatrix} \text{km/s} \quad , \quad (26.3)$$

$$\nabla_{\mathbf{r}}\nabla_{\mathbf{r}}v_{\text{ray}} = \begin{bmatrix} +6.9153621 \cdot 10^{-2} & +1.2790294 \cdot 10^{-1} & +1.3130920 \cdot 10^{-1} \\ +1.2790294 \cdot 10^{-1} & +2.2326886 \cdot 10^{-1} & +2.3242689 \cdot 10^{-1} \\ +1.3130920 \cdot 10^{-1} & +2.3242689 \cdot 10^{-1} & +2.3555783 \cdot 10^{-1} \end{bmatrix} \text{km/s} \quad . \quad (26.6)$$

- The mixed Hessian of the ray velocity magnitude is,

$$\nabla_{\mathbf{x}}\nabla_{\mathbf{r}}v_{\text{ray}} = \begin{bmatrix} -4.0882908 \cdot 10^{-3} & -1.1683970 \cdot 10^{-2} & -1.1895977 \cdot 10^{-2} \\ +1.3905876 \cdot 10^{-2} & +3.2945978 \cdot 10^{-2} & +3.5573825 \cdot 10^{-2} \\ -2.0344076 \cdot 10^{-2} & -4.0873696 \cdot 10^{-2} & -4.6773620 \cdot 10^{-2} \end{bmatrix} \text{s}^{-1} \quad . \quad (26.7)$$

- The ray velocity gradient wrt the model parameters, $\mathbf{m} = \{v_S, \gamma, \theta_{\text{ax}}, \psi_{\text{ax}}\}$, is,

$$\nabla_{\mathbf{m}}v_{\text{ray}} = \begin{bmatrix} +7.0814201 \cdot 10^{-1} & +2.118201 & -3.8685879 \cdot 10^{-2} & +7.8865746 \cdot 10^{-3} \end{bmatrix} \quad . \quad (26.8)$$

- The ray velocity Hessian wrt the model parameters is,

$$\nabla_{\mathbf{m}}\nabla_{\mathbf{m}}v_{\text{ray}} = \begin{bmatrix} +5.5511151 \cdot 10^{-16} & +1.4121344 & -2.5790586 \cdot 10^{-2} & +5.2577164 \cdot 10^{-3} \\ +1.4121344 & -4.2736428 & +7.8051907 \cdot 10^{-2} & -1.5911806 \cdot 10^{-2} \\ -2.5790586 \cdot 10^{-2} & +7.8051907 \cdot 10^{-2} & +4.8194597 \cdot 10^{-1} & -8.9427675 \cdot 10^{-2} \\ +5.2577164 \cdot 10^{-3} & -1.5911806 \cdot 10^{-2} & -8.9427675 \cdot 10^{-2} & +3.7910430 \cdot 10^{-2} \end{bmatrix} .$$

(26.9)

Validation

The relative errors of the finite-difference approximations are listed below.

- The relative errors of the spatial gradient and Hessian of the ray velocity magnitude are,

$$E(\nabla_{\mathbf{x}} v_{\text{ray}}) = \begin{bmatrix} -6.81 \cdot 10^{-11} & -4.98 \cdot 10^{-10} & -1.46 \cdot 10^{-10} \end{bmatrix} \quad , \tag{26.10}$$

$$E(\nabla_{\mathbf{x}} \nabla_{\mathbf{x}} v_{\text{ray}}) = \begin{bmatrix} +2.28 \cdot 10^{-7} & +2.82 \cdot 10^{-7} & +3.78 \cdot 10^{-8} \\ +2.82 \cdot 10^{-7} & -3.18 \cdot 10^{-6} & -1.32 \cdot 10^{-7} \\ +3.78 \cdot 10^{-8} & -1.32 \cdot 10^{-7} & +4.77 \cdot 10^{-7} \end{bmatrix} \quad . \tag{26.11}$$

- The relative errors of the directional gradient and Hessian of the ray velocity magnitude are,

$$E(\nabla_{\mathbf{r}} v_{\text{ray}}) = \begin{bmatrix} +9.63 \cdot 10^{-11} & -5.46 \cdot 10^{-11} & +3.55 \cdot 10^{-12} \end{bmatrix} \quad , \tag{26.12}$$

$$E(\nabla_{\mathbf{r}} \nabla_{\mathbf{r}} v_{\text{ray}}) = \begin{bmatrix} -1.68 \cdot 10^{-5} & +2.26 \cdot 10^{-5} & +8.30 \cdot 10^{-6} \\ +2.26 \cdot 10^{-5} & -1.55 \cdot 10^{-5} & -7.07 \cdot 10^{-8} \\ +8.30 \cdot 10^{-6} & -7.07 \cdot 10^{-8} & +3.90 \cdot 10^{-6} \end{bmatrix} \quad . \tag{26.13}$$

- The relative errors of the mixed Hessian of the ray velocity magnitude are,

$$E(\nabla_{\mathbf{x}} \nabla_{\mathbf{r}} v_{\text{ray}}) = \begin{bmatrix} -5.69 \cdot 10^{-6} & +5.47 \cdot 10^{-6} & +2.19 \cdot 10^{-6} \\ +2.51 \cdot 10^{-6} & +1.32 \cdot 10^{-6} & +1.72 \cdot 10^{-6} \\ +1.67 \cdot 10^{-6} & +1.57 \cdot 10^{-6} & +1.61 \cdot 10^{-6} \end{bmatrix} \quad . \tag{26.14}$$

- The relative errors of the ray velocity gradient wrt the model parameters are,

$$E(\nabla_{\mathbf{m}} v_{\text{ray}}) = \begin{bmatrix} +4.62 \cdot 10^{-13} & +1.26 \cdot 10^{-9} & +5.66 \cdot 10^{-11} & +3.42 \cdot 10^{-8} \end{bmatrix} \quad . \tag{26.15}$$

- The relative errors of the ray velocity Hessian wrt the model parameters are,

$$E\left(\nabla_{\mathbf{m}}\nabla_{\mathbf{m}}v_{\text{ray}}\right) = \begin{bmatrix} - & +1.65 \cdot 10^{-8} & +1.82 \cdot 10^{-7} & -3.01 \cdot 10^{-7} \\ +1.65 \cdot 10^{-8} & +8.24 \cdot 10^{-8} & +8.22 \cdot 10^{-8} & -2.03 \cdot 10^{-6} \\ +1.82 \cdot 10^{-7} & +8.22 \cdot 10^{-8} & -5.20 \cdot 10^{-8} & +6.10 \cdot 10^{-8} \\ -3.01 \cdot 10^{-7} & -2.03 \cdot 10^{-6} & +6.10 \cdot 10^{-8} & -4.65 \cdot 10^{-7} \end{bmatrix}. \quad (26.16)$$

## 27. MODEL 2. TTI AS A PARTICULAR CASE OF GENERAL ANISOTROPY

This section is similar to section 19, but model 2 is applied instead of model 1. We show that the same results obtained in this part (Part II) for the gradients and Hessians of the ray velocity (for the given inhomogeneous polar anisotropic model parameters, their spatial gradients and Hessians (Tables 10 and 11), and the given ray direction), can be obtained by applying the theory presented in Part I for general anisotropy. For this, as mentioned in Section 18, we first have to convert (transform) the polar anisotropic model parameters and their spatial gradients and Hessians, into "equivalent model parameters" described with the full set of the twenty-one elastic properties and their spatial gradients and Hessians. We perform the analysis for quasi-compressional waves only; the technique for the other wave types does not differ.

The computed crystal stiffness components (equation set 2.7b), and their relative gradients (equation set 19.4) are presented in Table 13, and the relative Hessians (equation set 19.5) in Tables 14 and 15, where the normalization factors are the reciprocals of the crystal stiffness components, $\hat{C}_{ij}^{-1}$.

Table 13. "Crystal" stiffness components of the polar anisotropy model 2 and their spatial gradients

| # | $\hat{C}_{ij}$ | Value, $(km/s)^2$ | Relative gradient, $km^{-1}$ (in global frame) | | |
|---|---|---|---|---|---|
| | | | $x_1$ | $x_2$ | $x_3$ |
| 1 | $\hat{C}_{11}$ | 6.3 | $-1.2658571 \cdot 10^{-1}$ | $+1.2478571 \cdot 10^{-1}$ | $+1.12 \cdot 10^{-2}$ |
| 2 | $\hat{C}_{13}$ | 6.8060753 | $+4.8161056 \cdot 10^{-3}$ | $-1.8491690 \cdot 10^{-1}$ | $+2.5037283 \cdot 10^{-1}$ |
| 3 | $\hat{C}_{33}$ | 9 | $-1.082 \cdot 10^{-1}$ | $+7.7 \cdot 10^{-2}$ | $+5.08 \cdot 10^{-2}$ |
| 4 | $\hat{C}_{44}$ | 2.25 | $-3.032 \cdot 10^{-1}$ | $+4.493 \cdot 10^{-1}$ | $-2.114 \cdot 10^{-1}$ |
| 5 | $\hat{C}_{66}$ | 1.125 | $-2.82 \cdot 10^{-1}$ | $+3.323 \cdot 10^{-1}$ | $-2.877 \cdot 10^{-1}$ |

Table 14. Spatial Hessians of "crystal" stiffness tensor (diagonal components), model 2

| Component | | Relative Hessian, $km^{-2}$ (in global frame) | | |
|---|---|---|---|---|
| # | $m_i$ | $x_1 x_1$ | $x_2 x_2$ | $x_3 x_3$ |
| 1 | $\hat{C}_{11}$ | $+3.8943143 \cdot 10^{-4}$ | $+1.1988064 \cdot 10^{-1}$ | $+5.2366960 \cdot 10^{-2}$ |
| 2 | $\hat{C}_{13}$ | $-1.4584210 \cdot 10^{-2}$ | $2.7041718 \cdot 10^{-3}$ | $-9.9483153 \cdot 10^{-2}$ |
| 3 | $\hat{C}_{33}$ | $-2.0946380 \cdot 10^{-2}$ | $+9.8764500 \cdot 10^{-2}$ | $+4.3490320 \cdot 10^{-2}$ |
| 4 | $\hat{C}_{44}$ | $-3.4548380 \cdot 10^{-2}$ | $+2.505987 \cdot 10^{-1}$ | $+2.2295080 \cdot 10^{-1}$ |
| 5 | $\hat{C}_{66}$ | $-7.1804060 \cdot 10^{-2}$ | $+5.3162500 \cdot 10^{-2}$ | $+3.0591044 \cdot 10^{-1}$ |

Table 15. Spatial Hessians of "crystal" stiffness tensor (off-diagonal components), model 2

| Component | | Relative Hessian, $\text{km}^{-2}$ (in global frame) | | |
|---|---|---|---|---|
| # | $m_i$ | $x_1 x_2$ | $x_2 x_3$ | $x_1 x_3$ |
| 1 | $\hat{C}_{11}$ | $+6.8348186 \cdot 10^{-2}$ | $-1.3453731 \cdot 10^{-1}$ | $-1.1946898 \cdot 10^{-1}$ |
| 2 | $\hat{C}_{13}$ | $+1.1867264 \cdot 10^{-1}$ | $+1.4634276 \cdot 10^{-3}$ | $-2.9130849 \cdot 10^{-3}$ |
| 3 | $\hat{C}_{33}$ | $+4.7034300 \cdot 10^{-2}$ | $-1.1064420 \cdot 10^{-1}$ | $-1.1694828 \cdot 10^{-1}$ |
| 4 | $\hat{C}_{44}$ | $-3.7663560 \cdot 10^{-2}$ | $-2.7962076 \cdot 10^{-1}$ | $-3.0608424 \cdot 10^{-1}$ |
| 5 | $\hat{C}_{66}$ | $+1.7836000 \cdot 10^{-2}$ | $-3.3786855 \cdot 10^{-1}$ | $-2.5853176 \cdot 10^{-1}$ |

We apply equations 19.11, 19.12, and 19.13 to compute the global stiffness components, their gradients, and their Hessians, respectively. The results are listed in Tables 16, 17, 18, where the units of the (density-normalized) stiffness components are $(\text{km}/\text{s})^2$, while the gradients and Hessians are presented in the "relative form", with the units $\text{km}^{-1}$ and $\text{km}^{-2}$, respectively.

Table 16. Stiffness components and their spatial gradients for polar anisotropy of model 2 presented as general (triclinic) anisotropy

| Component | | Value, $(\text{km/s})^2$ | Relative gradient, $\text{km}^{-1}$ (in global frame) | | |
| --- | --- | --- | --- | --- | --- |
| # | $C_{ij}$ | | $x_1$ | $x_2$ | $x_3$ |
| 1 | $C_{11}$ | +7.0400080 | $-1.6808611 \cdot 10^{-1}$ | $+2.0186365 \cdot 10^{-1}$ | $-4.4157486 \cdot 10^{-2}$ |
| 2 | $C_{12}$ | +4.9902779 | $-1.1154183 \cdot 10^{-2}$ | $-8.845387710^{-2}$ | $+1.9111833 \cdot 10^{-1}$ |
| 3 | $C_{13}$ | +5.5535457 | $+2.2243555 \cdot 10^{-2}$ | $-2.0337929 \cdot 10^{-1}$ | $+2.9897907 \cdot 10^{-1}$ |
| 4 | $C_{14}$ | $+9.6560206 \cdot 10^{-1}$ | $+2.7196262 \cdot 10^{-1}$ | $-8.894508 \cdot 10^{-1}$ | $+6.0723304 \cdot 10^{-1}$ |
| 5 | $C_{15}$ | $+9.5322989 \cdot 10^{-1}$ | $-2.4526929 \cdot 10^{-1}$ | $+2.6212815 \cdot 10^{-1}$ | $-2.5879453 \cdot 10^{-2}$ |
| 6 | $C_{16}$ | $+7.1492242 \cdot 10^{-1}$ | $-2.5819649 \cdot 10^{-1}$ | $+3.5511935 \cdot 10^{-1}$ | $-2.0713449 \cdot 10^{-1}$ |
| 7 | $C_{22}$ | +8.8169040 | $-1.1541624 \cdot 10^{-1}$ | $+9.4781477 \cdot 10^{-2}$ | $+1.3551763 \cdot 10^{-2}$ |
| 8 | $C_{23}$ | +5.1590883 | $+3.6870921 \cdot 10^{-2}$ | $-2.3560459 \cdot 10^{-1}$ | $+3.0269800 \cdot 10^{-1}$ |
| 9 | $C_{24}$ | +1.1413445 | $-7.8136734 \cdot 10^{-2}$ | $-1.0551131 \cdot 10^{-1}$ | $+3.0960389 \cdot 10^{-1}$ |
| 10 | $C_{25}$ | $+7.0980237 \cdot 10^{-2}$ | +1.1368019 | $-3.8903093$ | +3.0239231 |
| 11 | $C_{26}$ | $+4.1611518 \cdot 10^{-1}$ | $-3.2286423 \cdot 10^{-1}$ | $+4.2662057 \cdot 10^{-1}$ | $-1.3217761 \cdot 10^{-1}$ |
| 12 | $C_{33}$ | +9.6615655 | $-1.1166724 \cdot 10^{-1}$ | $+6.4585471 \cdot 10^{-2}$ | $+7.1432849 \cdot 10^{-2}$ |
| 13 | $C_{34}$ | $+3.0664657 \cdot 10^{-1}$ | $-2.6968781 \cdot 10^{-1}$ | $+7.775563 \cdot 10^{-1}$ | $-1.0188193$ |
| 14 | $C_{35}$ | $+1.4906431 \cdot 10^{-1}$ | $-5.0148811 \cdot 10^{-1}$ | $+1.2166970$ | $-1.2793458$ |
| 15 | $C_{36}$ | $-2.5108177 \cdot 10^{-1}$ | $-5.4431533 \cdot 10^{-1}$ | $+9.2899368 \cdot 10^{-1}$ | $-1.7141214 \cdot 10^{-1}$ |
| 16 | $C_{44}$ | $+7.3088929 \cdot 10^{-1}$ | $-7.3524658 \cdot 10^{-1}$ | $+1.3953900$ | $-8.4496734 \cdot 10^{-1}$ |
| 17 | $C_{45}$ | $-5.1414160 \cdot 10^{-1}$ | $-1.9773263 \cdot 10^{-1}$ | $+9.2546307 \cdot 10^{-2}$ | $+5.1521174 \cdot 10^{-2}$ |
| 18 | $C_{46}$ | $-2.7976620 \cdot 10^{-1}$ | $-1.0538391 \cdot 10^{-1}$ | $-1.4452145 \cdot 10^{-2}$ | $-2.0860524 \cdot 10^{-1}$ |
| 19 | $C_{55}$ | +1.5386221 | $-2.5607238 \cdot 10^{-1}$ | $+3.3823298 \cdot 10^{-1}$ | $-1.5292887 \cdot 10^{-1}$ |
| 20 | $C_{56}$ | $+2.4406654 \cdot 10^{-1}$ | $-4.0914502 \cdot 10^{-2}$ | $+1.4080937 \cdot 10^{-1}$ | $+3.1499622 \cdot 10^{-1}$ |
| 21 | $C_{66}$ | +1.3962499 | $-2.8271919 \cdot 10^{-1}$ | $+3.9222835 \cdot 10^{-1}$ | $-2.6664964 \cdot 10^{-1}$ |

Table 17. Hessians of polar anisotropy stiffness components presented as general (triclinic) anisotropy: diagonal elements, model 2

| Component | | Relative Hessian, $km^{-1}$ (in global frame) | | |
|---|---|---|---|---|
| # | $C_{ij}$ | $x_1 x_1$ | $x_2 x_2$ | $x_3 x_3$ |
| 1 | $C_{11}$ | $-2.5217007 \cdot 10^{-2}$ | $+1.4355551 \cdot 10^{-1}$ | $+2.3598753 \cdot 10^{-2}$ |
| 2 | $C_{12}$ | $+7.4140822 \cdot 10^{-3}$ | $+3.2735501 \cdot 10^{-2}$ | $-1.5896812 \cdot 10^{-1}$ |
| 3 | $C_{13}$ | $+9.8657806 \cdot 10^{-3}$ | $+3.0679626 \cdot 10^{-2}$ | $-4.3377992 \cdot 10^{-2}$ |
| 4 | $C_{14}$ | $-3.0155432 \cdot 10^{-2}$ | $-3.5600743 \cdot 10^{-1}$ | $-1.4861618 \cdot 10^{-1}$ |
| 5 | $C_{15}$ | $-1.9234469 \cdot 10^{-1}$ | $-1.9144448 \cdot 10^{-1}$ | $-2.4720615 \cdot 10^{-1}$ |
| 6 | $C_{16}$ | $-1.9604585 \cdot 10^{-1}$ | $-6.9374878 \cdot 10^{-2}$ | $-2.9185101 \cdot 10^{-1}$ |
| 7 | $C_{22}$ | $-1.4154664 \cdot 10^{-2}$ | $+1.1341023 \cdot 10^{-1}$ | $+6.0838041 \cdot 10^{-4}$ |
| 8 | $C_{23}$ | $+3.9988965 \cdot 10^{-3}$ | $+2.1873359 \cdot 10^{-2}$ | $-7.1977109 \cdot 10^{-2}$ |
| 9 | $C_{24}$ | $-4.9435247 \cdot 10^{-2}$ | $+1.1743376 \cdot 10^{-2}$ | $+1.0598852 \cdot 10^{-1}$ |
| 10 | $C_{25}$ | $-1.2077583$ | $-6.2171055$ | $-1.5083418$ |
| 11 | $C_{26}$ | $-1.9919628 \cdot 10^{-1}$ | $+1.0413095 \cdot 10^{-2}$ | $-2.1866688 \cdot 10^{-1}$ |
| 12 | $C_{33}$ | $-1.8127525 \cdot 10^{-2}$ | $+8.6997135 \cdot 10^{-2}$ | $+2.7497673 \cdot 10^{-2}$ |
| 13 | $C_{34}$ | $-2.1863344 \cdot 10^{-1}$ | $+3.7060673 \cdot 10^{-1}$ | $-7.3245408 \cdot 10^{-1}$ |
| 14 | $C_{35}$ | $-2.7756207 \cdot 10^{-1}$ | $+1.0094907$ | $-2.9311318 \cdot 10^{-1}$ |
| 15 | $C_{36}$ | $+5.6595233 \cdot 10^{-2}$ | $+8.9039331 \cdot 10^{-1}$ | $+1.4935043 \cdot 10^{-1}$ |
| 16 | $C_{44}$ | $-7.1555157 \cdot 10^{-2}$ | $+4.6992388 \cdot 10^{-1}$ | $+8.8701940 \cdot 10^{-1}$ |
| 17 | $C_{45}$ | $-2.2461040 \cdot 10^{-1}$ | $-4.851339 \cdot 10^{-1}$ | $-4.4627518 \cdot 10^{-1}$ |
| 18 | $C_{46}$ | $-2.9434881 \cdot 10^{-1}$ | $-5.8251629 \cdot 10^{-1}$ | $-6.1932426 \cdot 10^{-1}$ |
| 19 | $C_{55}$ | $+3.9593281 \cdot 10^{-4}$ | $+2.0201302 \cdot 10^{-1}$ | $+3.8098269 \cdot 10^{-1}$ |
| 20 | $C_{56}$ | $+2.3572450 \cdot 10^{-1}$ | $+3.4014649 \cdot 10^{-1}$ | $+4.1660650 \cdot 10^{-1}$ |
| 21 | $C_{66}$ | $-2.6949765 \cdot 10^{-2}$ | $+2.2013427 \cdot 10^{-1}$ | $+3.0077695 \cdot 10^{-1}$ |

Table 18. Hessians of polar anisotropy stiffness components presented

as general (triclinic) anisotropy: off-diagonal elements, model 2

| Component | | Relative Hessian, $\text{km}^{-1}$ (in global frame) | | |
|---|---|---|---|---|
| # | $C_{ij}$ | $x_1 x_2$ | $x_2 x_3$ | $x_1 x_3$ |
| 1 | $C_{11}$ | $+1.4046489 \cdot 10^{-2}$ | $-1.3362002 \cdot 10^{-1}$ | $-8.9484845 \cdot 10^{-2}$ |
| 2 | $C_{12}$ | $+1.1143068 \cdot 10^{-1}$ | $+4.4173639 \cdot 10^{-2}$ | $-3.4899066 \cdot 10^{-2}$ |
| 3 | $C_{13}$ | $+1.5315024 \cdot 10^{-1}$ | $-2.6299025 \cdot 10^{-2}$ | $-9.7131122 \cdot 10^{-3}$ |
| 4 | $C_{14}$ | $+3.0689364 \cdot 10^{-1}$ | $+1.0652913 \cdot 10^{-1}$ | $-2.8941059 \cdot 10^{-3}$ |
| 5 | $C_{15}$ | $+6.7231472 \cdot 10^{-2}$ | $+1.7425876 \cdot 10^{-1}$ | $-1.570753 \cdot 10^{-1}$ |
| 6 | $C_{16}$ | $-5.7735958 \cdot 10^{-2}$ | $+9.7091642 \cdot 10^{-2}$ | $-1.1270939 \cdot 10^{-1}$ |
| 7 | $C_{22}$ | $+4.2858194 \cdot 10^{-2}$ | $-1.1096614 \cdot 10^{-1}$ | $-1.2506246 \cdot 10^{-1}$ |
| 8 | $C_{23}$ | $+1.3141530 \cdot 10^{-1}$ | $-8.8555138 \cdot 10^{-3}$ | $+4.2471284 \cdot 10^{-2}$ |
| 9 | $C_{24}$ | $+1.2492918 \cdot 10^{-1}$ | $-7.3212741 \cdot 10^{-2}$ | $-1.3150462 \cdot 10^{-1}$ |
| 10 | $C_{25}$ | $+2.7641566$ | $+3.2199762$ | $+6.4081316 \cdot 10^{-1}$ |
| 11 | $C_{26}$ | $-1.1490330 \cdot 10^{-1}$ | $+6.6302601 \cdot 10^{-2}$ | $-3.2183826 \cdot 10^{-2}$ |
| 12 | $C_{33}$ | $+6.7305519 \cdot 10^{-2}$ | $-9.9799979 \cdot 10^{-2}$ | $-1.2714903 \cdot 10^{-1}$ |
| 13 | $C_{34}$ | $-7.8583562 \cdot 10^{-1}$ | $+5.1162243 \cdot 10^{-2}$ | $+8.5146917 \cdot 10^{-2}$ |
| 14 | $C_{35}$ | $-1.1825760$ | $-5.3959285 \cdot 10^{-1}$ | $+4.8718245 \cdot 10^{-1}$ |
| 15 | $C_{36}$ | $-1.9031972 \cdot 10^{-1}$ | $-3.7261288 \cdot 10^{-1}$ | $-4.0171345 \cdot 10^{-1}$ |
| 16 | $C_{44}$ | $-1.7419612 \cdot 10^{-1}$ | $-6.8933524 \cdot 10^{-1}$ | $-5.9262649 \cdot 10^{-1}$ |
| 17 | $C_{45}$ | $+1.9227142 \cdot 10^{-1}$ | $+3.5832830 \cdot 10^{-1}$ | $-1.2957257 \cdot 10^{-1}$ |
| 18 | $C_{46}$ | $+9.1433584 \cdot 10^{-2}$ | $+3.0765028 \cdot 10^{-1}$ | $+4.1552149 \cdot 10^{-2}$ |
| 19 | $C_{55}$ | $+5.3613438 \cdot 10^{-2}$ | $-3.4570020 \cdot 10^{-2}$ | $-3.8139665 \cdot 10^{-1}$ |
| 20 | $C_{56}$ | $+3.9373203 \cdot 10^{-1}$ | $-1.8517986 \cdot 10^{-1}$ | $-9.8498727 \cdot 10^{-1}$ |
| 21 | $C_{66}$ | $-1.8525098 \cdot 10^{-2}$ | $-3.6276162 \cdot 10^{-1}$ | $-3.1987921 \cdot 10^{-1}$ |

The data in Tables 16, 17 and 18 make it possible to compute the ray velocity and its derivatives for a general triclinic medium applying the workflow and technique of Part I.

The gradients and Hessians of the reference Hamiltonian

- The slowness gradient and Hessian are,

$$H_{\mathbf{p}}^{\bar{\tau}}(P) = [+1.5166714 \quad +1.2780939 \quad -1.7765505] \text{ km/s} \quad , \quad (27.1)$$

$$H_{\mathbf{pp}}^{\bar{\tau}}(P) = \begin{bmatrix} +8.2513733 \cdot 10^{-1} & -3.5936251 & +1.1136010 \cdot 10^{+1} \\ -3.5936251 & +6.7229251 & +1.4142784 \cdot 10^{+1} \\ +1.1136010 \cdot 10^{+1} & +1.4142784 \cdot 10^{+1} & +8.9619569 \end{bmatrix} (\text{km/s})^2 \quad . \quad (27.2)$$

- The spatial gradient and Hessian are,

$$H_{\mathbf{x}}^{\bar{\tau}}(P) = [-6.4413638 \cdot 10^{-2} \quad +5.9193417 \cdot 10^{-2} \quad +1.3477602 \cdot 10^{-2}] \text{ km}^{-1} \quad , \quad (27.3)$$

$$H_{\mathbf{xx}}^{\bar{\tau}}(P) = \begin{bmatrix} -2.3359340 \cdot 10^{-2} & +6.2310053 \cdot 10^{-2} & -5.9045123 \cdot 10^{-2} \\ +6.2310053 \cdot 10^{-2} & +5.8444300 \cdot 10^{-2} & -1.0498619 \cdot 10^{-1} \\ -1.1994543 \cdot 10^{-2} & -1.0498619 \cdot 10^{-1} & +8.2031297 \cdot 10^{-2} \end{bmatrix} \text{ km}^{-2} \quad . \quad (27.4)$$

- The mixed Hessian is,

$$H_{\mathbf{xp}}^{\bar{\tau}}(P) = \begin{bmatrix} +2.4180773 \cdot 10^{-1} & +1.0552098 \cdot 10^{-1} & -5.6442094 \cdot 10^{-1} \\ -2.5525685 \cdot 10^{-1} & +6.0778214 \cdot 10^{-2} & +1.1286547 \\ +5.0489656 \cdot 10^{-2} & -2.6507915 \cdot 10^{-1} & -1.0294983 \end{bmatrix} \text{ km}^{-2} \quad . \quad (27.5)$$

The gradients and Hessians of the arclength-related Hamiltonian

- The slowness gradient and Hessian are,

$$H_{\mathbf{p}}(P) = [0.5696 \quad 0.48 \quad -0.6672] = \mathbf{r} \quad , \quad (27.6)$$

$$H_{\mathbf{pp}}(P) = \begin{bmatrix} +4.0256052 & +1.9821015 & +4.7604599 \cdot 10^{-1} \\ +1.9821015 & +5.5014458 & +1.9534099 \\ +4.7604599 \cdot 10^{-1} & +1.9534099 & +6.9500392 \end{bmatrix} \text{km/s} \quad . \quad (27.7)$$

- The inverse of this matrix is,

$$H_{\mathbf{pp}}^{-1}(P) = \begin{bmatrix} +3.0273776 \cdot 10^{-1} & -1.1298551 \cdot 10^{-1} & +1.1020071 \cdot 10^{-2} \\ -1.1298551 \cdot 10^{-1} & +2.4408944 \cdot 10^{-1} & -6.0865906 \cdot 10^{-2} \\ +1.1020071 \cdot 10^{-2} & -6.0865906 \cdot 10^{-2} & +1.6023651 \cdot 10^{-1} \end{bmatrix} \text{s/km} \quad . \quad (27.8)$$

- The spatial gradient and Hessian are,

$$H_{\mathbf{x}}(P) = \begin{bmatrix} -2.4191138 \cdot 10^{-2} & +2.2230636^{-2} & +5.0616382 \cdot 10^{-3} \end{bmatrix} \text{s/km}^2 \quad , \quad (27.9)$$

$$H_{\mathbf{xx}}(P) = \begin{bmatrix} +1.4928374 \cdot 10^{-3} & +1.0786867 \cdot 10^{-2} & -1.7903156 \cdot 10^{-2} \\ +1.0786867 \cdot 10^{-2} & +3.6464051 \cdot 10^{-2} & -4.2688646 \cdot 10^{-2} \\ -1.7903156 \cdot 10^{-2} & -4.2688646 \cdot 10^{-2} & +2.1351968 \cdot 10^{-1} \end{bmatrix} \text{s/km}^3 \quad . \quad (27.10)$$

The mixed Hessian is,

$$H_{\mathbf{xp}}(P) = \begin{bmatrix} -1.0894744 \cdot 10^{-1} & -1.3722342 \cdot 10^{-1} & -5.4292379 \cdot 10^{-3} \\ +1.6259604 \cdot 10^{-1} & +2.4845444 \cdot 10^{-1} & +1.4635067 \cdot 10^{-1} \\ -9.0398797 \cdot 10^{-2} & -1.8992910 \cdot 10^{-1} & -2.5279577 \cdot 10^{-1} \end{bmatrix} \text{s/km}^3 \quad . \quad (27.11)$$

Finally, we compute the two gradients, $\nabla_{\mathbf{x}} \mathbf{v}_{\text{ray}}(\mathbf{x},\mathbf{r})$, $\nabla_{\mathbf{r}} \mathbf{v}_{\text{ray}}(\mathbf{x},\mathbf{r})$, and the three Hessians, $\nabla_{\mathbf{x}} \nabla_{\mathbf{x}} \mathbf{v}_{\text{ray}}(\mathbf{x},\mathbf{r})$, $\nabla_{\mathbf{r}} \nabla_{\mathbf{r}} \mathbf{v}_{\text{ray}}(\mathbf{x},\mathbf{r})$, $\nabla_{\mathbf{x}} \nabla_{\mathbf{r}} \mathbf{v}_{\text{ray}}(\mathbf{x},\mathbf{r})$ of the ray velocity magnitude. As expected, the resulted ray velocity derivatives fully coincide with the derivatives listed in equations 14.40-14.44 for compressional waves, computed directly for polar anisotropic media (without its conversion to general anisotropy). We define the discrepancies $E$ between the two methods as,

$$E = \frac{T - P}{P} \quad , \quad (27.12)$$

where $T$ and $P$ stand for the triclinic and polar anisotropies, respectively. These discrepancies are listed below.

- For the inverted slowness vector,

$$E(\mathbf{p}) = \begin{bmatrix} 0 & +5.43 \cdot 10^{-16} & -3.34 \cdot 10^{-16} \end{bmatrix} \quad , \tag{27.13}$$

- For the ray velocity magnitude,

$$E(v_{\text{ray}}) = -1.67 \cdot 10^{-15} \quad . \tag{27.14}$$

- For the spatial gradient and Hessian,

$$E(\nabla_{\mathbf{x}} v_{\text{ray}}) = \begin{bmatrix} +5.44 \cdot 10^{-16} & +2.17 \cdot 10^{-15} & -6.07 \cdot 10^{-15} \end{bmatrix} \quad , \tag{27.15}$$

$$E(\nabla_{\mathbf{x}} \nabla_{\mathbf{x}} v_{\text{ray}}) = \begin{bmatrix} +3.60 \cdot 10^{-14} & -1.01 \cdot 10^{-15} & +5.86 \cdot 10^{-16} \\ -1.01 \cdot 10^{-15} & 0 & +5.63 \cdot 10^{-16} \\ +5.86 \cdot 10^{-16} & +5.63 \cdot 10^{-16} & +1.07 \cdot 10^{-15} \end{bmatrix} \quad . \tag{27.16}$$

- For the directional gradient and Hessian,

$$E(\nabla_{\mathbf{r}} v_{\text{ray}}) = \begin{bmatrix} -1.79 \cdot 10^{-14} & +1.17 \cdot 10^{-14} & -1.12 \cdot 10^{-15} \end{bmatrix} \quad , \tag{27.17}$$

$$E(\nabla_{\mathbf{r}} \nabla_{\mathbf{r}} v_{\text{ray}}) = \begin{bmatrix} -3.76 \cdot 10^{-15} & -1.14 \cdot 10^{-15} & +2.97 \cdot 10^{-15} \\ -6.35 \cdot 10^{-16} & +1.46 \cdot 10^{-16} & -1.40 \cdot 10^{-15} \\ +1.24 \cdot 10^{-15} & -2.10 \cdot 10^{-15} & +2.48 \cdot 10^{-15} \end{bmatrix} \quad . \tag{27.18}$$

- For the mixed Hessian,

$$E(\nabla_{\mathbf{x}} \nabla_{\mathbf{r}} v_{\text{ray}}) = \begin{bmatrix} -5.68 \cdot 10^{-16} & -5.77 \cdot 10^{-16} & -4.14 \cdot 10^{-16} \\ -2.59 \cdot 10^{-15} & 0 & -7.57 \cdot 10^{-16} \\ -4.29 \cdot 10^{-15} & +1.49 \cdot 10^{-15} & -6.51 \cdot 10^{-16} \end{bmatrix} \quad . \tag{27.19}$$

As expected, the discrepancies between the results of the two approaches are "almost" zero, where the worst relative discrepancy does not exceed $E_{\max} = 3.6 \cdot 10^{-14}$. Note that both methods are "almost" analytic and exact. We say "almost", because computing the ray velocity derivatives includes the slowness vector inversion, which is a numerical procedure. In addition to the round-off errors (which are the main source of the tiny discrepancies in equations 27.15-27.19), the governing equations for this inversion are formulated differently for polar and triclinic anisotropies.

## 28. CONCLUSIONS

In this paper (Part II) we apply the theory and workflow, presented in Part I of this study for computing analytically the gradients and Hessians of the ray velocity magnitudes in general anisotropic media, specifically for polar anisotropic (TTI) media. The derivatives are formulated and computed wrt the ray locations and directions and wrt the model (material and geometric) parameters, considering both the coupled qP-qSV and the decoupled SH waves. The particular case of acoustic approximation for qP waves is considered as well.

The polar anisotropic model presented in this part is particularly important since it is widely used in the seismic method for describing the propagation of waves/rays in structural compacted (and also unconsolidated) rock layers. A similar approach can be applied to lower symmetry materials, such as tilted orthorhombic or tilted monoclinic; however, these topics are beyond the scope of this study and will be presented in our future research.

The obtained set of ray velocity partial derivatives constructs the traveltime and amplitude sensitivity kernels (matrices) which are the corner stones in many seismic anisotropic modeling and inversion applications, in particular, two-point ray bending methods and tomographic applications. Our formulae are obtained directly in the ray (group) angle domain (unlike previous works in which the ray velocity (first) derivatives are formulated in the phase angle domain).

The correctness of the analytic ray velocity derivatives was tested against numerical finite differences derivatives, for two different sets of model parameters. The first model characterizes compacted sedimentary shale/sand layering and the second one characterizes a special (extreme) case where the polar anisotropic material has a large negative anellipticity, and for the selected ray direction, the slowness inversion yields a shear wave triplication. We also analyzed the sensitivity and the correctness of the ray velocity derivatives for two approximation (simplification) cases (which are carried out in many practical applications): a) When only the spatial derivatives of the material axial compressional velocity are accounted for (ignoring all other model parameters derivatives), and b) When the spatial derivatives of all the material parameters are accounted for (ignoring only the spatial derivatives of the symmetry axis direction angles). The accuracy of the derivatives in both cases considerably decreases; the latter (case b) can be still acceptable, however, the former (case a) can yield wrong results. We further provide more insight on the feasibility and the correctness of the acoustic approximation for qP waves in polar anisotropic media, which has an excellent accuracy for the inverted slowness vector and the ray velocity magnitude, but insufficient accuracy for the ray velocity derivatives, especially for the second derivatives. Finally, we show that the same results for the ray velocity derivatives obtained directly for the polar anisotropic model, can be obtained by transforming the model parameters into the twenty-one stiffness tensor components of a general anisotropic (triclinic) medium and applying the theory described in Part I.

**ACKNOWLEDGMENT**

The authors are grateful to Emerson for the financial and technical support of this study and for the permission to publish its results. Gratitude is extended to Anne-Laure Tertois, Alexey Stovas, Yuriy Ivanov, Yury Kligerman, Michael Slawinski, Mikhail Kochetov, and Beth Orshalimy for valuable remarks and comments that helped to improve the content and the style of this paper.

**APPENDIX A. SLOWNESS VECTOR INVERSION FOR POLAR ANISOTROPY**

In this appendix we briefly review the existing studies on the slowness inversion for polar anisotropy, in particular, the approach proposed by Ravve and Koren (2021b) which is used in the present work. Vavryčuk (2006) suggested an inversion approach, based on the cofactors of the difference between the Christoffel matrix and the identity matrix, $\mathbf{\Gamma} - \mathbf{I}$, to compute the slowness vector in anisotropic elastic media characterized by the following symmetries: general anisotropy, orthorhombic, and transverse isotropy with a vertical axis of symmetry (VTI). The Hamiltonian in this approach is the unit eigenvalue of the Christoffel matrix (i.e., the solution of the Christoffel equation, rather than the equation itself). This method leads to a set of three polynomial equations of degree six, which means a higher algebraic complexity, $6 \times 6 \times 6 = 216$, than the classical approach suggested and used by Musgrave, 1954a, 1954b, 1970; Fedorov, 1968; Helbig, 1994; Grechka, 2017, with the algebraic complexity $5 \times 5 \times 6 = 150$. In a particular case of coupled qP-qSV waves in polar anisotropy, and under a commonly accepted assumption that the angle between the slowness and the ray direction vectors, $\mathbf{n}$ and $\mathbf{r}$, respectively, cannot exceed $90^o$, $0 < \mathbf{n} \cdot \mathbf{r} = v_{\text{phs}} / v_{\text{ray}} \leq 1$ (where $v_{\text{phs}}$ is the phase velocity magnitude), the method suggested by Vavryčuk (cited above) reduces to a set of two quartic equations (algebraic complexity 16) and a constraint that cuts off one half of the roots. For polar anisotropic media (TTI), the method has been later improved and extended by Zhang & Zhou (2018), where the axial and the normal (in the plane normal to the axial direction) slowness components were replaced by implicit functions of the phase angle between direction of the symmetry axis $\mathbf{k}$ and the phase direction $\mathbf{n}$. Song & Every (2000) derived approximate formulae for the reciprocal ray velocity magnitude ($v_{\text{ray}}^{-1} = \mathbf{p} \cdot \mathbf{r}$), for qP and qS waves in weakly anisotropic orthorhombic media vs. the ray direction $\mathbf{r}$. The approximations are arranged as a truncated series in $\mathbf{r}$, where the coefficients depend on the elastic properties of the media.

The inverse solutions for qP and SH waves in polar anisotropy are unique. However, for qSV waves, the inversion is challenging due to qSV triplications; the effect of qSV triplications was studied by Dellinger (1991), Thomsen and Dellinger (2003), Schoenberg and Daley (2003), Vavryčuk (2003), and Roganov and Stovas (2010). Stovas (2016), Xu and Stovas (2018), and Xu et al. (2021) studied triplications in orthorhombic media. Xu et al. (2020) explored triplications of converted waves in TTI media. Stovas et al. (2021) studied phase and ray velocity surfaces and triplications in elastic and acoustic models of different symmetry classes.

Ravve and Koren (2021b) proposed an alternative method to solve the slowness vector inversion in polar anisotropic media. Taking into account that the three vectors: the ray velocity, the slowness and the polar axis direction are coplanar, and applying the common constraint that the angle between the phase and ray velocity vectors does not exceed $90^\text{o}$, Ravve and Koren (2021b) reduced the slowness inversion equation set 6.1 in Part I (for general anisotropy) to a sixth-degree univariate polynomial (for polar anisotropy) with a single real root for qP waves and either one or three real roots for qSV waves, depending on the triplications (e.g., Thomsen and Dellinger, 2003; Grechka, 2013; Zhang and Zhou, 2018),

$$d_6 c^6 + d_5 c^5 + d_4 c^4 + d_3 c^3 + d_2 c^2 + d_1 c + d_o = 0 \quad , \tag{A1}$$

where the unknown parameter $c$ is related to the phase angle $\vartheta_{phs}$ between the slowness vector and the polar axis of symmetry direction $\mathbf{k}$,

$$c = \cot \vartheta_{phs} \quad . \tag{A2}$$

The coefficients of the polynomial depend on the ray angle $\vartheta_{ray}$ between the ray velocity direction $\mathbf{r}$ and the axis of symmetry direction,

$$d_o = -(1+2\varepsilon)(1-f)(f+2\varepsilon)^2 m^2$$
$$d_1 = +2(f+2\varepsilon)^2 \left[1+\varepsilon-(1+\delta)f\right] m\sqrt{1-m^2}$$
$$d_2 = -4\varepsilon^2 (1-f)\left[1-(1+2f)m^2\right]$$
$$\quad -2\varepsilon f \left\{ 2(1-f)+(f+2\delta)f - \left[4-\delta(8-10f)-3(2-f)f\right]m^2 \right\}$$
$$\quad -f\left\{(1+2\delta)f(1-2\delta-f)+\left[f+8\delta-4(2-\delta)\delta f-(1-2\delta)f^2\right]m^2\right\}$$
$$d_3 = -4f\left\{2(1+\varepsilon)(\varepsilon-2\delta)-\left[1+3\varepsilon-2\delta(2+\delta)\right]f+(1+\varepsilon)f^2\right\}m\sqrt{1-m^2}$$
$$d_4 = -2(1-f)f\left[2(2\delta-\varepsilon)+f\right]-f\left\{2\varepsilon(2-f)-\left(4\delta^2+1-f\right)f-2\delta\left[4-2(2+\varepsilon)f+f^2\right]\right\}m^2$$
$$d_5 = +2f^2 \left[1+\varepsilon-(1+\delta)f\right] m\sqrt{1-m^2}$$
$$d_6 = -(1-f)f^2 \left(1-m^2\right)$$

$$\tag{A3}$$

where $v_P(\mathbf{x}), f(\mathbf{x}), \delta(\mathbf{x})$ and $\varepsilon(\mathbf{x})$ are the medium properties, and,

$$m = \mathbf{k} \cdot \mathbf{r} = \cos \vartheta_{ray} \quad , \quad \sin \vartheta_{ray} = \sqrt{1-m^2} \quad , \quad 0 \leq \vartheta_{ray} \leq \pi \quad . \tag{A4}$$

The real and complex roots of the univariate polynomial are computed as the eigenvalues of a non-symmetric companion matrix (e.g., Johnson & Riess, 1982; Edelman & Murakami, 1995; Press et at., 2002).

Applying the acoustic approximation, we reduce the resolving equation to a fourth-degree polynomial equation,

$$c^4 - \frac{m}{\sqrt{1-m^2}}(1+2\delta)c^3 - 4(\varepsilon-\delta)c^2 - (1+2\delta)\frac{\sqrt{1-m^2}}{m}c + (1+2\varepsilon)^2 = 0 \quad , \quad (A5)$$

where only the "compressional" root is used.

After the phase angle has been established, we compute the unit vector $\mathbf{n}$ of the phase (slowness) direction. The ray and phase direction vectors and angles, and the polar axis direction vector $\mathbf{k}$ satisfy the constraint,

$$\sin\vartheta_{ray}\mathbf{n} - \sin\vartheta_{phs}\mathbf{r} = \sin(\vartheta_{ray} - \vartheta_{phs})\mathbf{k} \quad , \quad (A6)$$

which yields the phase direction,

$$\mathbf{n} = \frac{\sin(\vartheta_{ray} - \vartheta_{phs})\mathbf{k} + \sin\vartheta_{phs}\mathbf{r}}{\sin\vartheta_{ray}} \quad . \quad (A7)$$

Thus, to establish the slowness vector, we compute separately its magnitude $p$ and its direction vector $\mathbf{n}$. The magnitude is delivered by the Christoffel equation adjusted for polar anisotropy; in terms of the Thomsen (1986) parameters, it accepts the following form,

$$\left[1-f+2(\varepsilon-f\delta)\sin^2\vartheta_{phs}-2f(\varepsilon-\delta)\sin^4\vartheta_{phs}\right]v_P^4 p^4$$
$$-\left(2-f+2\varepsilon\sin^2\vartheta_{phs}\right)v_P^2 p^2+1=0 \quad , \tag{A8}$$

which is equivalent to equation 2.6.

For SH waves, an analytic solution exists where the slowness vector is an explicit function of the medium properties and the ray direction,

$$\mathbf{p}=\frac{2\gamma m\mathbf{k}+\mathbf{r}}{v_P\sqrt{1-f}\sqrt{1+2\gamma}\sqrt{1+2\gamma m^2}}=\frac{2\gamma m\mathbf{k}+\mathbf{r}}{v_S\sqrt{1+2\gamma}\sqrt{1+2\gamma m^2}} \quad , \tag{A9}$$

while the magnitudes of the ray and phase velocities read,

$$\frac{v_{phs}}{v_P}=\frac{\sqrt{1-f}\sqrt{1+2\gamma}\sqrt{1+2\gamma m^2}}{\sqrt{1+4\gamma(1+\gamma)m^2}} \quad , \quad \frac{v_{ray}}{v_P}=\frac{\sqrt{1-f}\sqrt{1+2\gamma}}{\sqrt{1+2\gamma m^2}} \quad , \tag{A10}$$

or equivalently,

$$\frac{v_{phs}}{v_S}=\frac{\sqrt{1+2\gamma}\sqrt{1+2\gamma m^2}}{\sqrt{1+4\gamma(1+\gamma)m^2}} \quad , \quad \frac{v_{ray}}{v_S}=\frac{\sqrt{1+2\gamma}}{\sqrt{1+2\gamma m^2}} \quad . \tag{A11}$$

**APPENDIX B. SIGNS OF THE HAMILTONIANS**

To check for the right sign, consider an isotropic case where the Thomsen parameters vanish; due to continuity wrt these parameters, the conclusions will be valid for anisotropic media as well. For isotropic media, equation 2.15 separates into compressional and shear factors,

$$H^{\bar{\tau}}_{\text{PSV}}(\mathbf{x},\mathbf{p}) = -\left(p^2 v_P^2 - 1\right)\left(p^2 v_S^2 - 1\right) = 0 \quad , \tag{B1}$$

where $v_P$ and $v_S$ are compressional and shear velocities of isotropic media, and $p$ is the slowness magnitude. Of course, each factor in equation B1 can be now considered separately, but let's pretend that the medium is very close to isotropic, with the infinitesimal Thomsen parameters, so we keep both multipliers as a limit case of equation 2.15 that converges to equation B1. The slowness gradient of the PSV reference Hamiltonian becomes,

$$H^{\bar{\tau}}_{\text{PSV},\mathbf{p}}(\mathbf{x},\mathbf{p}) = 2\left(v_P^2 + v_S^2 - 2p^2 v_P^2 v_S^2\right)\mathbf{p} \quad . \tag{B2}$$

The arclength-related Hamiltonian, $H_{\text{PSV}}(\mathbf{x},\mathbf{p})$, is obtained from the reference Hamiltonian, $H^{\bar{\tau}}_{\text{PSV}}(\mathbf{x},\mathbf{p})$,

$$H_{\text{PSV}}(\mathbf{x},\mathbf{p}) = \frac{H^{\bar{\tau}}_{\text{PSV}}(\mathbf{x},\mathbf{p})}{d\bar{\tau}/ds} = \frac{H^{\bar{\tau}}_{\text{PSV}}(\mathbf{x},\mathbf{p})}{\sqrt{H^{\bar{\tau}}_{\text{PSV},\mathbf{p}} \cdot H^{\bar{\tau}}_{\text{PSV},\mathbf{p}}}} = -\frac{1}{2p} \frac{\left(p^2 v_P^2 - 1\right)\left(p^2 v_S^2 - 1\right)}{v_P^2 + v_S^2 - 2p^2 v_P^2 v_S^2} \quad , \tag{B3}$$

where $d\bar{\tau}/ds$ is the metric, and the slowness gradient of the "isotropic" reference Hamiltonian is given by,

$$H_{\text{PSV},\mathbf{p}}(\mathbf{x},\mathbf{p}) = \frac{2 v_P^4 v_S^4 p^6 - v_P^2 v_S^2 \left(v_P^2 + v_S^2\right) p^4 + \left[\left(v_P^2 + v_S^2\right)^2 - 2 v_P^2 v_S^2\right] p^2 + v_P^2 + v_S^2}{2 p^3 \left(v_P^2 + v_S^2 - 2 v_P^2 v_S^2 p^2\right) \left|v_P^2 + v_S^2 - 2 v_P^2 v_S^2 p^2\right|} \mathbf{p} \quad . \tag{B4}$$

For compressional waves in isotropic media,

$$p = \frac{1}{v_P} \quad , \quad \mathbf{p} = \frac{\mathbf{r}}{v_P} \quad , \quad H_{P,\mathbf{p}} = +v_P \mathbf{p} = +\mathbf{r} \quad . \tag{B5}$$

For shear waves in isotropic media,

$$p = \frac{1}{v_S} \quad , \quad \mathbf{p} = \frac{\mathbf{r}}{v_S} \quad , \quad H_{SV,\mathbf{p}} = -v_S \mathbf{p} = -\mathbf{r} \quad . \tag{B6}$$

Compute the scaler of the reference Hamiltonian,

$$\alpha_{sc}(\bar{\tau}) = \frac{d\bar{\tau}}{d\tau} = \frac{1}{\mathbf{p} \cdot H_{PSV,\mathbf{p}}^{\bar{\tau}}} \quad , \tag{B7}$$

$$\alpha_{sc} = \frac{1}{2(v_P^2 + v_S^2 - 2p^2 v_P^2 v_S^2)p^2} \quad , \quad \alpha_{sc} = \begin{cases} +\dfrac{v_P^2}{2(v_P^2 - v_S^2)} = +\dfrac{1}{2f} & qP \\[1em] -\dfrac{v_S^2}{2(v_P^2 - v_S^2)} = -\dfrac{1-f}{2f} & qSV \end{cases} \tag{B8}$$

Thus, we see that the sign of the Hamiltonian in equation 2.15 is correct for compressional waves, but should be changed for the shear waves,

$$H_P^{\bar{\tau}} = +H_{PSV}^{\bar{\tau}} \quad , \quad H_{SV}^{\bar{\tau}} = -H_{PSV}^{\bar{\tau}} = -H_P^{\bar{\tau}} \quad , \tag{B9}$$

where $H_{PSV}^{\bar{\tau}}$ is the Hamiltonian of the coupled qP-qSV waves listed in equation 2.15.

With this correction, for both wave types, $H_{P,\mathbf{p}} = +\mathbf{r}$ and $H_{SV,\mathbf{p}} = +\mathbf{r}$, while the scaler $\alpha_{sc}$ is positive as well. Introduction of additional correction multipliers $1/(2f)$ and $(1-f)/(2f)$ for qP and qSV waves, respectively, results in $\alpha_{sc} \to 1$ when the anisotropic parameters become infinitesimal. However, this does not lead to apparent advantages, so we account only for the sign. In the further derivations, the sign for the Hamiltonian $H_{\text{PSV}}$ and all types of its derivatives will be specified for qP waves only. This sign should be changed to the opposite for qSV waves.

Next, consider the sign of the SH Hamiltonian. In the case of isotropic media, equation 2.16 converges to,

$$H_{\text{SH}}^{\bar{\tau}}(\mathbf{x},\mathbf{p}) = p^2 v_S^2 - 1 \ , \quad H_{\text{SH},\mathbf{p}}^{\bar{\tau}}(\mathbf{x},\mathbf{p}) = 2\mathbf{p} v_S^2 \quad , \tag{B10}$$

which results in,

$$H_{\text{SH}}(\mathbf{x},\mathbf{p}) = \frac{p^2 v_S^2 - 1}{2 p v_S^2} \ , \quad H_{\text{SH},\mathbf{p}}(\mathbf{x},\mathbf{p}) = \mathbf{p} v_S \quad , \tag{B11}$$

and finally leads to,

$$p = \frac{1}{v_S} \ , \quad \mathbf{p} = \frac{\mathbf{r}}{v_S} \ \rightarrow \ H_{\text{SH},\mathbf{p}} = \mathbf{r} \quad . \tag{B12}$$

Thus, the sign of the SH Hamiltonian $H_{\text{SH}}^{\bar{\tau}}$ in equation 2.16 is correct.

## APPENDIX C. KELVIN-FORM ROTATION OF STIFFNESS TENSOR

In this appendix, we recall the technique that makes it possible to rotate the coordinate axes and obtain the components of the fourth-order stiffness tensor in a new frame, using only matrix algebra. We emphasize that the reference frame is rotated, not the tensor itself; "rotating a tensor" is just a (convenient) slang. Generally, given the three Euler angles, one can compute the rotation matrix that relates vector or tensor components between the two frames. Note that different sets of Euler angles may exist. In this study, we apply the *ZYZ* scheme shown in Figure 1, with the Euler angles $\{\theta_{\text{rot}}, \psi_{\text{rot}}, \delta_{\text{rot}}\}$ (zenith, azimuth spin), where the first two angles define the orientation of the local vertical axis in the global frame, and the third angle defines an additional rotation of the local frame about this axis.

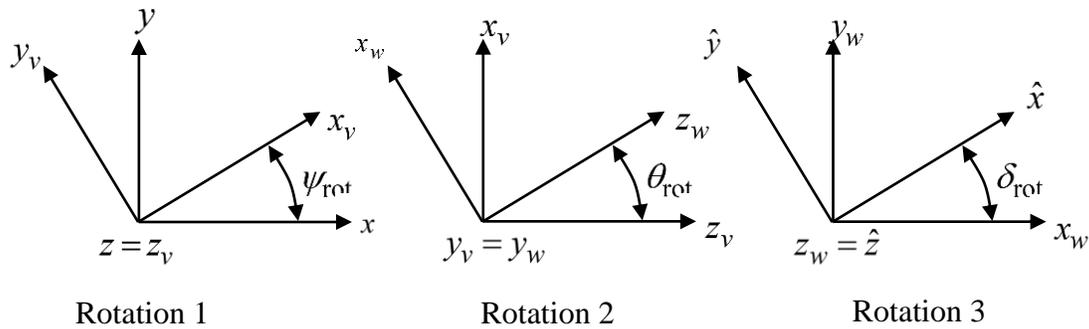

Figure 1: Euler angles and their corresponding rotations

In Figure 1, $xyz$ and $\hat{x}\hat{y}\hat{z}$ are the global and local (crystal) reference frames, respectively, while $x_v y_v z_v$ and $x_w y_w z_w$ are the two intermediate frames, where $z = z_v$, $y_v = y_w$ and $z_w = \hat{z}$. The global-to-local rotation is performed in three stages,

- About global axis $z$ for azimuth $\psi_{\text{rot}}$
- About axis $y_v$ of the first intermediate frame for zenith (tilt) $\theta_{\text{rot}}$
- About axis $z_w$ of the second intermediate frame for spin $\delta_{\text{rot}}$

The three rotations that convert the global axes to the local are described by the three matrices.

Combining these matrices, one can compute the resulting local-to-global rotation matrix $\mathbf{A}_{\text{rot}}$, whose columns are,

$$\mathbf{A}_{\text{rot},1} = \begin{bmatrix} +\cos\theta_{\text{rot}}\cos\psi_{\text{rot}}\cos\delta_{\text{rot}} - \sin\psi_{\text{rot}}\sin\delta_{\text{rot}} \\ -\cos\theta_{\text{rot}}\cos\psi_{\text{rot}}\sin\delta_{\text{rot}} - \sin\psi_{\text{rot}}\cos\delta_{\text{rot}} \\ \sin\theta_{\text{rot}}\cos\psi_{\text{rot}} \end{bmatrix}, \quad (C1)$$

$$\mathbf{A}_{\text{rot},2} = \begin{bmatrix} +\cos\theta_{\text{rot}}\sin\psi_{\text{rot}}\cos\delta_{\text{rot}} + \cos\psi_{\text{rot}}\sin\delta_{\text{rot}} \\ -\cos\theta_{\text{rot}}\sin\psi_{\text{rot}}\sin\delta_{\text{rot}} + \cos\psi_{\text{rot}}\cos\delta_{\text{rot}} \\ \sin\theta_{\text{rot}}\sin\psi_{\text{rot}} \end{bmatrix}, \quad (C2)$$

$$\mathbf{A}_{\text{rot},3} = \begin{bmatrix} -\sin\theta_{\text{rot}}\cos\delta_{\text{rot}} \\ \sin\theta_{\text{rot}}\sin\delta_{\text{rot}} \\ \cos\theta_{\text{rot}} \end{bmatrix}. \quad (C3)$$

In the case where the local axes are the crystal frame of polar anisotropy, the zenith and azimuth angles are the polar angles of the axis of symmetry, and the spin angle may be assumed zero, $\theta_{\text{rot}} = \theta_{\text{ax}}$, $\psi_{\text{rot}} = \psi_{\text{ax}}$, $\delta_{\text{rot}} = 0$, thus, the rotation matrix simplifies to,

$$\mathbf{A}_{\text{rot}} = \begin{bmatrix} +\cos\theta_{\text{ax}}\cos\psi_{\text{ax}} & +\cos\theta_{\text{ax}}\sin\psi_{\text{ax}} & -\sin\theta_{\text{ax}} \\ -\sin\psi_{\text{ax}} & +\cos\psi_{\text{ax}} & 0 \\ +\sin\theta_{\text{ax}}\cos\psi_{\text{ax}} & +\sin\theta_{\text{ax}}\sin\psi_{\text{ax}} & +\cos\theta_{\text{ax}} \end{bmatrix}. \quad (C4)$$

The classical formula that allows the frame rotation for the fourth-order tensor can be written the component-wise form,

$$C_{ijkl}^{\text{loc}} = A_{\text{rot},im} A_{\text{rot},jn} A_{\text{rot},ko} A_{\text{rot},lp} C_{mnop}^{\text{glb}}, \quad (C5)$$

where $\mathbf{A}_{\text{rot}}$ is the $3\times 3$ rotation matrix (equation C4) that converts the tensor components to the local reference frame. However, it is easier to rotate the frame for the matrix-form presentations of the stiffness tensor, applying only matrix algebra, rather than for the actual fourth-order tensor.

Rotation of a reference frame for Voigt-form stiffness tensor was originally proposed by Bond (1943). It was later "upgraded" for more suitable Kelvin-form stiffness tensor (e.g., Bona et al., 2008; Slawinski, 2015),

$$\mathbf{C} = \begin{bmatrix} C_{11} & C_{12} & C_{13} & \sqrt{2}\,C_{14} & \sqrt{2}\,C_{15} & \sqrt{2}\,C_{16} \\ C_{12} & C_{22} & C_{23} & \sqrt{2}\,C_{24} & \sqrt{2}\,C_{25} & \sqrt{2}\,C_{26} \\ C_{13} & C_{23} & C_{33} & \sqrt{2}\,C_{34} & \sqrt{2}\,C_{35} & \sqrt{2}\,C_{36} \\ \sqrt{2}\,C_{14} & \sqrt{2}\,C_{24} & \sqrt{2}\,C_{34} & 2C_{44} & 2C_{45} & 2C_{46} \\ \sqrt{2}\,C_{15} & \sqrt{2}\,C_{25} & \sqrt{2}\,C_{35} & 2C_{45} & 2C_{55} & 2C_{56} \\ \sqrt{2}\,C_{16} & \sqrt{2}\,C_{26} & \sqrt{2}\,C_{36} & 2C_{46} & 2C_{56} & 2C_{66} \end{bmatrix}, \qquad (C6)$$

and in this appendix we use the Kelvin-form rotation only. Assume the "global to local" rotation matrix $\mathbf{A}_{\text{rot}}$ has been defined or computed,

$$\mathbf{A}_{\text{rot}} = \begin{bmatrix} a_{11} & a_{12} & a_{13} \\ a_{21} & a_{22} & a_{23} \\ a_{31} & a_{32} & a_{33} \end{bmatrix}, \qquad (C7)$$

where in a general case the spin angle does not vanish. The original $6 \times 6$ Bond rotation matrix consists of four $3 \times 3$ blocks,

$$\mathbf{R}_{\text{rot}} = \begin{bmatrix} \mathbf{R}_A & \mathbf{R}_B \\ \mathbf{R}_C & \mathbf{R}_D \end{bmatrix}, \qquad (C8)$$

where each block is related to the components of the standard $3 \times 3$ rotation matrix $\mathbf{A}_{\text{rot}}$. The blocks read as follows,

$$\mathbf{R}_A = \begin{bmatrix} a_{11}^2 & a_{12}^2 & a_{13}^2 \\ a_{21}^2 & a_{22}^2 & a_{23}^2 \\ a_{31}^2 & a_{32}^2 & a_{33}^2 \end{bmatrix} \quad , \tag{C9}$$

$$\mathbf{R}_B = \begin{bmatrix} \sqrt{2}a_{12}a_{13} & \sqrt{2}a_{11}a_{13} & \sqrt{2}a_{11}a_{12} \\ \sqrt{2}a_{22}a_{23} & \sqrt{2}a_{21}a_{23} & \sqrt{2}a_{21}a_{22} \\ \sqrt{2}a_{32}a_{33} & \sqrt{2}a_{31}a_{33} & \sqrt{2}a_{31}a_{32} \end{bmatrix} \quad , \tag{C10}$$

$$\mathbf{R}_C = \begin{bmatrix} \sqrt{2}a_{21}a_{31} & \sqrt{2}a_{22}a_{32} & \sqrt{2}a_{23}a_{33} \\ \sqrt{2}a_{11}a_{31} & \sqrt{2}a_{12}a_{32} & \sqrt{2}a_{13}a_{33} \\ \sqrt{2}a_{11}a_{21} & \sqrt{2}a_{12}a_{22} & \sqrt{2}a_{13}a_{23} \end{bmatrix} \quad , \tag{C11}$$

$$\mathbf{R}_D = \begin{bmatrix} a_{22}a_{33} + a_{23}a_{32} & a_{21}a_{33} + a_{23}a_{31} & a_{21}a_{32} + a_{22}a_{31} \\ a_{12}a_{33} + a_{13}a_{32} & a_{11}a_{33} + a_{13}a_{31} & a_{11}a_{32} + a_{12}a_{31} \\ a_{12}a_{23} + a_{13}a_{22} & a_{11}a_{23} + a_{13}a_{21} & a_{11}a_{22} + a_{12}a_{21} \end{bmatrix} \quad . \tag{C12}$$

This stiffness tensor rotation reads,

$$\mathbf{C}_{\text{glb}} = \mathbf{R}_{\text{rot}}^T \mathbf{C}_{\text{loc}} \mathbf{R}_{\text{rot}} \quad , \quad \mathbf{C}_{\text{loc}} = \mathbf{R}_{\text{rot}} \mathbf{C}_{\text{glb}} \mathbf{R}_{\text{rot}}^T \quad . \tag{C13}$$

We emphasize that rotation matrices $\mathbf{A}_{\text{rot}}$ and $\mathbf{R}_{\text{rot}}$ represent "global to local" rotations. If the strain and stress tensors are presented in the vector form of factor $\sqrt{2}$ for off-diagonal components, the same $6 \times 6$ rotation matrix can be used to transform these vectors,

$$\boldsymbol{\varepsilon}_{\text{glb}} = \mathbf{R}_{\text{rot}}^T \boldsymbol{\varepsilon}_{\text{loc}} \quad , \quad \boldsymbol{\varepsilon}_{\text{loc}} = \mathbf{R}_{\text{rot}} \boldsymbol{\varepsilon}_{\text{glb}} \quad ,$$
$$\boldsymbol{\sigma}_{\text{glb}} = \mathbf{R}_{\text{rot}}^T \boldsymbol{\sigma}_{\text{loc}} \quad , \quad \boldsymbol{\sigma}_{\text{loc}} = \mathbf{R}_{\text{rot}} \boldsymbol{\sigma}_{\text{glb}} \quad ,$$
(C14)

where,

$$\boldsymbol{\varepsilon} = \begin{bmatrix} \varepsilon_{11} & \varepsilon_{22} & \varepsilon_{33} & \sqrt{2}\varepsilon_{23} & \sqrt{2}\varepsilon_{13} & \sqrt{2}\varepsilon_{12} \end{bmatrix}^T \quad ,$$
$$\boldsymbol{\sigma} = \begin{bmatrix} \sigma_{11} & \sigma_{22} & \sigma_{33} & \sqrt{2}\sigma_{23} & \sqrt{2}\sigma_{13} & \sqrt{2}\sigma_{12} \end{bmatrix}^T \quad .$$
(C15)

Of course, the Hook's law holds in either frame, and can be arranged in the matrix (rather than tensor) form,

$$\boldsymbol{\sigma} = \mathbf{C} \boldsymbol{\varepsilon} \quad , \tag{C16}$$

where $\boldsymbol{\sigma}$ and $\boldsymbol{\varepsilon}$ are vectors of stress and strain (equation C14), and $\mathbf{C}$ is the Kelvin-form stiffness matrix (equation C6).

The Kelvin mapping has a number of advantages over the Voigt mapping. The Kelvin mapping preserves the tensor character of vectors and matrices, while the Voigt mapping does not (Nagel et al., 2016). In particular, the Kelvin-form rotation matrix $\mathbf{R}_{\text{rot}}$ is orthogonal, $\mathbf{R}_{\text{rot}}^{-1} = \mathbf{R}_{\text{rot}}^T$. In addition, the 6 × 6 compliance matrix $\mathbf{S} = \mathbf{C}^{-1}$ can be rotated in the same way as the stiffness matrix. Both properties do not hold for the Voigt-form rotation.

# LIST OF TABLES

## LIST OF FIGURES